\documentclass[aps,prd,showpacs,eqsecnum,twocolumn,superscriptaddress,a4paper,nofootinbib]
{revtex4-1}
\usepackage{amsmath,amssymb,graphicx,color}
\usepackage{bm}
\usepackage{epsf}

\begin{document}


\title{Mass ejection from disks surrounding a low-mass black hole:
  Viscous neutrino-radiation hydrodynamics simulation in full general
  relativity}

\author{Sho Fujibayashi} 
\affiliation{Max Planck Institute for
  Gravitational Physics (Albert Einstein Institute), Am Mühlenberg 1,
  Potsdam-Golm 14476, Germany}

\author{Masaru Shibata} 
\affiliation{Max Planck Institute for
  Gravitational Physics (Albert Einstein Institute), Am Mühlenberg 1,
  Potsdam-Golm 14476, Germany}
\affiliation{Center for Gravitational Physics, Yukawa Institute for Theoretical
  Physics, Kyoto University, Kyoto, 606-8502, Japan}

\author{Shinya Wanajo} 
\affiliation{Max Planck Institute for
  Gravitational Physics (Albert Einstein Institute), Am Mühlenberg 1,
  Potsdam-Golm 14476, Germany}
  \affiliation{Interdisciplinary Theoretical Science (iTHES) Research Group, RIKEN,
Wako, Saitama 351-0198, Japan}

\author{Kenta Kiuchi}
\affiliation{Max Planck Institute for
  Gravitational Physics (Albert Einstein Institute), Am Mühlenberg 1,
  Potsdam-Golm 14476, Germany}
\affiliation{Center for Gravitational Physics, Yukawa Institute for Theoretical
  Physics, Kyoto University, Kyoto, 606-8502, Japan}

\author{Koutarou Kyutoku} 
\affiliation{Department of Physics, Kyoto University, Kyoto 606-8502, 
Japan}
\affiliation{Department of Particle and Nuclear Physics, the Graduate University
for Advanced Studies (Sokendai), Tsukuba 305-0801, Japan}
\affiliation{Interdisciplinary Theoretical Science (iTHES) Research Group, RIKEN,
Wako, Saitama 351-0198, Japan}
\affiliation{Center for Gravitational Physics, Yukawa Institute for Theoretical Physics, 
Kyoto University, Kyoto, 606-8502, Japan} 

\author{Yuichiro Sekiguchi} \affiliation{Department of Physics, Toho
  University, Funabashi, Chiba 274-8510, Japan}

\date{\today}
\newcommand{\beq}{\begin{equation}}
\newcommand{\eeq}{\end{equation}}
\newcommand{\beqn}{\begin{eqnarray}}
\newcommand{\eeqn}{\end{eqnarray}}
\newcommand{\pa}{\partial}
\newcommand{\vp}{\varphi}
\newcommand{\varep}{\varepsilon}
\newcommand{\ep}{\epsilon}
\newcommand{\comp}{(M/R)_\infty}
\begin{abstract}

New viscous neutrino-radiation hydrodynamics simulations are performed for
accretion disks surrounding a spinning black hole with low mass
$3M_\odot$ and dimensionless spin 0.8 or 0.6 in full general
relativity, aiming at modeling the evolution of a merger remnant of
massive binary neutron stars or low-mass black hole-neutron star
binaries. We reconfirm the following results found by previous studies
of other groups: 15--30\% of the disk mass is ejected from the system
with the average velocity of $\sim 5$--10\% of the speed of light for
the plausible profile of the disk as merger remnants.  In addition, we
find that for the not extremely high viscous coefficient case, the
neutron richness of the ejecta does not become very high, because 
weak interaction processes enhance the electron fraction during the
viscous expansion of the disk before the onset of the mass ejection, resulting in the suppression of the
lanthanide synthesis.  For high-mass disks, the viscous expansion
timescale is increased by a longer-term neutrino emission, and hence,
the electron fraction of the ejecta becomes even higher.  We also
confirm that the mass distribution of the electron fraction depends
strongly on the magnitude of the given viscous coefficient. This
demonstrates that a first-principle magnetohydrodynamics simulation is
necessary for black hole-disk systems with sufficient grid resolution
and with sufficiently long timescale (longer than seconds) to clarify
the nucleosynthesis and electromagnetic signals from them.

\end{abstract}


\maketitle

\section{Introduction}\label{sec1}

The first direct detection of gravitational waves from the final stage
of an inspiraling binary neutron star system (GW170817) by advanced
LIGO and advanced VIRGO~\cite{GW170817} was accompanied with a wide
variety of the follow-up observations of electromagnetic
counterparts~\cite{GW170817a}. This event heralded the opening of the
era of the multi-messenger astronomy composed of gravitational-wave
and electromagnetic-counterpart observations, and demonstrated that the
observation of electromagnetic signals plays a key role for
understanding the merger and subsequent mass ejection processes of
neutron-star binaries, which cannot be understood only from the 
gravitational-wave observation. 

A popular interpretation for the merger and post-merger evolution of
binary neutron stars in GW170817 is as follows~(e.g., see
Refs.~\cite{EM2017,MM2017,shibata17,Perego17}).  After the merger of a
binary neutron star, a hypermassive neutron star was formed with an
accretion disk around it. Subsequently, the hypermassive neutron star
survived for $\sim 0.1$--1\,s, and eventually collapsed to a black
hole surrounded by a disk of mass 0.1--$0.2M_\odot$, which might be
the central engine of a gamma-ray burst associated with
GW170817~\cite{GRB,radio}. At the merger and during the post-merger
stage, ejection of matter with mass $\sim 0.05M_\odot$ occurred. At
the merger, a neutron-rich material, which can synthesize heavy
$r$-process elements, was ejected in the dynamical process, while in the
post-merger stage, the mass ejection occurred from a disk (or torus)
surrounding the remnant hypermassive neutron star and the black hole
subsequently formed. The observational results suggest that the
neutron richness of the post-merger ejecta is unlikely to be very high.
A canonical interpretation for this is that the neutrino irradiation
from the hypermassive neutron star to the ejecta is strong enough to
reduce the neutron richness. However, physically well-modeled and
well-resolved numerical simulations taking into account full general
relativity, neutrino transport, and angular momentum transport by
magnetohydrodynamics or viscous hydrodynamics effects have not been
performed yet. Thus it is not clear whether the canonical interpretation is
really correct.

Although the electromagnetic observation for GW170817 provides us rich
information for the neutron-star merger, it will not be always the
case that the next events have the similar feature for the
electromagnetic counterparts as the GW190425 event
suggests~\cite{GW190425}. For GW170817, the total mass of two neutron
stars was not so large that the remnant could form a massive neutron
star at least temporarily. On the other hand, for more massive binary
neutron stars or for the case of black hole-neutron star binaries, a
black hole surrounded by a disk is the expected remnant. For this
case, the neutrino emission as strong as that from hypermassive
neutron stars is absent, and the properties of the post-merger ejecta
can be different from those for
GW170817~\cite{MF2013,MF2014,Perego14,MF2015,Just2015,SM17,Fujiba2018,FTQFK18,Janiuk19,Fujiba2019,FTQFK19,Miller19}.
It is worthy to explore in detail the ejecta properties for the case
that a black hole is the immediate remnant of neutron-star mergers.

There are a lot of previous work for exploring the ejecta from
the system composed of a black hole and a disk surrounding it.
However, the previous work has been performed in some simplifications.
In the pioneer simulation work by Fern\'andez, Metzger, and their
collaborators~\cite{MF2013,MF2014,MF2015}, they only qualitatively took into account the general relativistic effects and radiation transfer effects of neutrinos. In particular, the
spacetime structure around the black hole was only qualitatively
considered.  In Refs.~\cite{Perego14,Just2015}, the authors took into
account the neutrino irradiation effect carefully, but again they did
not or only qualitatively did take into account the general
relativistic effects and black hole spacetime.  In
Refs.~\cite{SM17,FTQFK18,Janiuk19,FTQFK19,Miller19}, the authors
performed a magnetohydrodynamics simulation taking into account the
general relativistic effect with a fixed Kerr black hole as the
background spacetime. However, in
Refs.~\cite{SM17,FTQFK18,Janiuk19,FTQFK19}, the neutrino transfer
effect was only approximately taken into account and/or the equation
of state employed was an approximate one (e.g., effects of heavy
nuclei are not incorporated, and in some work, degenerate pressure of
electrons, which is the key in the dense disk, is not taken into
account).  In Ref.~\cite{Miller19}, numerical simulations are
performed only for a short term, $\sim 0.13$\,s (i.e., an optimistic
magnetic field for enhancing the mass ejection is initially prepared),
although general relativity (background spacetime of a black hole), a
realistic equation of state, and a detailed radiation transport are
incorporated together for the first time.

In the present work, we perform a long-term fully general relativistic
viscous neutrino-radiation hydrodynamics simulation for black
hole-disk systems approximately taking into account the neutrino
irradiation effect. By focusing on the viscous evolution process, this
work can provide a complementary aspect for the evolution of
black-hole accretion disks to the magnetohydrodynamics work.  In
particular, we carefully resolve the black hole spacetime and the
inner part of the disk, and take into account the effect of the
self-gravity of the disk and the black-hole evolution by matter
accretion self-consistently.  Particular emphasis is put on the point
that we resolve the vicinity of the black hole including the inner
region of the disk with the resolution higher than those in the
previous simulations. This setting enables us to follow accurate
viscous evolution of the disk.

The paper is organized as follows. In Sec.~\ref{sec2}, we briefly
summarize the basic equations employed in the present simulation
study, and then, we describe the method to prepare the initial
condition composed of the equilibrium state of disks.  Dependence of
the property of the disk surrounding a black hole on the equation of
state and velocity profile is also presented.  Section~\ref{sec3}
presents numerical results for the simulations, focusing on the
properties of the ejecta and nucleosynthesis in the matter ejected
from the disks.  Section~\ref{sec4} is devoted to a summary.
Throughout this paper, $G$, $c$, and $k$ denote the gravitational
constant, speed of light, and Boltzmann's constant, respectively.

\section{Methods of numerical computation}\label{sec2}

\subsection{Basic equations}\label{sec2-1}

We evolve black hole-disk systems in the framework of full general
relativity.  For the case that the disk mass is much smaller than the
black-hole mass, we are allowed to perform a simulation in a fixed
background of black-hole spacetime.  However, for the case that the
disk mass is a substantial fraction of the black-hole mass, such
assumption may break down. In the last decade, the
numerical-relativity community has established methods to evolve
systems composed of black holes for a long timescale stably and
accurately (e.g., Ref.~\cite{shibata16} for a review).  Thus, we do
not have to assume the fixed background. In the present
work, we numerically solve both Einstein's equation and matter-field
equations self-consistently.

The major purpose of this paper is to clarify the viscous evolution of
a system composed of a self-gravitating disk surrounding a spinning
low-mass black hole, which is the plausible outcome formed after the
merger of a low-mass black hole-neutron star binary or a binary
neutron star of high total mass. For the disk evolution in nature, the
magnetohydrodynamics or viscous heating/angular momentum transport is
one of the key processes. In addition, the neutrino cooling of the
disk and neutrino irradiation to the matter are key processes for the
disk evolution and for determining the property of the matter ejected
from the system.  In this work, we choose basic equations as follows: 
Einstein's equation, the viscous-hydrodynamics equations, the
evolution equation for the viscous tensor, the evolution equations for
the lepton fractions including the electron fraction, and 
neutrino-radiation transfer equations. Here, for the equations for the
lepton fractions, we take into account electron and positron capture,
electron-positron pair annihilation, nucleon-nucleon bremstrahlung,
and plasmon decay~\cite{Fujiba2018}.  For solving Einstein's equation,
we employ the Baumgarte-Shapiro-Shibata-Nakamura formalism~\cite{BSSN}
together with the puncture formulation~\cite{puncture}, Z4c constraint
propagation prescription~\cite{Z4c}, and 5th-order Kreiss-Oliger
dissipation.  In reality, the viscous angular momentum transport is
likely to be induced effectively by a magnetohydrodynamics
process~\cite{BH98}. In the present work, this process is approximated
by viscous hydrodynamics.  All these basic equations are the same as
those in Ref.~\cite{Fujiba2018} in which we performed a simulation for
the system of a massive neutron star and disk.

One crucial difference of the present simulation from the previous
ones~\cite{Fujiba2018,Fujiba2019} is that we have to evolve a spinning
black hole located at the center stably and accurately for a long
timescale, at least $\sim 3$--5\,sec.  This can be achieved only by
employing a very high grid resolution around the central region. We
find that to follow the black hole with dimensionless spin of
$\chi=0.8$ accurately, i.e., to get the evolution of mass and spin for
the black hole in a reasonable accuracy, the grid spacing in the
central region, $\Delta x_0$, should be smaller than $\sim 0.02GM_{\rm
  BH}/c^2$ (see Sec.~\ref{sec3}) where $M_{\rm BH}$ is the initial
mass of the black hole: For $M_{\rm BH}=3M_\odot$, the grid spacing
has to be smaller than $\sim 80$\,m.  Otherwise, the area and
dimensionless spin of the black hole, respectively, increase and
decrease spuriously and significantly.  In the present work, we employ
$\Delta x_0=0.016GM_{\rm BH}/c^2$. We also performed simulations with
$\Delta x_0=0.0133GM_{\rm BH}/c^2$ and $0.020GM_{\rm BH}/c^2$ for a
particular model (referred to as K8 model in Table~\ref{table1}), and
show the convergence property in Sec.~\ref{sec3-6}. We note that such
a high resolution helps accurately resolving the inner region of the
disk.

As mentioned above, a sufficiently high grid resolution is the key for
a reliable simulation.  Furthermore, we have to evolve the disk for
the timescale of $\agt 2\times 10^5GM_{\rm BH}/c^3 \sim 3$\,s, because
the viscous timescale of the disk is much longer than the dynamical
timescale (the typical orbital period) of the system.  With these
requirements (high-resolution and long-term simulation is required),
however, a three-dimensional simulation under no assumption of
symmetry is still quite expensive. Thus, as in
Refs.~\cite{Fujiba2018,Fujiba2019}, we assume the axial symmetry of
the spacetime as well as the reflection symmetry with respect to the
equatorial plane: We employ a cartoon method~\cite{cartoon,cartoon2}
to impose the symmetry for solving Einstein's equation; i.e., we
employ the Cartesian coordinates $(x, y, z)$ and evolve the system
only in the $(x, z)$ plane (the $y=0$ plane).

As already mentioned, in addition to Einstein's equation, we solve the
viscous-hydrodynamics equation, evolution equations for the viscous
tensor and lepton fractions, and radiation transfer equation. These
equations are solved in the cylindrical coordinates composed of $(x,
z)$.

For the $x$ and $z$ directions, the following non-uniform grid is used
for the present numerical simulation: For $x \leq x_0=0.8GM_{\rm
  BH}/c^2$, a uniform grid is used, and for $x > x_0$, the grid
spacing $\Delta x$ is increased uniformly as $\Delta
x_{i+1}=1.01\Delta x_i$ where the subscript $i$ denotes the $i$th grid
with $x=0$ at $i=0$. For $z$, the same grid structure as for $x$ is
used. The black-hole horizon is always located in the uniform grid
zone.  The location of the outer boundaries along each axis is chosen
to be $\approx 1400GM_{\rm BH}/c^2 \approx 6100(M_{\rm
  BH}/3M_\odot)$\,km in this study.

\subsection{Initial condition}

We prepare an axisymmetric equilibrium state for the black hole-disk
system in the framework of full general relativity as the initial
condition for our numerical simulation.  For the gravitational field
equations, we employ a puncture formulation developed in
Ref.~\cite{Shibata2007} with the following line element: 
\beqn
ds^2=-\alpha^2 c^2 dt^2 &+& \psi^4 [e^{2\eta}(dr^2+r^2d\theta^2)
\nonumber \\
&&~~ + r^2 \sin^2\theta(\beta^\varphi dt + d\varphi)^2].
\eeqn
Here, we employ spherical polar coordinates and $\alpha$, $\psi$,
$\beta^\varphi$, and $\eta$ are functions of $r$ and $\theta$.  Note that $\eta$
and $\beta$ vanish for the non-spinning black hole.  In the following, 
we denote the dimensionless spin parameter by $\chi$, which is chosen
to be 0.8 or 0.6.

We consider the case that the fluid four velocity has the form of
$u^r=0=u^\theta$ and $u^\varphi=\Omega u^t$, and the fluid is
isentropic. Here $\Omega$ is the angular velocity which is also a
function of $r$ and $\theta$. Then, under the assumption of the
isentropy, the Euler equation is integrated to give the first integral
in the form
\beq
{h \over u^t}+{1 \over c^2} \int h u_\varphi d\Omega=C, \label{integ1}
\eeq
where $h$ is the specific enthalpy and $C$ is a constant.  Using the
rest-mass density $\rho$, the specific internal energy $\varep$, and
the pressure $P$, $h$ is written as $c^2+\varep+P/\rho$.  Using the
normalization relation for $u^\mu$, $u^\mu u_\mu=-c^2$, $u^t$ is written
as
\beq
u^t =\left(\alpha^2-c^{-2}\gamma_{\varphi\varphi} (\Omega + \beta^\varphi)^2\right)^{-1/2},
\label{ut}
\eeq
where $\gamma_{\varphi\varphi}=\psi^4 r^2 \sin^2\theta$. 

In this paper, we simply assume that $j=c^{-2}h u_\varphi$ is a function of 
$\Omega$ in the form of 
\beq
j=A_n \Omega^{-n}, \label{eqj}
\eeq
where $A_n$ and $n$ are constants (see Ref.~\cite{Poland} for more
careful choice).  In the Newtonian limit, $j \approx u_\varphi \approx
\varpi^2 \Omega$ with $\varpi=r\sin\theta$, and hence, $\Omega \propto
\varpi^{-2/(n+1)}$. Thus, for $n=1/3$, the Keplerian angular velocity
is recovered in the Newtonian limit and for $n=0$, the specific
angular momentum is constant.  Since the angular velocity profile of
the remnant disk of neutron-star mergers is close to the Keplerian, we
should try to employ the value of $n$ as close as $1/3$.  We find that
the disk mass in the equilibrium states with $1/6 \leq n < 1/3$ in our
present setting becomes quite small, if the radius of the outer edge
of the disk is a reasonable value for a remnant of neutron-star
mergers as $\alt 50GM_{\rm BH}/c^2$ (see discussion below). Thus, we
employ the values of $n \leq 1/7$.

For a given value of $n$ and an equation of state, we have to
determine $\Omega$, $h$, $A_n$, and $C$. In the following, we describe
the basic equations for this procedure.  First, Eq.~(\ref{eqj}) is
rewritten as
\beq
\Omega^n (\Omega + \beta^\varphi) h u^t \gamma_{\varphi\varphi}=A_n c^2, \label{eqj2}
\eeq
where we used
\beq
u_{\varphi}=u^t(\Omega+\beta^\varphi)\gamma_{\varphi\varphi}. \label{eqvarphi}
\eeq
Using Eq.~(\ref{eqj}), Eq.~(\ref{integ1}) is rewritten as
\beq
{h \over u^t} + {A_n \Omega^{1-n} \over 1-n}=
h\left({1 \over u^t} + {u_\varphi \Omega \over c^2(1-n)}\right)=C. \label{integ2}
\eeq
From Eqs.~(\ref{ut}) and (\ref{eqj2}), we also obtain
\beq
{A_n \Omega^{-n}[c^2 \alpha^2-\gamma_{\varphi\varphi}(\Omega+\beta^\varphi)^2] 
\over \gamma_{\varphi\varphi}(\Omega+\beta^\varphi)}
=\left(C - {A_n \Omega^{1-n} \over 1-n}\right). \label{eq2.7}
\eeq
This algebraic equation is used to determine $\Omega$ for given values
of $A_n$ and $C$ with gravitational fields computed.

$A_n$ and $C$ are determined by choosing the inner and outer edges 
of the disk in the equatorial plane. Since the values of $h$ are 
identical at such edges, Eq.~(\ref{integ2}) gives
\beqn
\left({1 \over u^t} + {u_\varphi \Omega \over c^2(1-n)}\right)_{\rm in}= 
\left({1 \over u^t} + {u_\varphi \Omega \over c^2(1-n)}\right)_{\rm out},
\label{cond1}
\eeqn
where ``in'' and ``out'' in the subscripts indicate the quantities at the
inner and outer edges, respectively.  Here, $u^t$ and $u_\varphi$ are given by
Eqs.~(\ref{ut}) and (\ref{eqvarphi}), respectively. 

In addition, Eq.~(\ref{eqj}) is written to $(j\Omega^n)_{\rm
  in}=(j\Omega^n)_{\rm out}$, which leads to
\beq
(u_\varphi\Omega^n)_{\rm in}=(u_\varphi \Omega^n)_{\rm out}. \label{cond2}
\eeq 
Equations~(\ref{cond1}) and (\ref{cond2}) constitute simultaneous
equations for $\Omega_{\rm in}$ and $\Omega_{\rm out}$, and hence, by
solving these equations, we first determine these angular
velocities. Subsequently, using Eqs.~(\ref{eqj2}) and (\ref{integ2}),
$A_n$ and $C$ are determined for the value of $h$ at $\rho=\rho_{\rm
  min}$, where the minimum value of $h$ and $\rho_{\rm min}$ are found
for a given tabulated equation of state. Then, $\Omega$ at each point
is determined by solving Eq.~(\ref{eq2.7}).  Once $A_n$, $C$, and
$\Omega$ at each point are determined, $h$ is determined from
Eq.~(\ref{integ2}).

\begin{figure}[t]
\includegraphics[width=85mm]{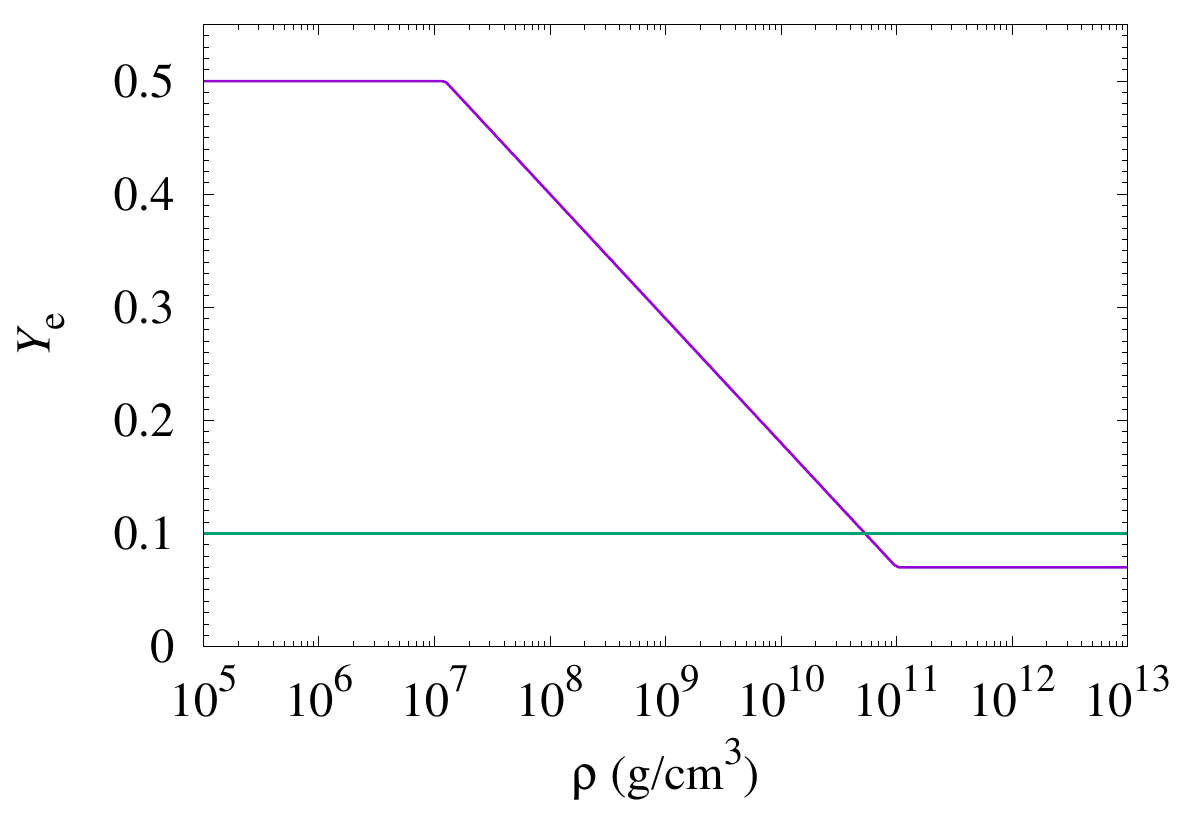} \\ 
\includegraphics[width=85mm]{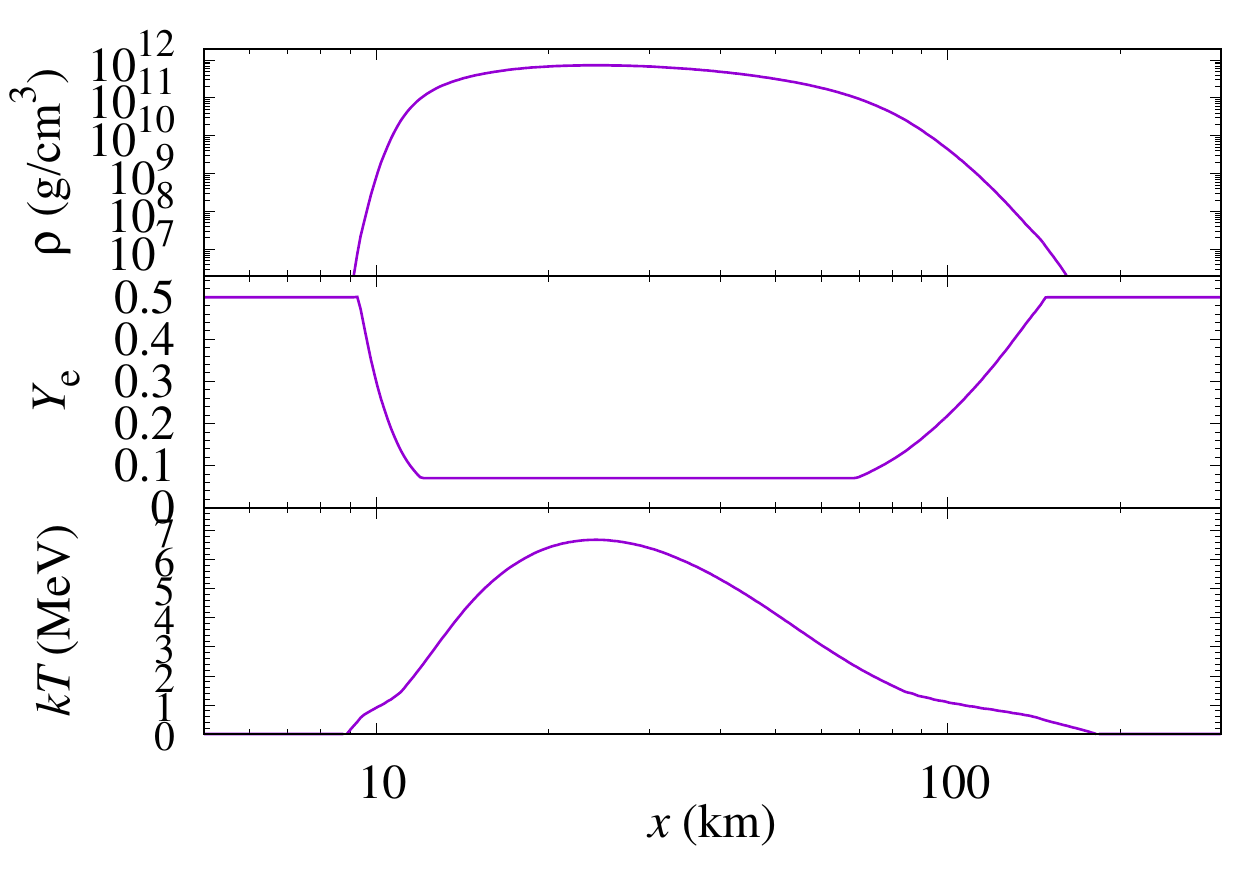} 
\caption{Top: Two relations for $Y_e$ employed for constructing
  equilibrium states of disks surrounding the black hole.  Bottom:
  Density, electron fraction, and temperature ($k T$) as
  functions of the radius in the equatorial plane for a disk model
  (model K8 in Table~\ref{table1}).
\label{fig1}}
\end{figure}

For a given value of $h$, thermodynamical variables, $\rho$, $\varep$,
and $P$, are determined through the equation of state given. In this
work, we employ a tabulated equation of state based on the DD2
equation of state~\cite{DD2} for a relatively high-density part and the 
Timmes equation of state for the low-density part~\cite{Timmes}.  We
choose the lowest rest-mass density to be $0.1\,{\rm g/cm^3}$ in the
table.  In this equation of state, $\varep$, $P$, and $h$ are
functions of $\rho$, $Y_e$, and $T$ where $Y_e$ and $T$ are the
electron fraction and matter temperature, respectively. Thus, to
determine $\rho$, $Y_e$, and $T$ from $h$, we need to employ two
conditions for relating these three variables.  One condition is just
the assumption that the specific entropy $s$ is constant because
Eq.~(\ref{integ1}) is derived under this condition.  For the other, we
adopt two relations.  In one case, we assume a relation between $Y_e$
and $\rho$ in the form $\rho(Y_e)$ (see Fig.~\ref{fig1}) and in the
other case, we simply set $Y_e=0.1$.  For the former case, we derive
an approximate relation of $\rho(Y_e)$ from our numerical results for
the remnant of binary neutron star mergers~\cite{Sekig} (see also,
e.g., Refs.~\cite{Foucart16,Radice16,Bovard}), for which the value of
$Y_e$ in the major part of the disk is approximately determined by
weak interaction processes. For this model, the value of $Y_e$ is
larger for the smaller density, because the effect of the electron
degeneracy, compared to the thermal effects, becomes weaker.  We
basically employ this as the fiducial model, and the equation of state
is simply referred to by specifying the value of $s/k$.  The equation
of state with $Y_e=0.1$ could be an approximate model for the remnant
of black hole-neutron star mergers~\cite{Foucart,Kyutoku}. When we
refer to this model, we always mention $Y_e=0.1$.

\begin{figure*}[t]
\includegraphics[width=86mm]{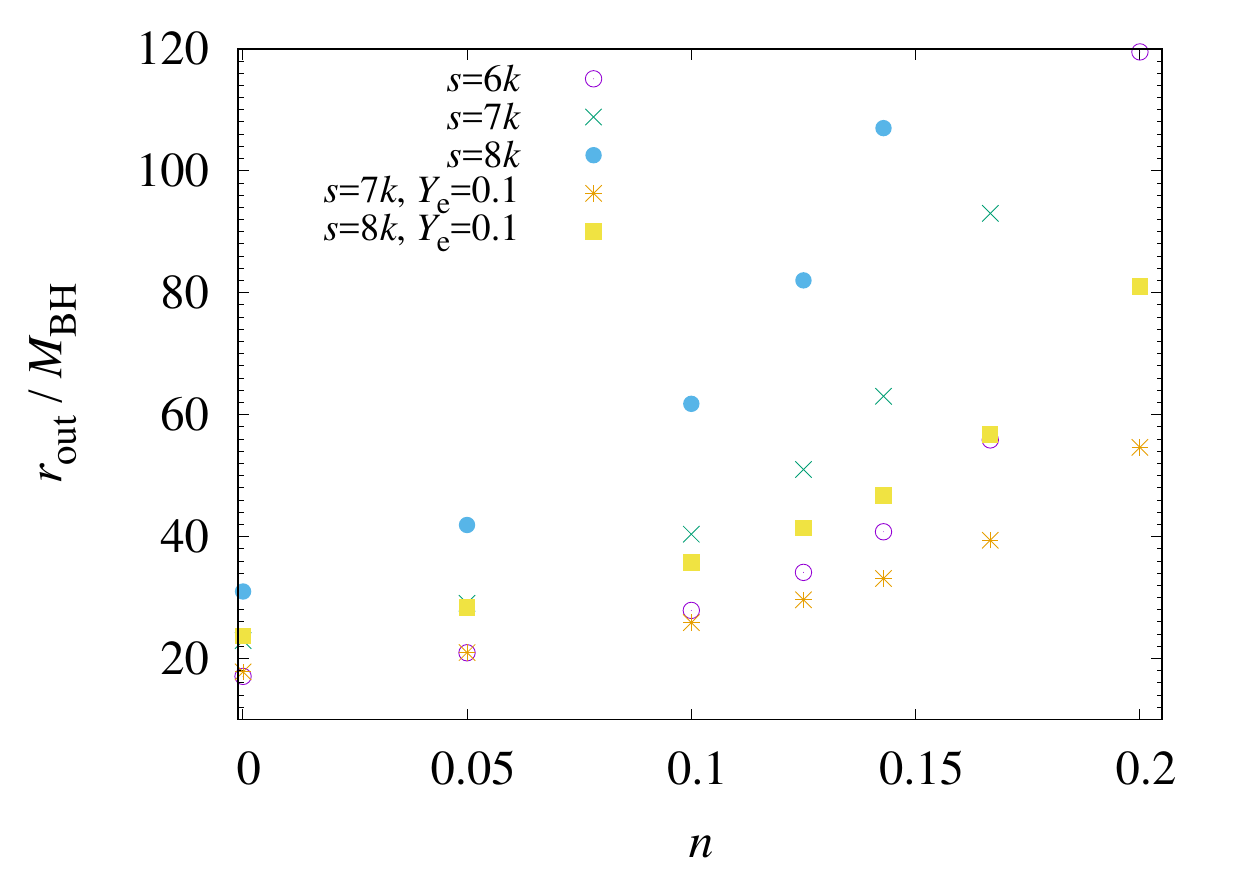}~~
\includegraphics[width=86mm]{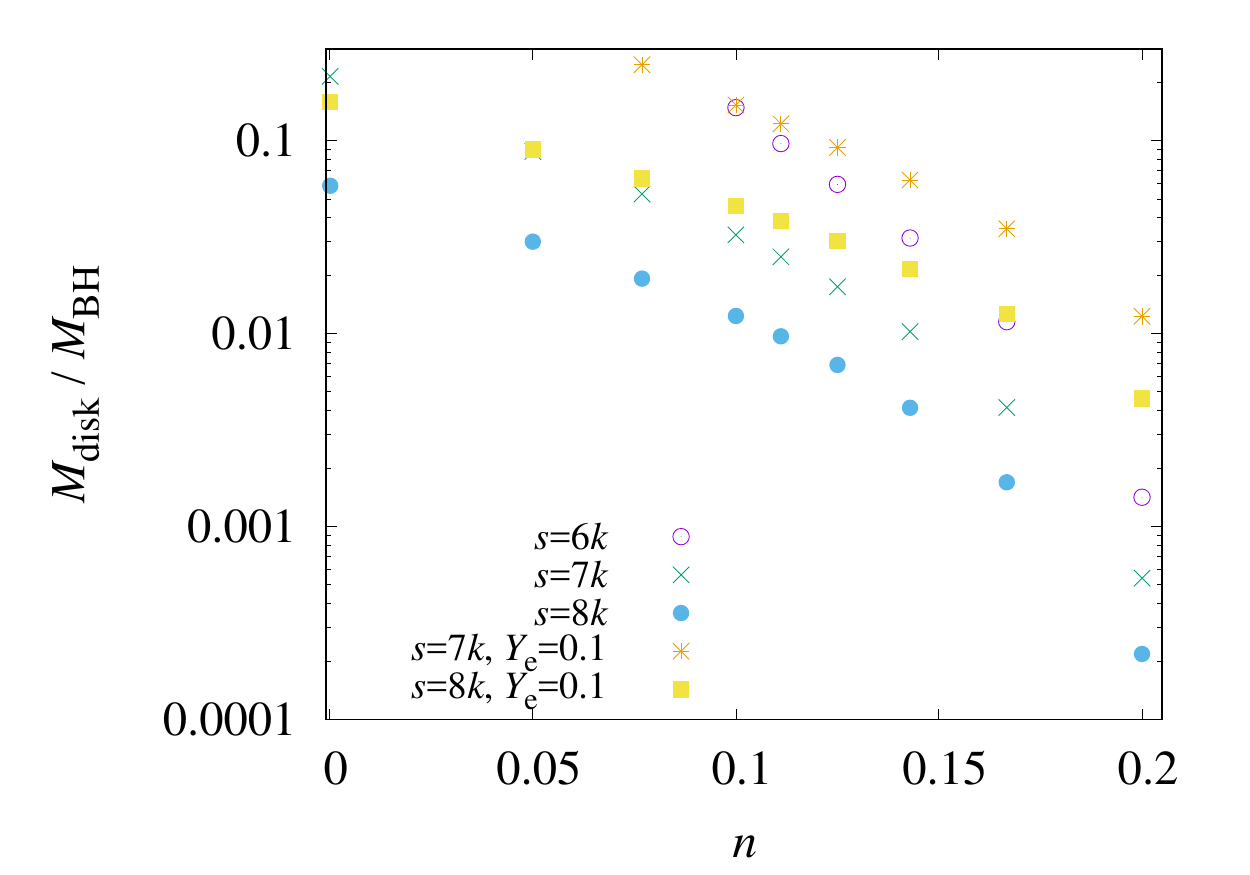} 
\caption{Left: $r_{\rm out}/(GM_{\rm BH}/c^2)$ as a function of $n$ for the
  case of $M_{\rm disk}/M_{\rm BH}=1/30$.  Right: $M_{\rm disk}/M_{\rm
    BH}$ as a function of $n$ for $r_{\rm out}=40GM_{\rm BH}/c^2$. For
  both panels, $r_{\rm in}=2.0GM_{\rm BH}/c^2$ and $\chi=0.8$.
\label{fig2}}
\end{figure*}

\begin{figure*}[t]
\includegraphics[width=86mm]{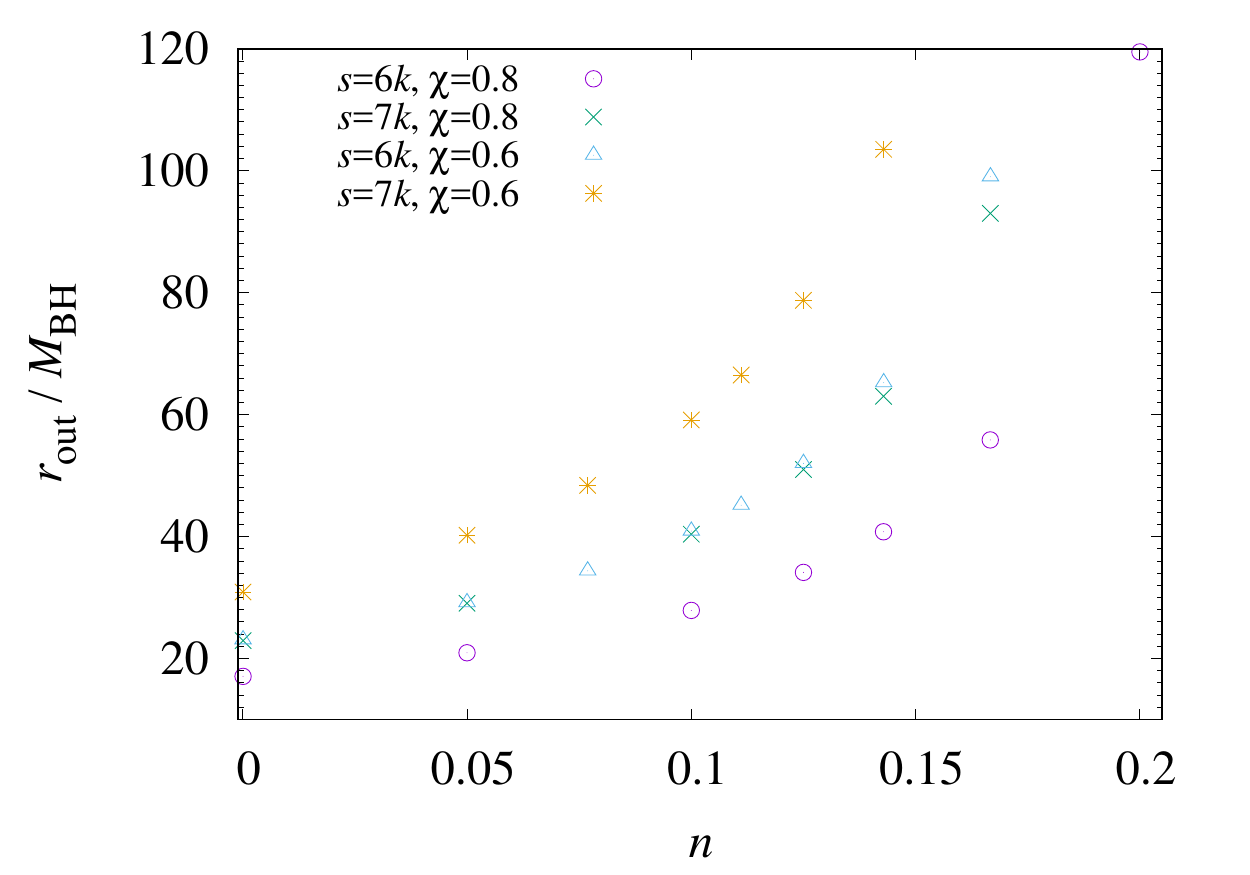}~~ 
\includegraphics[width=86mm]{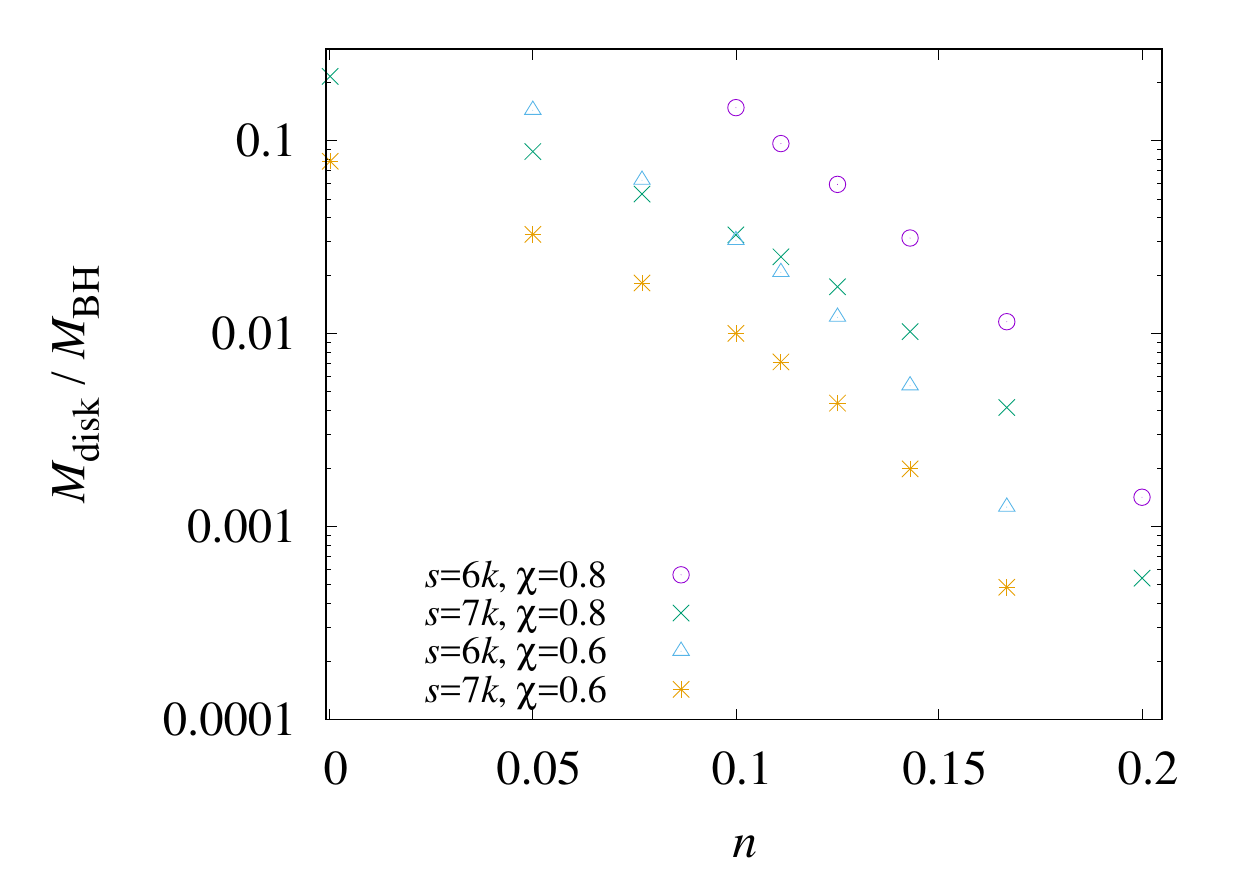} 
\caption{The same as Fig.~\ref{fig2} but for the comparison between
  $\chi=0.8$ and $0.6$.  For $\chi=0.8$ and 0.6, $r_{\rm in}=2.0GM_{\rm
    BH}/c^2$ and $2.8GM_{\rm BH}/c^2$, respectively.
\label{fig3}}
\end{figure*}


The bottom panel of Fig.~\ref{fig1} displays the rest-mass density,
electron fraction, and temperature as functions of the
radius in the equatorial plane for a typical equilibrium model
employed in this paper as an initial condition for the simulation
(model K8 in Table~\ref{table1}). For this model, $M_{\rm
  BH}=3M_\odot$, $\chi=0.8$, $M_{\rm disk}=0.1M_\odot$, $r_{\rm
  in}=2.0GM_{\rm BH}/c^2$, $r_{\rm out} \approx 41GM_{\rm BH}/c^2$,
$s/k=6$, and $n=1/7$. We find that the maximum density is $\alt
10^{12}\,{\rm g/cm^3}$ and the maximum temperature (in terms of $kT$)
is $\sim 7$\,MeV.  In the dense region, the electron degeneracy is
high, and as a result, the electron fraction is low $\leq 0.1$. This
is a typical structure of the disk around low-mass black holes with
$M_{\rm disk}=O(0.1M_\odot)$.

Figure~\ref{fig2} displays several relations for the radius of the
outer edge, $r_{\rm out}$, and baryon mass, $M_{\rm disk}$, of the
disk for a variety of the values of $n$ and $s/k$. In this plot, we
fix $M_{\rm BH}=3M_\odot$, $\chi=0.8$, and $r_{\rm in}=2.0GM_{\rm
  BH}/c^2$.  On the other hand, we employ a variety of the equations
of state with $s/k=6$--8 and a wide range of $n$.  The left panel
shows the outer edge of the disk, $r_{\rm out}$, as a function of $n$
for $M_{\rm disk}=0.1M_\odot$.  For the larger values of $n$ toward
$1/3$, the velocity profile of the disk approaches the Keplerian
profile and the disk becomes geometrically thin. As a result, to
preserve a given value of $M_{\rm disk}$, the extent of the disk
(i.e., $r_{\rm out}$) needs to be increased with $n$. It is also found
that for the higher value of $s/k$, the extent of the disk has to be
larger.  The reason for this is that for the higher value of $s/k$,
the pressure for a given value of the density is larger, and hence,
the overall density becomes relatively small.

The right panel of Fig.~\ref{fig2} shows $M_{\rm disk}/M_{\rm BH}$ as
a function of $n$ for $r_{\rm out}=40GM_{\rm BH}/c^2$. This clarifies
that for the larger values of $n$, the disk mass is smaller for the
given extent of the disk, and for the smaller values of $s/k$ the
disk mass is larger. It is also found that for the $Y_e=0.1$ case, the
disk mass is larger for the given value of $s/k$. This is reasonable
because for the smaller value of $Y_e$, the electron degeneracy
pressure is less important, and hence, to enhance the pressure, a
higher value of the rest-mass density is needed.

In Fig.~\ref{fig3}, we compare the results for $\chi=0.6$ and 0.8.
For $\chi=0.6$, the inner edge of the disk cannot be as small as that
for $\chi=0.8$ because the location of the innermost stable circular
orbit around the black hole is closer to the black hole for the larger
value of $\chi$ for given black-hole mass.  Here, we set $r_{\rm
  in}=2.8GM_{\rm BH}/c^2$ for $\chi=0.6$ while it is $2.0GM_{\rm
  BH}/c^2$ for $\chi=0.8$.  It is found for $\chi=0.6$ that the value
of $r_{\rm out}$ becomes larger for given parameters of $s$, $n$, and
$M_{\rm disk}$ than for $\chi=0.8$.  It is also found that the disk
mass is smaller for the given values of $s$, $n$, and $r_{\rm
  out}=40GM_{\rm BH}/c^2$ than for $\chi=0.8$. The reason for these
results is that for the smaller values of $\chi$, the gravitational
potential of the black hole is shallower, and hence, the amount of the
bounded material becomes smaller.  Thus, to obtain a model with the same
disk mass and same value of $r_{\rm out}/(GM_{\rm BH}/c^2)$ for
$\chi=0.6$ as for $\chi=0.8$, we need to prepare a smaller value of
$s$ or a smaller value of $n$.

The simulations are performed for several models as the initial
conditions. First, we employ models of $s/k=6$ and $n=1/7$ for the
fiducial $Y_e$ case and $s/k=8$ and $n=1/8$ for the $Y_e=0.1$ case
with $\chi=0.8$ (see models K8 and Y8 in
Table~\ref{table1}). For both cases, we set $(r_{\rm in}, r_{\rm out})
\approx (2GM_{\rm BH}/c^2, 41GM_{\rm BH}/c^2)$ and the mass of the
black hole and the baryon rest-mass of the disk to be $M_{\rm
  BH}=3M_\odot$ and $M_{\rm disk}=0.1M_\odot$, respectively.  We also
employ a compact disk model (D8) for which $s/k=6$, $n=1/10$, $(r_{\rm
  in}, r_{\rm out}) \approx (2GM_{\rm BH}/c^2, 29GM_{\rm BH}/c^2)$
with the same mass and spin, $M_{\rm BH}=3M_\odot$, $\chi=0.8$, and
$M_{\rm disk}=0.1M_\odot$, as those for models K8 and Y8.  We will show
that the slight difference in $s$ and $n$ does not drastically change
overall dynamics of the disk and the properties of the ejecta,
although for initially more compact disks, the ejecta mass becomes
smaller as a natural consequence.  In addition, we employ one model
with $\chi=0.6$, $M_{\rm BH}=3M_\odot$, and $M_{\rm disk}=0.1M_\odot$
in which $n=1/10$, $s=6k$, and $(r_{\rm in}, r_{\rm out}) \approx
(2.8GM_{\rm BH}/c^2, 41GM_{\rm BH}/c^2)$ to explore the dependence of
the numerical results on the black-hole spin.

We note that the outer edge of the disk should not be very large for
modeling the merger remnant of neutron-star binaries;
numerical-relativity simulations have shown that $r_{\rm out}$ is
between 100--200\,km. For this reason, we fiducially set $r_{\rm
  out} \approx 40GM_{\rm BH}/c^2$ which is $\approx 180$\,km for
$M_{\rm BH}=3M_\odot$.  For the compact model (D8), $r_{\rm out}
\approx 130$\,km, which is also a reasonable value



In addition to these models, we employ initial conditions with $M_{\rm
  disk}=0.5M_\odot$ and $r_{\rm out} \approx 58GM_{\rm BH}/c^2$, and
with $M_{\rm disk}=0.03M_\odot$ and $r_{\rm out} \approx 29GM_{\rm
  BH}/c^2$.  For both models, we employ the fiducial $Y_e$ equation of
state, with $\chi=0.8$, $M_{\rm BH}=3M_\odot$, $r_{\rm BH}=2GM_{\rm
  BH}/c^2$, $n=1/7$, and $s=6k$, which are the same as those for model
K8 (see Table~\ref{table1}). By performing simulations for these
initial conditions, we examine the effect of the mass (i.e., the
effects by the density and temperature) of the disk on the evolution
of the system. We note that the initial condition with $M_{\rm
  disk}=0.03M_\odot$ is a good model for neutron-star mergers, but
that with $0.5M_\odot$ is not. Rather, such a high-mass disk may be a
good model for the remnant of the massive stellar core collapse to a
black hole.


Because the mass ratio, $M_{\rm disk}/M_{\rm BH}$, of the initial
conditions employed here is fairly large (1/100--1/6) and hence the
disks are weakly self-gravitating, they may be subject to
non-axisymmetric deformation even if the angular velocity profile is
close to the Keplerian. The previous work in general relativity (e.g.,
Refs.~\cite{Hawley91,Oleg11,Kiuchi11}) shows that if the self gravity
of the disks is not extremely large, spiral arms are formed and
contribute to angular momentum transport by the gravitational torque
exerted by the non-axisymmetric structure.  In our simulation, such
non-axisymmetric effects cannot be taken into account, but the angular
momentum transport is incorporated through the viscous
hydrodynamics. Moreover, the previous 
work~\cite{Hawley91,Oleg11,Kiuchi11} shows that the density
enhancement in the spiral arms is not very strong and the angular
momentum transport is much less efficient than that in the viscous
hydrodynamics with the alpha viscous parameter~\cite{SS73},
$\alpha_\nu=O(10^{-2})$. Therefore we suppose that the
non-axisymmetric deformation effects would not be very important for
the models that we employ here.

\subsection{Method for analysing ejecta}

\begin{table*}[t]
\caption{Initial conditions for the numerical simulation. Described
  are the model name, black hole mass, disk mass, black-hole
  dimensionless spin, the radii at the inner and outer edges of the disk ($r_{\rm
    in}$ and $r_{\rm out}$), entropy per baryon ($s/k$), $n$, and electron fraction ($Y_e$)
  for the disk, and $\alpha_{\nu}H$.  The units of the mass are
  $M_\odot$ and the units of $r_{\rm in}$ and $r_{\rm out}$ are
  $GM_{\rm BH}/c^2 \approx 4.43$\,km. The last column shows whether the neutrino
  irradiation is switched on or off.}
 \begin{tabular}{ccccccccccc} \hline
~Model~ & ~$M_{\rm BH}$~ & ~$M_{\rm disk}$~ & ~~$\chi$~~ 
& ~~$r_{\rm in}$~~ & ~~$r_{\rm out}$~~ & ~$s/k$~
& ~~~~$n$~~~~ & ~~$Y_e$~~& ~~$\alpha_\nu H$\,(km)~~ & neutrino irradiation\\
 \hline \hline
K8 &  3.0 & 0.10  & 0.8 & 2.0 & 41 & 6 & 1/7  & 0.07--0.5 & 0.45 & yes\\
K8h&  3.0 & 0.10  & 0.8 & 2.0 & 41 & 6 & 1/7  & 0.07--0.5 & 0.90 & yes\\
K8s&  3.0 & 0.10  & 0.8 & 2.0 & 41 & 6 & 1/7  & 0.07--0.5 & 1.35 & yes\\
K8n&  3.0 & 0.10  & 0.8 & 2.0 & 41 & 6 & 1/7  & 0.07--0.5 & 0.45 & no \\
D8 &  3.0 & 0.10  & 0.8 & 2.0 & 29 & 6 & 1/10 & 0.07--0.5 & 0.45 & yes\\
Y8 &  3.0 & 0.10  & 0.8 & 2.0 & 41 & 8 & 1/8  & 0.1       & 0.45 & yes\\
K6 &  3.0 & 0.10  & 0.6 & 2.8 & 41 & 6 & 1/10 & 0.07--0.5 & 0.45 & yes\\
C8 &  3.0 & 0.50  & 0.8 & 2.0 & 58 & 6 & 1/7  & 0.07--0.5 & 0.45 & yes\\
E8 &  3.0 & 0.03  & 0.8 & 2.0 & 29 & 6 & 1/7  & 0.07--0.5 & 0.45 & yes\\
 \hline
 \end{tabular}
 \label{table1}
\end{table*}

Here, we briefly summarize how we identify the matter as ejecta.  The
unbound matter should be considered as ejecta. In this work, we employ
the following condition for identifying matter in an unbound orbit:
$|h u_t| > h_{\rm min}c^2$ where $h_{\rm min}$ denotes the minimum value
of the specific enthalpy $h$ in the chosen tabulated equation of
state, which is $\approx 0.9987c^2$, and $u_t$ is a negative quantity.
The reason why $h_{\rm min}$ is smaller than $c^2$ is that the effect
of the binding energy of nucleus is present in the equation of state.

To analyse the ejecta, we first extract the outgoing component of the
matter at the radius of $r_{\rm ext}=2000$--4000\,km and identify the
ejecta. Here, by changing the extraction radius, we examine the
convergence of the ejecta mass. In addition, we analyze the matter
located within a sphere of $r=r_{\rm ext}$ and the component with $|h
u_t| > h_{\rm min}c^2$ is identified as the ejecta. By summing up these two
components, we determine the quantities of the ejecta.

For the ejecta component escaping from a sphere of $r=r_{\rm ext}$, we
define the ejection rates of the rest mass and total energy at a given
radius by
\beqn
\dot M_{\rm eje}&:=&c^{-1}\oint \rho \sqrt{-g} u^i dS_i, \label{ejectrate}\\
\dot E_{\rm eje}&:=&c^{-1}\oint \rho \hat e \sqrt{-g} u^i dS_i,
\eeqn
where $g$ denotes the determinant of the spacetime metric and $\hat
e:=h \alpha u^t-P/(\rho \alpha u^t)$. The surface integral is
performed at $r=r_{\rm ext}$ with $dS_i=\delta_{ir}r_{\rm
  ext}^2d\theta d\varphi$ for the ejecta component.  Here, we note
that $\rho \sqrt{-g} u^t$ obeys the continuity equation of the rest
mass and $\rho \hat e \sqrt{-g} u^t$ obeys the energy conservation
equations in the absence of gravity. Hence, for the region far from
the central object, the time integration of these quantities are
conserved.  Thus, by performing the time integration, the total rest
mass and energy of the ejecta (which escape away from a sphere of
$r=r_{\rm ext}$) are calculated by
\beqn
M_{\rm eje,esc}(t)&:=&\int^t \dot M_{\rm eje} dt,\\
E_{\rm eje,esc}(t)&:=&\int^t \dot E_{\rm eje} dt. 
\eeqn
In addition, we add the rest mass for the ejecta component located
inside a sphere of $r=r_{\rm ext}$, $M_{\rm eje,in}(t)$, giving the total
ejecta mass of the ejecta, $M_{\rm eje}=M_{\rm eje,esc}+M_{\rm eje,in}$. 

We note that far from the central object, $E_{\rm eje,esc}$ is
approximated by 
\beq
E_{\rm eje,esc}\approx M_{\rm eje,esc} c^2 + U + T_{\rm kin}+{GM_{\rm BH} M_{\rm
  eje,esc} \over r_{\rm ext}}, \label{Eeje}
\eeq
where $U$ and $T_{\rm kin}$ are the values of the internal energy and
kinetic energy of the ejecta at $r_{\rm ext}\rightarrow \infty$,
respectively. The last term of Eq.~(\ref{Eeje}) approximately denotes
the contribution of the potential binding energy to $E_{\rm eje,esc}$,
which cannot be neglected for $r_{\rm ext} \alt 10^3GM_{\rm BH}/c^2
\approx 4500$\,km because the ejecta velocity, $v_{\rm eje}$, is $\sim
0.05c$ and $(v_{\rm eje}/c)^2$ is of the order of $10^{-3}$.  Since
the ratio of the internal energy to the kinetic energy of the ejecta
decreases with its expansion, we may approximate $U/T_{\rm kin}
\approx 0$, and hence, $E_{\rm eje,esc}$ by $E_{\rm eje,esc} \approx
M_{\rm eje,esc}c^2 + T_{\rm kin}+GM_{\rm BH} M_{\rm eje,esc}/r_{\rm
  ext}$ for the region far from the central object.  We then define
the average velocity of the ejecta (for the component that escapes
from a sphere of $r=r_{\rm ext}$) by
\beq
v_{\rm eje}:=\sqrt{{2(E_{\rm eje,esc}-M_{\rm eje,esc}c^2-GM_{\rm BH} M_{\rm
    eje,esc}/r_{\rm ext}) \over M_{\rm eje,esc}}}. 
\eeq
We note that the correction of the gravitational potential energy
$GM_{\rm BH} M_{\rm eje,esc}/r_{\rm ext}$ is important for $r_{\rm
  ext} \alt 10^4$\,km, and just by taking into account this
correction, the values of $v_{\rm eje}$ become only weakly dependent
on the extraction radius, $r_{\rm ext}$.

\section{Numerical Results}\label{sec3}

\subsection{Setting}

Numerical computations are performed for the black hole-disk systems
summarized in the previous section (see also Table~\ref{table1}).  For
the viscous-hydrodynamics, we need to input the viscous coefficient
$\nu$. In this work, we set $\nu=\alpha_\nu h c_s H/c^2$ where
$\alpha_\nu$ is the dimensionless viscous coefficient (the so-called
alpha parameter), $c_s$ is the sound velocity, and $H$ is a scale
height. We basically employ $\alpha_\nu=0.05$ taking into account the
result of recent magnetohydrodynamics simulations of
Refs.~\cite{FTQFK18,FTQFK19}, which indicates that the magnitude of
the effective viscous parameter is high with $\alpha_\nu \approx
0.05$--0.1 in the vicinity of spinning black holes.

For the fiducial model, we set $H=9\,{\rm km} \approx 2GM_{\rm
  BH}/c^2$. That is, we set it approximately equal to the radius at
the innermost stable circular orbit around the Kerr black hole of
$\chi=0.8$.  For the outer part of the disk, the value of the scale
height might be larger than $H \approx 9$\,km, because it could be
approximately $c_s/\Omega$ in the standard accretion disk
theory~\cite{SS73}. However, for the non-stationary system, this is
not likely to be the case, if we suppose that the viscosity is
effectively enhanced by the turbulence caused by the
magneto-rotational instability (MRI)~\cite{BH98}, because the
exponential growth rate of the MRI is proportional to the local
angular velocity, $\Omega$ (for the Keplerian angular velocity
profile, the growth rate is $3\Omega/4$). That is, to establish the
turbulence by enhancing the magnetic-field strength by several orders
of magnitude until the saturation of the magnetic-field growth, it
takes a timescale of $\agt 10/(3\Omega/4)$.  Here, the rotational
period of the disk is $2\pi/\Omega \approx 310$\,ms for the radius of
$r=1000$\,km with the black-hole mass of $3M_\odot$ assuming the
Kepler motion of the disk, while the timescale of the mass ejection,
which is triggered by the viscous heating in the inner part of the
disk, is $\sim 0.5(\alpha_\nu/0.05)^{-1}(H/9\,{\rm km})^{-1}$\,s (see
Sec.~\ref{sec3-2}). Thus, it would not be realistic to employ a large
value of it for the outer part of the disk. Employing
$c_s/\Omega$ as the scale height could result in an overestimated
value of the scale height for a large value of $c_s$, and the mass
ejection process could be induced from an outer part of the disk in an
unrealistic manner for the early evolution stage of disks with $t \ll
1$\,s.

To examine the dependence of the numerical results on the value of
$\nu$, we change $\alpha_\nu H$ from $0.45$\,km to $0.90$\,km and
$1.35$\,km for the model of K8 series (see Table~\ref{table1}). We
note that in our setting of $\nu$, varying $H$ is equivalent to
varying $\alpha_\nu$; numerical results for $\alpha_\nu=0.05$ and
$H=9$\,km are the same as those, e.g., for $\alpha_\nu=0.01$ and
$H=45$\,km. However for simplicity, in the following, the viscous
coefficients with $\alpha_\nu H=0.90$\,km and $1.35$\,km are referred
to as $\alpha_\nu=0.10$ and $\alpha_\nu=0.15$ with $H=9$\,km,
respectively.

The viscous timescale (for heating and angular momentum transport) is
written approximately by
\beqn
\tau_{\rm vis}:={R^2 \over \nu} 
&\approx& 0.37\,{\rm s}(hc^{-2})^{-1}\left({\alpha_\nu \over 0.05}\right)^{-1}
\left({c_s \over 0.05c}\right)^{-1}
\nonumber \\
&&\times 
\left({H \over 9\,{\rm km}}\right)^{-1}
\left({R \over 50\,{\rm km}}\right)^{2},\label{tvis}
\eeqn
where $R$ denotes the cylindrical radius of the disk. As we show in
Sec.~\ref{sec3-2}, the evolution timescale for our choice of the
viscous coefficient is indeed of the order of 0.1\,s.
We note that the dynamical timescale of the disk is approximately
\beqn
\tau_{\rm dyn}&:=&2\pi\sqrt{{R^{3} \over GM_{\rm BH}}} \nonumber \\
&=&3.5\,{\rm ms} \left({R \over 50\,{\rm km}}\right)^{3/2}
\left({M \over 3M_\odot}\right)^{-1/2},\label{tdyn}
\eeqn
and hence, $\tau_{\rm vis}$ is much longer than $\tau_{\rm dyn}$.
Thus, if the system evolves by the viscous process, the evolution 
should proceed in a quasi-stationary manner.

For many models in this paper, the simulations are performed taking
into account the neutrino irradiation effect. For one model, K8n, we
switch off the neutrino irradiation effect to examine whether it is important or not. In this work, we do not
incorporate a heating effect by the neutrino pair
annihilation~\cite{Fujiba2018}, because the neutrino luminosity is not
very high for most of the evolution stage of the system in the models
employed. Only in the very early stage with $t \alt 10$\,ms, the
neutrino luminosity is high ($\agt 10^{53}\,{\rm erg/s}$ in total),
and thus, the neutrino pair annihilation heating may play a role in
the evolution of the disk and associated mass ejection.  However, in
this paper we should not consider mass ejection in such an early stage
because the system is initially in a spuriously varying phase due to a
rather artificial initial condition (composed of a stationary
equilibrium state of the disk of the perfect fluid), and thus, it is
not very clear whether the numerical results show some physical
phenomena or not for such an early stage. In one model with high disk
mass (model C8: see Table~\ref{table1}), the neutrino pair
annihilation heating may be important, but we do not take into account
also for this model because model C8 is employed just for the
comparison with low-disk mass models and the comparison should be done
in the same simulation setting.

As described in Sec.~\ref{sec2-1}, numerical simulations are carried
out typically with the grid spacing of $\Delta x=0.016M_{\rm BH}$ near
the black hole.  This grid resolution is higher than the resolution in
previous general relativistic magnetohydrodynamics
simulations~\cite{SM17,FTQFK18,FTQFK19,Miller19}, and this setting
enables us to perform a well-resolved simulation for the inner part of
the disk (i.e., the thermal and geometrical structure of the disk can
be well resolved). For model K8, we perform simulations varying the
grid resolution as $\Delta x=0.0133M_{\rm BH}$ and $0.020M_{\rm BH}$
to confirm the only weak dependence of the numerical results on the grid
resolution.

\subsection{Viscous hydrodynamics of disks for $M_{\rm disk}=0.1M_\odot$}\label{sec3-2}

\begin{figure*}[t]
\includegraphics[width=85mm]{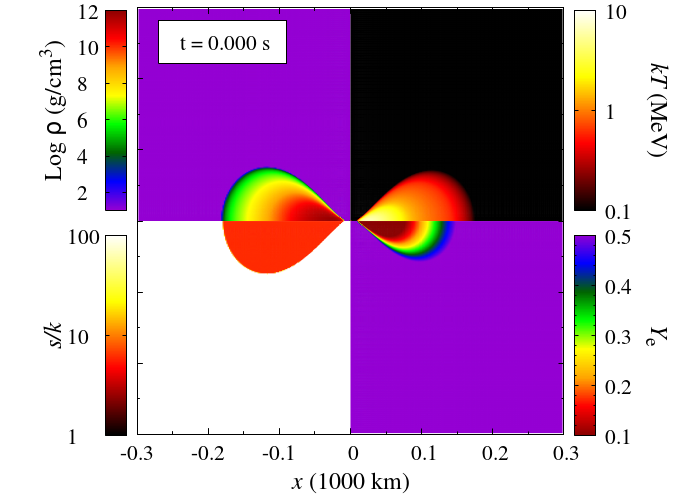} 
\includegraphics[width=85mm]{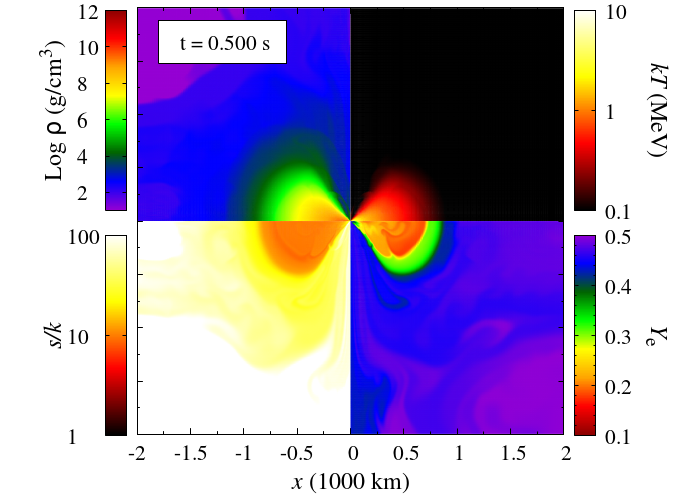} \\
\includegraphics[width=85mm]{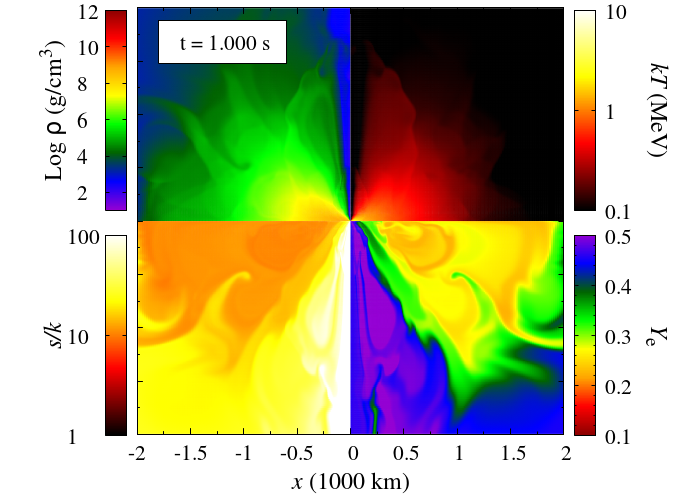} 
\includegraphics[width=85mm]{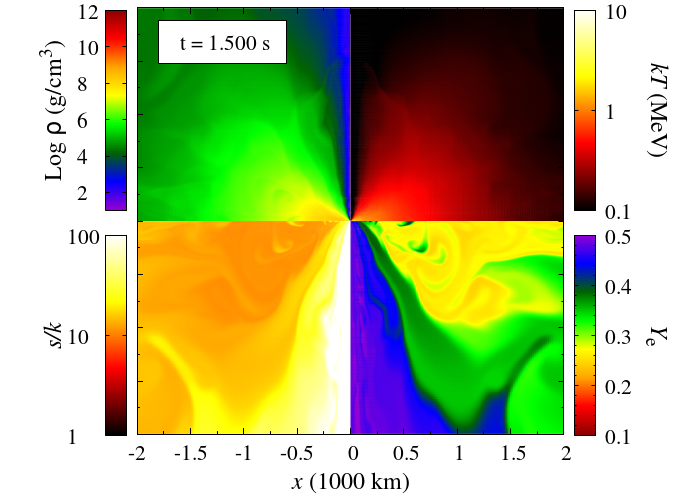} \\
\includegraphics[width=85mm]{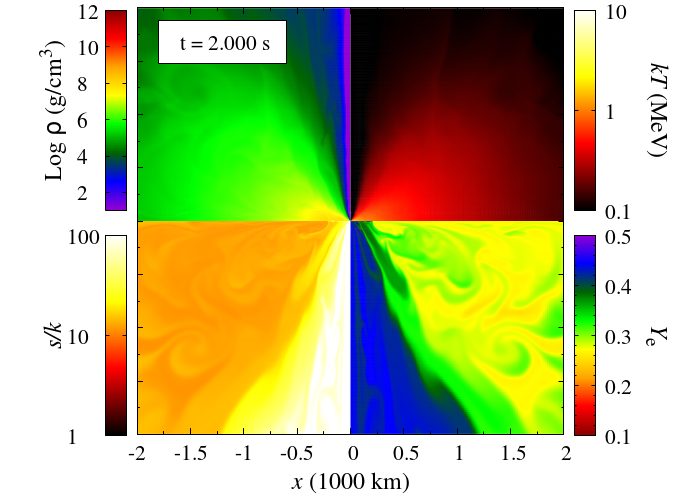} 
\includegraphics[width=85mm]{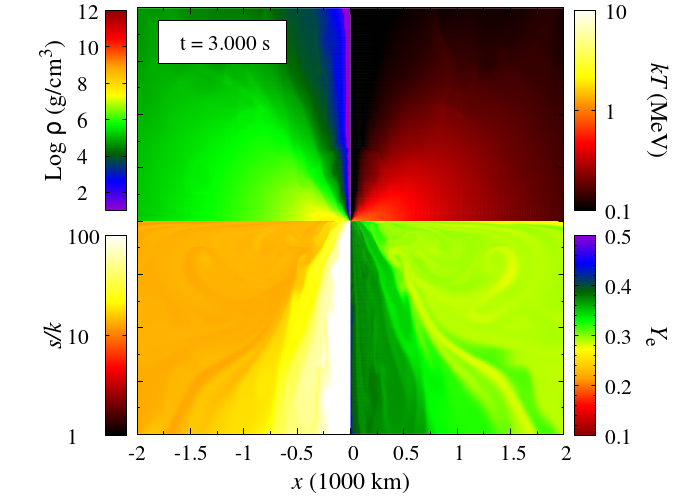} 
\caption{Snapshots for the rest-mass density in units of ${\rm
    g/cm^3}$, temperature ($k T$) in units of MeV, specific entropy
  per baryon in units of $k$, and electron fraction $Y_e$ for selected
  time slices for model K8. Only for the first panel ($t=0$) the
  plotted region is 300\,km$\times 300$\,km, and for others, it is
  2000\,km$\times 2000$\,km. We note that the rest-mass density, the
  value of $Y_e$, and temperature of the atmosphere artificially added
  is $\approx 10\,{\rm g/cm^3}$, 0.5, and $\approx 0.036$\,MeV$/k$,
  respectively (cf. the first panel).
\label{fig4}}
\end{figure*}

\begin{figure*}[t]
\includegraphics[width=85mm]{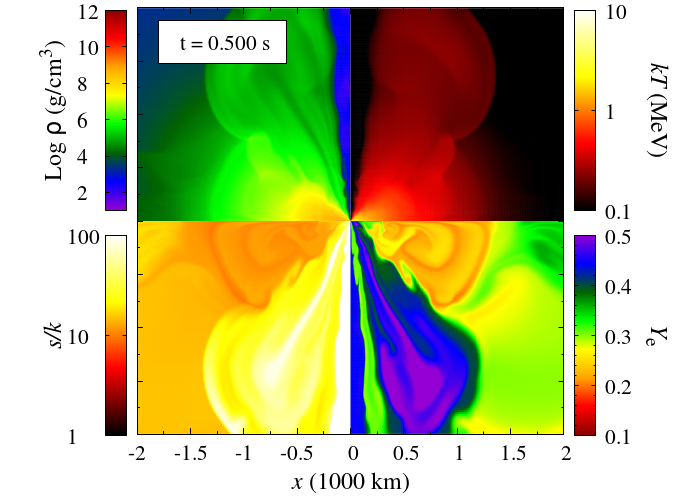} 
\includegraphics[width=85mm]{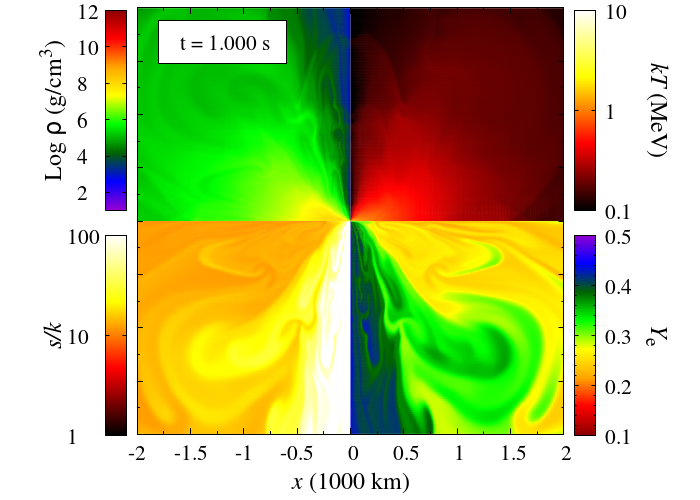} \\
\includegraphics[width=85mm]{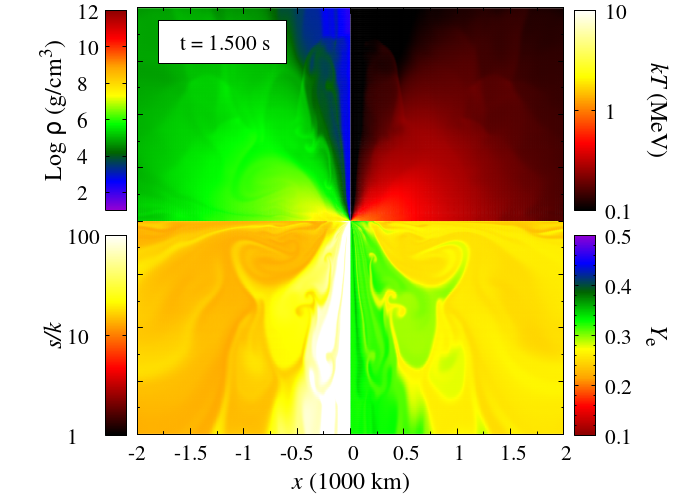} 
\includegraphics[width=85mm]{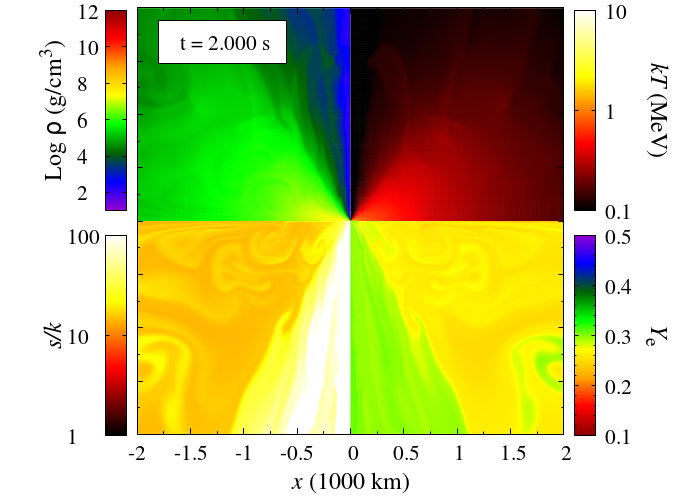}
\caption{The same as Fig~\ref{fig4} but for model K8h.
\label{fig5}}
\end{figure*}


This subsection focuses on presenting the results for the models with
$M_{\rm disk}=0.1M_\odot$, except for model D8.  Because the disk is
more compact for model D8 than for others, the fraction of the disk
matter that falls into the black hole is larger and the ejecta mass is
smaller than those for the other less-compact disk models. Besides
this difference, the numerical results depend only weakly on the
initial disk compactness. Thus, we only briefly summarize the results
for model D8 in Appendix B, comparing them with those for model K8,
and in this subsection, we only show the results for other models. 

\begin{figure*}[t]
(a)\includegraphics[width=84mm]{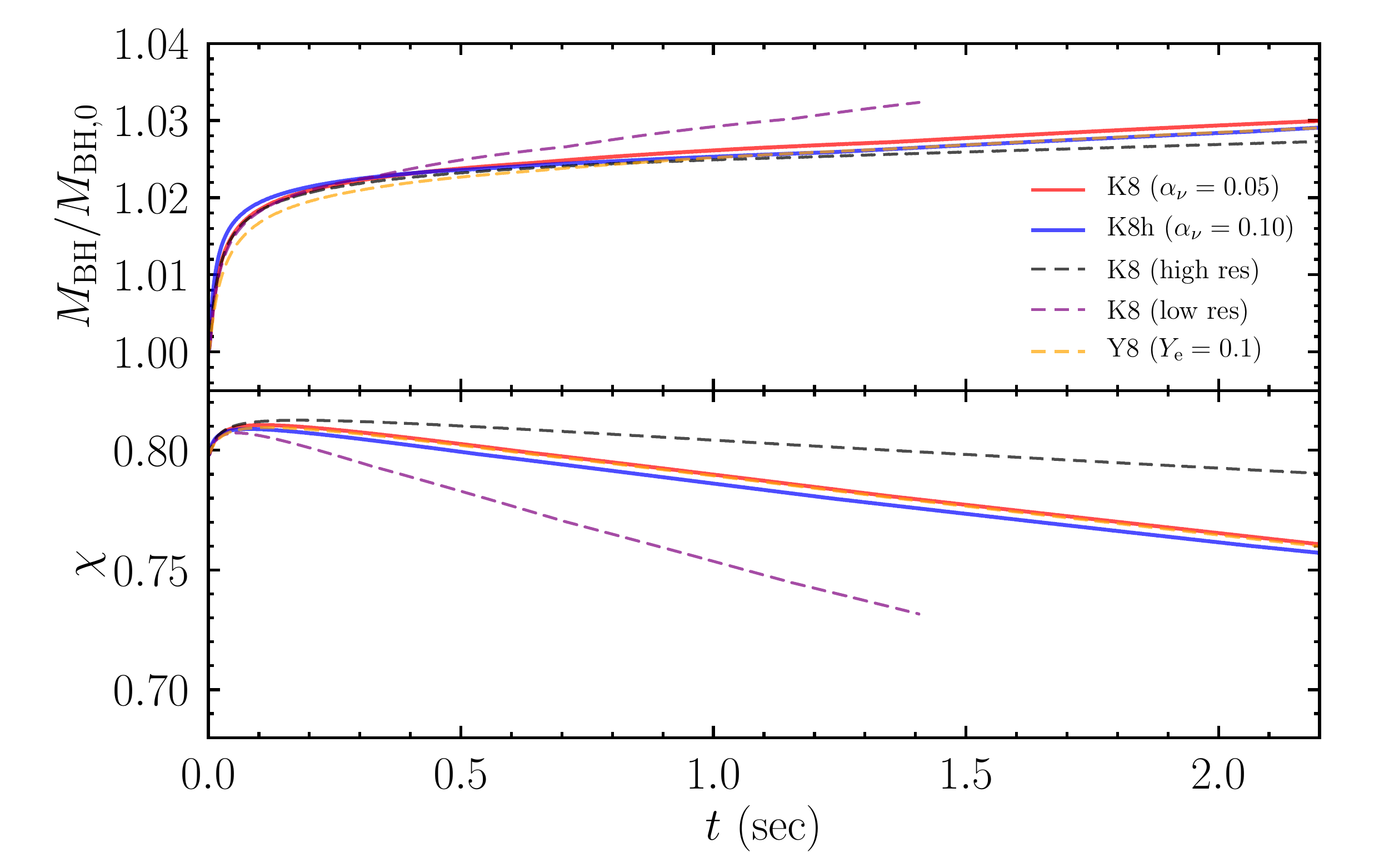} 
(b)\includegraphics[width=84mm]{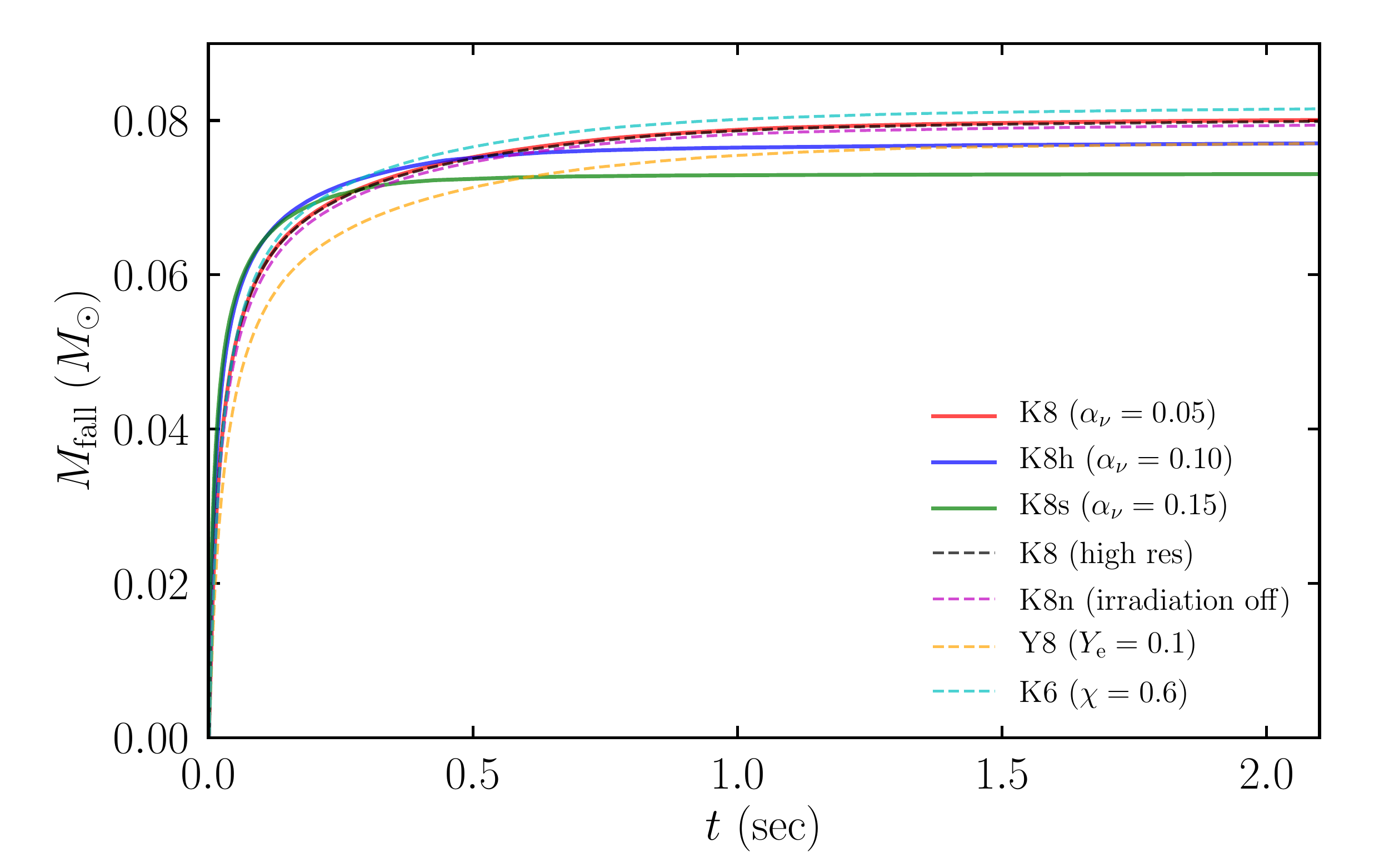} 
\caption{Left: Evolution of the black-hole mass and dimensionless spin
  for models K8, K8h, and Y8. For model K8, the results with three
  different grid resolutions are shown. The gradual decrease of $\chi$
  is due to the insufficient grid resolution.  Right: Total amount of
  the rest mass swallowed into the black hole for models K8 (with two
  grid resolutions), K8h, K8s, K8n, Y8, and K6.
\label{fig6}}
\end{figure*}

Figures~\ref{fig4} and \ref{fig5} display the evolution of the
profiles for the rest-mass density, temperature, specific entropy per
baryon, and electron fraction for models K8 and K8h.
Figure~\ref{fig6} plots (a) the evolution of the black-hole mass and
dimensionless spin for models K8 with three different grid resolution,
K8h, and Y8, and (b) the total rest mass swallowed by the black hole
for all the models with the initial disk mass $M_{\rm
  disk}=0.1M_\odot$. The mass and dimensionless spin of the black hole
during the evolution are approximately determined in the same method
as in Ref.~\cite{K10}: We calculate the area, $A_{\rm AH}$, and
circumferential radii around the equatorial and meridian planes, $c_e$
and $c_p$, for the apparent horizon of the black hole, and then
estimate the mass and dimensionless spin, assuming that $A_{\rm AH}$,
$c_e$, and $c_p$ are written as functions of the mass, $M_{\rm BH}$,
and dimensionless spin, $\chi$, of Kerr black holes as in the vacuum
black-hole case.

\begin{figure*}[t]
(a)\includegraphics[width=84mm]{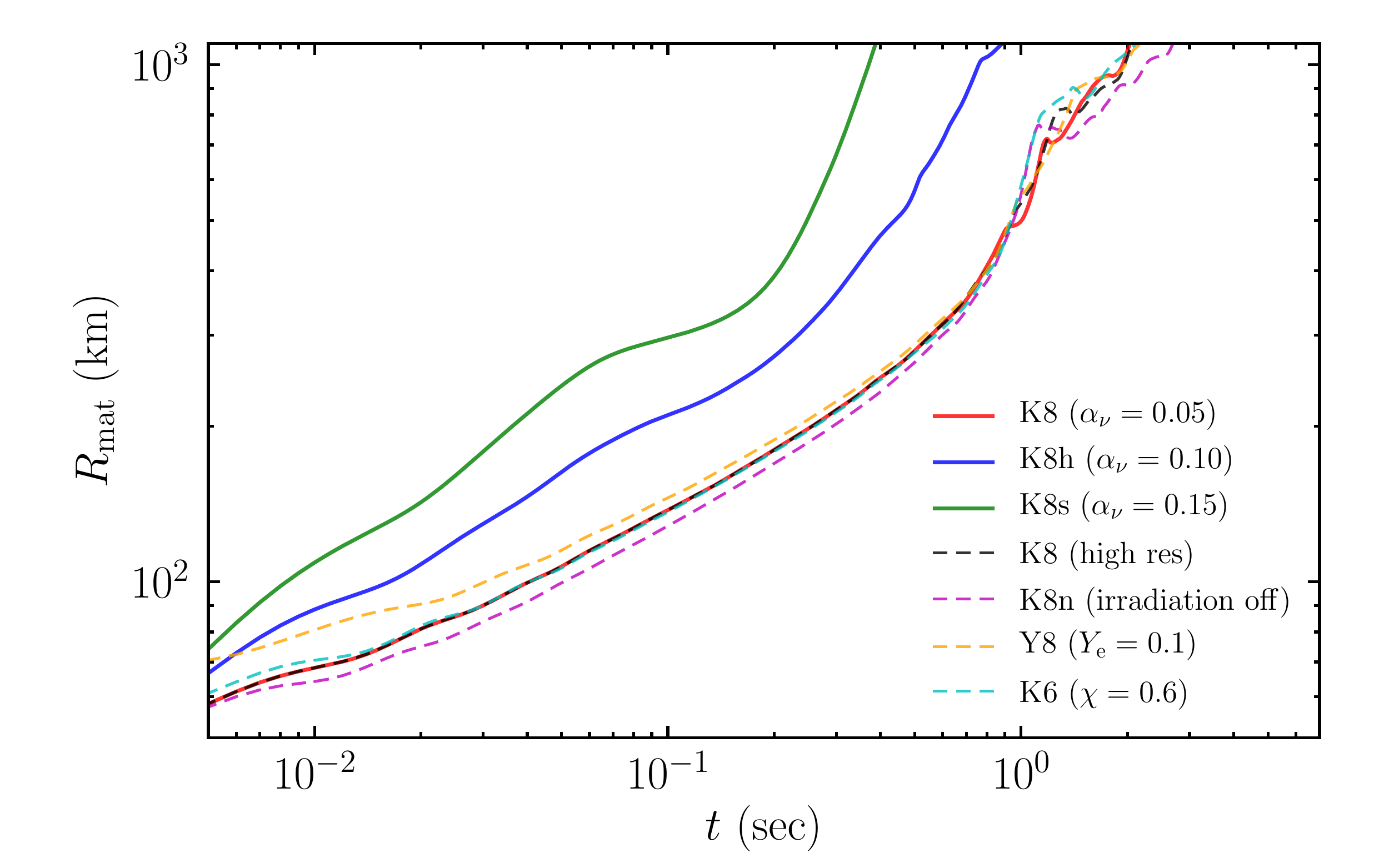}~~
(b)\includegraphics[width=84mm]{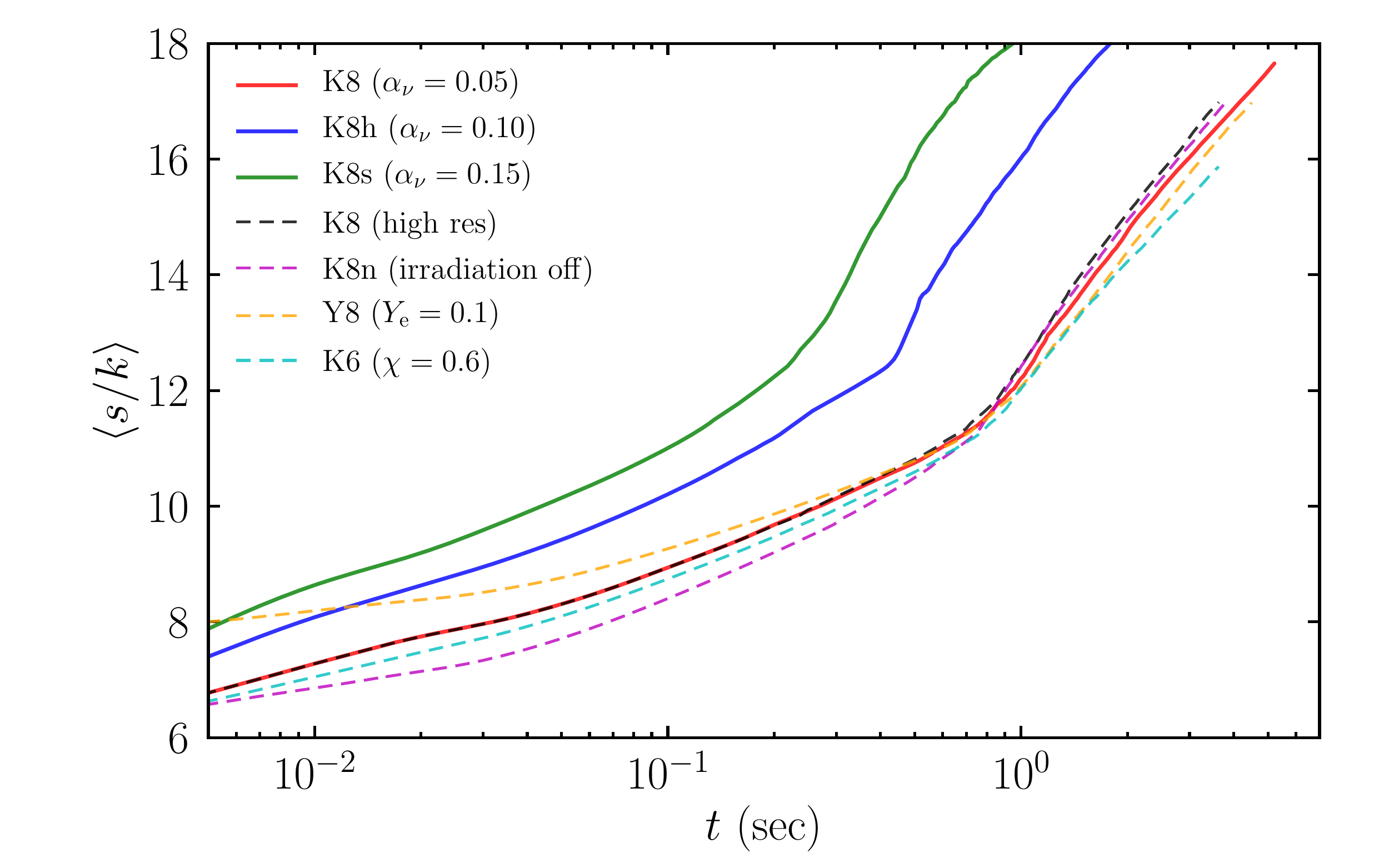} \\
(c)\includegraphics[width=84mm]{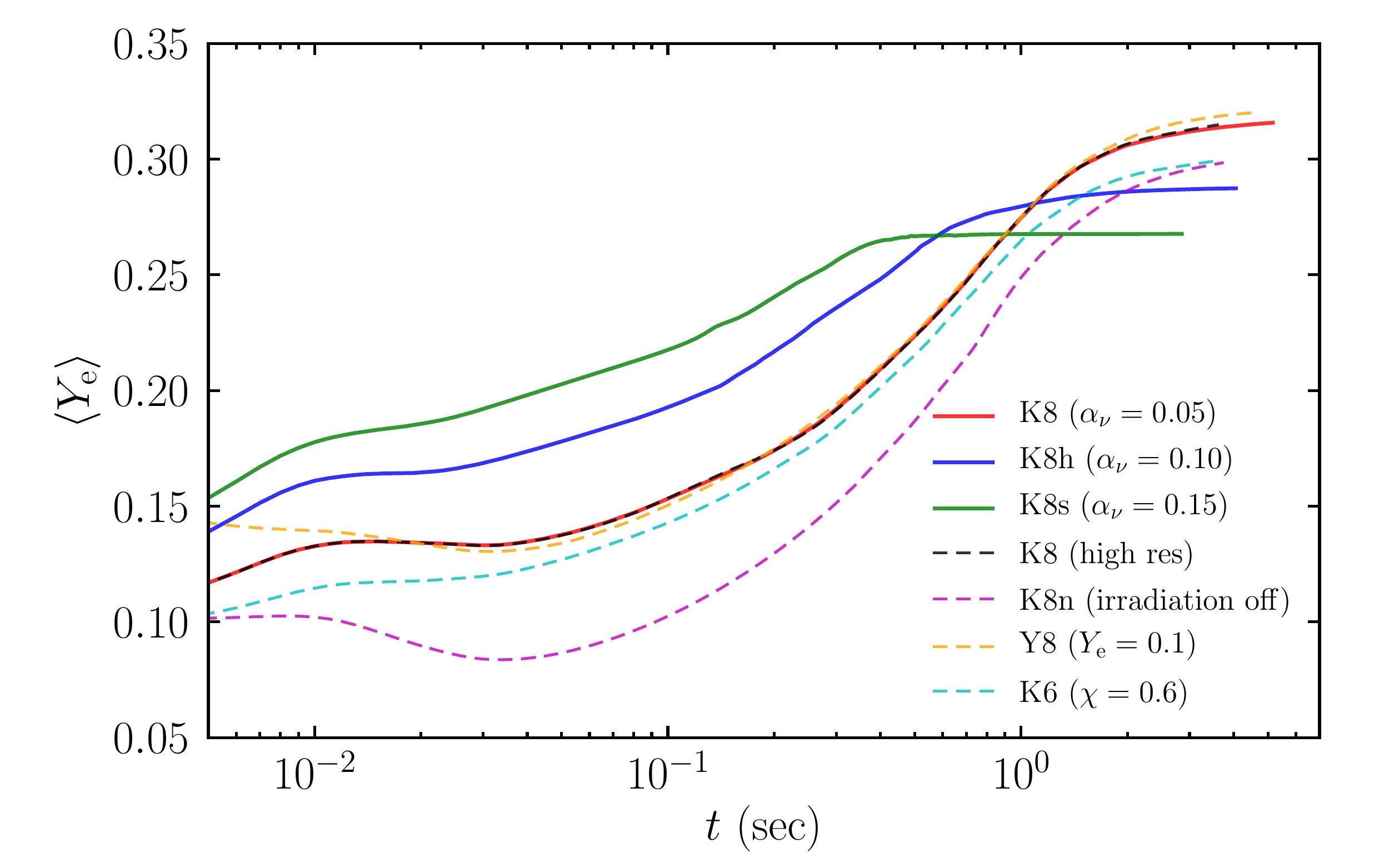}~~
(d)\includegraphics[width=84mm]{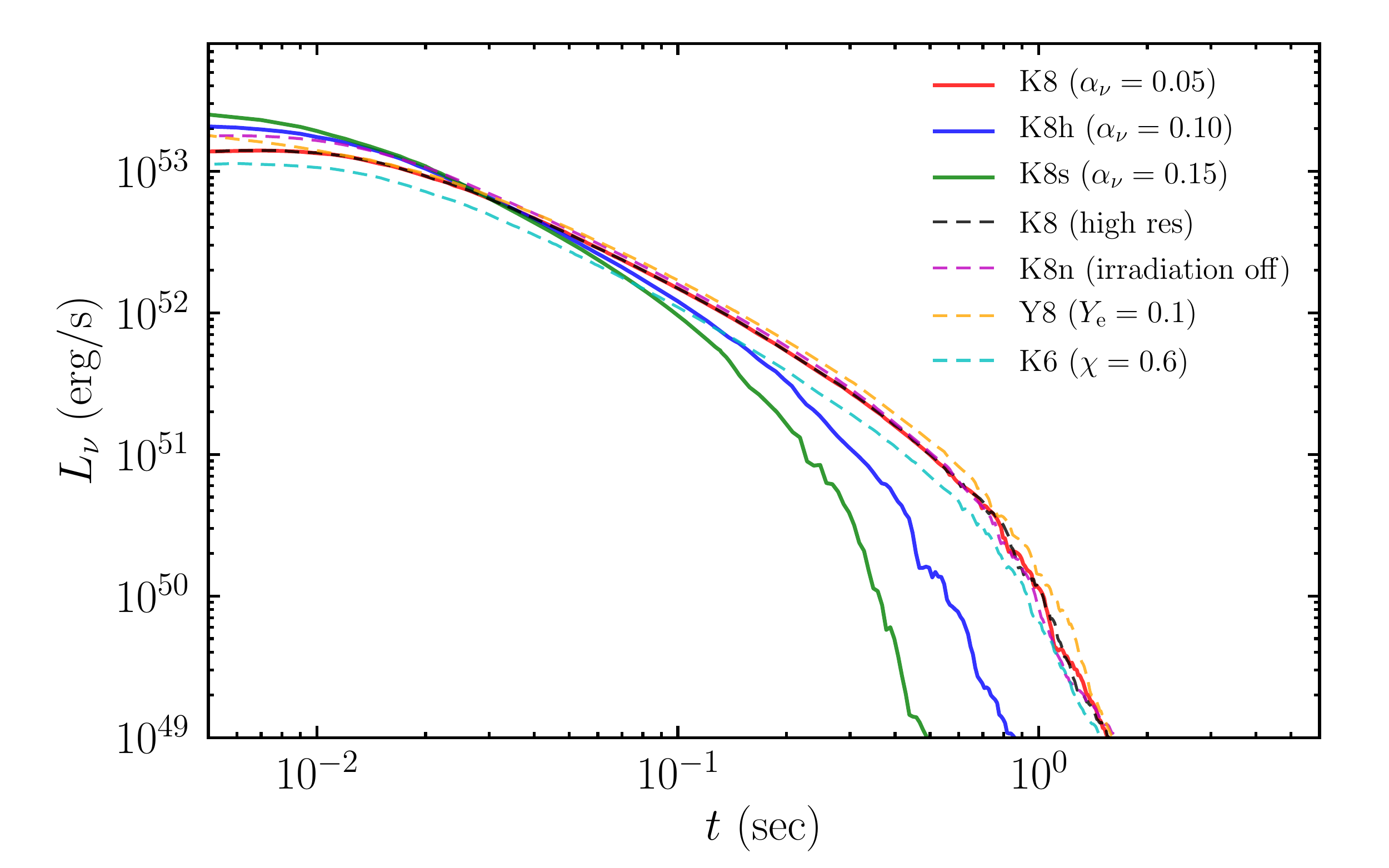} \\
(e)\includegraphics[width=84mm]{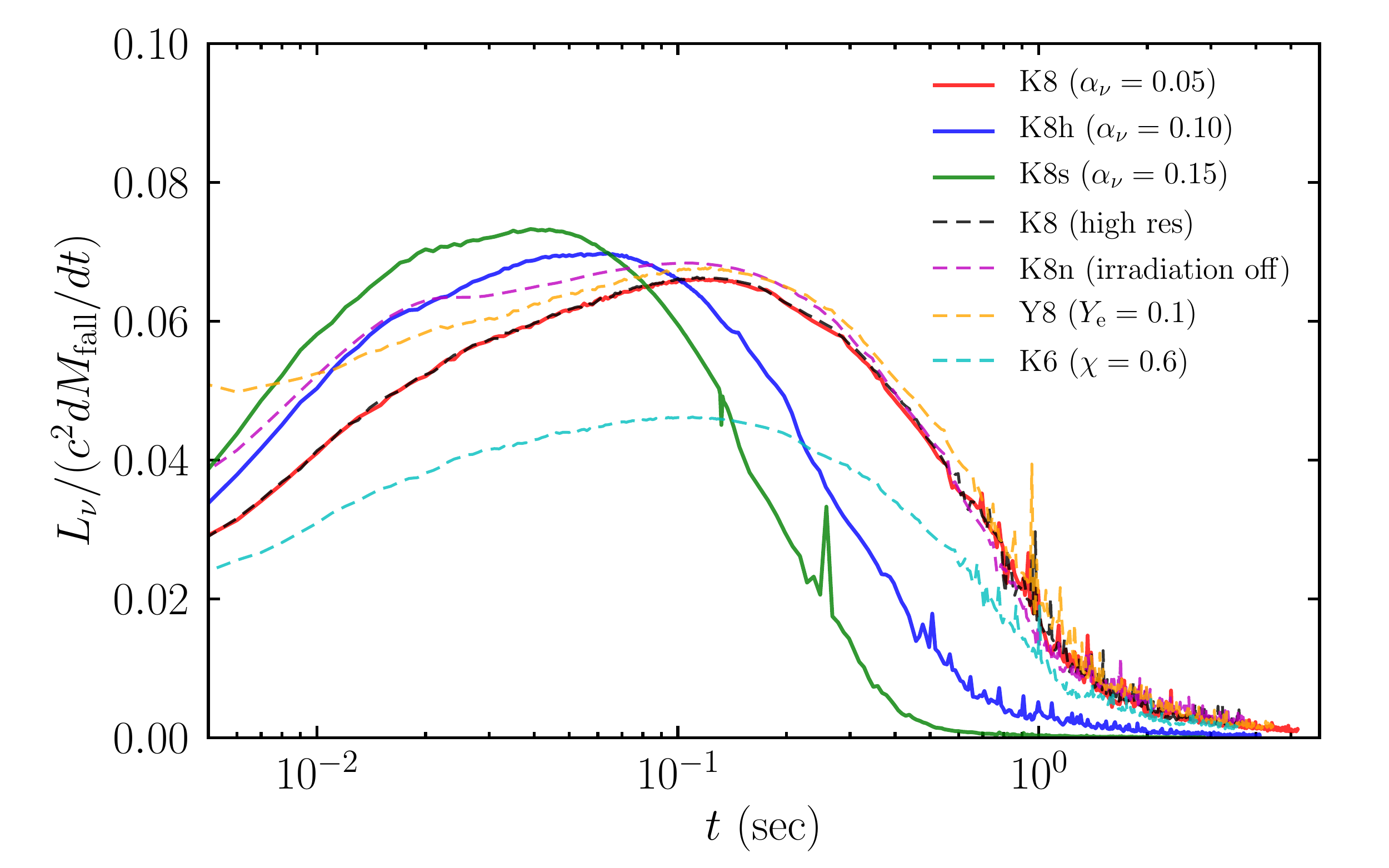}~~
(f)\includegraphics[width=84mm]{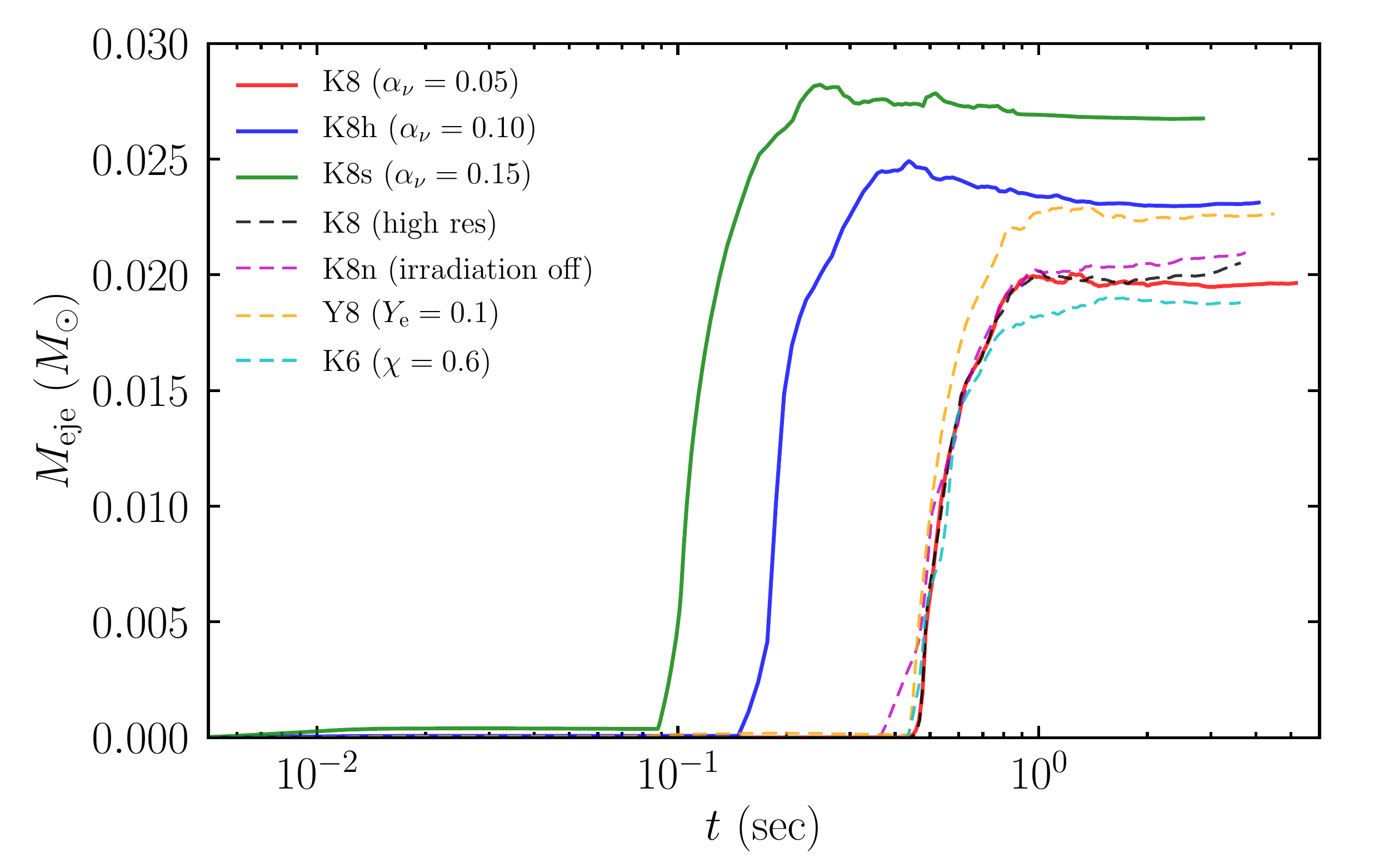} 
\caption{Several quantities for models K8, K8h, K8n, Y8, and K6.
  Average cylindrical radius (top left), average specific entropy (top
  right), and average value of $Y_e$ (middle left) of the matter
  located outside the black hole as functions of time. The middle
  right panel shows the time evolution of the total neutrino
  luminosity, $L_\nu$. The bottom two panels show 
an efficiency of the neutrino emission defined by the total neutrino
luminosity, $L_\nu$, divided by the rest-mass energy accretion rate of
the matter into the black hole, $c^2 dM_{\rm fall}/dt$ (left) and
total mass of the ejecta component $M_{\rm eje}$ (bottom right) as
functions of time.  For model K8, the results with a higher resolution
run are also plotted and show their weak dependence on the grid
resolution.
\label{fig7}}
\end{figure*}

The first three panels of Fig.~\ref{fig7} show the evolution of the
average cylindrical radius $R_{\rm mat}$, average value of the
specific entropy $\langle s \rangle$, and average value of $Y_e$,
$\langle Y_e \rangle$, for the matter located outside the black
hole. Here, these average quantities are defined by
\beqn
R_{\rm mat}&:=&\sqrt{I \over M_{\rm mat}}, \\
\langle s \rangle &:=&{1 \over M_{\rm mat}} \int_{\rm out} \rho_* s \,d^3x, \\
\langle Y_e \rangle &:=&{1 \over M_{\rm mat}} \int_{\rm out} \rho_* Y_e \,d^3x, 
\eeqn
where $\rho_*:=\rho \sqrt{-g} u^t$, and 
$I$ and $M_{\rm mat}$ denote a moment of inertia and rest mass of the 
matter located outside the black hole, defined by 
\beqn
I&:=& \int_{\rm out} \rho_* (x^2+y^2) \,d^3x, \\
M_{\rm mat}&:=& \int_{\rm out} \rho_* \,d^3x. 
\eeqn
$\int_{\rm out}$ implies that the volume integral is performed for the
matter outside the black hole.  The last three panels of
Fig.~\ref{fig7} show the total neutrino luminosity $L_\nu$, an
efficiency of the neutrino emission defined by the total neutrino
luminosity, $L_\nu$, divided by the rest-mass energy accretion rate of
the matter into the black hole, $c^2 dM_{\rm fall}/dt$, and the total
ejecta mass $M_{\rm eje}$ as functions of time. Note that the neutrino
luminosity is defined by the total neutrino emission rate minus the
neutrino absorption rate, both of which are calculated by the volume
integral.

As Figs.~\ref{fig4}--\ref{fig7} show, the disk evolves approximately
on the viscous timescale defined by Eq.~(\ref{tvis}).  We note that
the typical value of $c_s$ is $0.05c$ at $R \sim 100$\,km and $c_s$ is
a decreasing function of $R$.  In the early stage of the evolution of
the disk with the timescale less than 200\,ms, a substantial fraction of
the inner part of the disk with small values of $R$ falls into the
black hole. Irrespective of the models and grid resolutions, $\sim
60$--70\% of the initial disk mass, $M_{\rm disk}$, falls into the
black hole during this early stage, and as a result, the black-hole
mass increases by $\sim 0.6$--$0.7M_{\rm disk}$ as found in
Fig.~\ref{fig6}(a).  Subsequently, the infall rate significantly
decreases (see Fig.~\ref{fig6}(b)) and for $t \agt 1$\,s the accretion
rate onto the black hole becomes smaller than the mass ejection rate
(mass outflow rate measured for a sphere of $r=2000$--4000\,km), which
is typically $\dot M_{\rm eje} \sim 10^{-2}M_\odot$/s at the peak.
Figure~\ref{fig6}(a) also shows that for $t \alt 200$\,ms, the
dimensionless spin slightly increases due to the matter accretion onto
the black hole. We note that the subsequent gradual decrease of $\chi$
is due to the insufficient grid resolution (cf.~Sec.~\ref{sec3-6}).

As Fig.~\ref{fig4} and Fig.~\ref{fig7}(a) show, the outer part of the
disk expands outwards spending more than hundreds milliseconds by the
viscous effect.  Since the local viscous timescale for the outer part
of the disk with a large value of $R$ is longer than the inner part,
the expansion timescale there could be much longer than the timescale
of the matter infall to the black hole. However, because the viscous
heating/angular momentum transport in the inner part of the disk 
also contributes to the disk expansion, the outer part of the disk
expands on the timescale much shorter than the local viscous timescale
of Eq.~(\ref{tvis}). For $\alpha_\nu=0.05$, an outer edge of the disk
in the equatorial plane with its rest-mass density $\sim 10^6\,{\rm g/cm^3}$
reaches $\sim 10^3$\,km at $t \sim 0.7$\,s. Simulations for models K8h
and K8s show that this timescale is approximately proportional to
$\alpha_\nu^{-1}$ as predicted from Eq.~(\ref{tvis}) (e.g., compare
Figs.~\ref{fig4} and \ref{fig5}, and see Fig.~\ref{fig7}).


For $\alpha_\nu=0.05$, the mass ejection sets in (or strictly
speaking, the ejecta component appears) at $\sim 0.5$\,s after the
onset of the viscous evolution. Before this time, the mass ejection is
not activated as the second panel of Fig.~\ref{fig4} indicates (in
this panel, no region with $\rho \agt 10^3\,{\rm g/cm^3}$ is found for
$r \agt 10^3$\,km). A signal of the mass ejection is identified by the
evolution of $R_{\rm mat}$ in Fig.~\ref{fig7}(a), which shows a steep
increase of this quantity from $\sim 300$\,km to higher values
irrespective of the models. The primary driving force of the mass
ejection is the viscous heating in the inner region of the disk under
the negligible neutrino cooling.  For $t \agt 0.5$\,s, the disk has
already expanded by the gradual viscous heating/angular momentum
transport enough to decrease its temperature ($k T$) below $\sim
2$\,MeV (see Fig.~\ref{fig4}). As a result, the neutrino cooling does
not play an important role (the neutrino cooling timescale becomes
longer than the viscous timescale: see Appendix A), and thus, the
viscous heating is fully used for the disk heating and mass
ejection~\cite{MF2013,Just2015}. Indeed, the mass ejection efficiency
is enhanced when the neutrino luminosity decreases below a certain
threshold as observed in Fig.~\ref{fig7}(d) and (f). By contrast, in
the early disk evolution, in particular for $t \alt 100$\,ms during
which the total neutrino luminosity is $\agt 10^{52}\,{\rm erg/s}$,
the viscous heating is mostly consumed by the neutrino emission, and
thus, the viscous effect cannot have power to eject matter. Indeed,
the viscous heating efficiency, $\sim \nu M_{\rm disk} \Omega^2$, is
of the same order of $L_\nu$ for the typical values of $\nu$, $M_{\rm
  disk}$, and $\Omega$ of the disk.

The viscous heating is always most efficient in the innermost region
of the disk (i.e., the region closest to the black hole). The
enhancement of the specific entropy then triggers convective motion
from the innermost region toward outer regions, after the neutrino
cooling becomes inefficient.  We evaluated the Solberg-Holland
frequency for the convective instability~\cite{Tassoul,LM81}, and
found that the innermost region of the disk at $r \sim
100$--150\,km near the equatorial plane is indeed unstable to the
convective motion. The frequency of the convective instability in the
innermost region is $\sim (10\,{\rm ms})^{-1}$, and hence, the
timescale for the convective motion is much shorter than the viscous
timescale.

However, the convective activity is suppressed by the centrifugal
force, which tends to stabilize the convective motion. This
stabilization effect is strong in the vicinity of the main body of the
disk with the highest density.  As a result, the matter of the high
specific entropy produced in the innermost region goes along a
high-latitude region of the disk. Nevertheless, the high-entropy
convective blob eventually brings the thermal energy into the outer
part of the disk with a large cylindrical radius and with a high
latitude. By the increase of the thermal energy there, the matter in
the outer part of the disk obtains the energy enough to escape from
the system as ejecta. Therefore the onset time of the mass ejection is
determined approximately by the viscous heating timescale for the
innermost region of the disk at the moment that the neutrino emission
timescale becomes as long as the viscous heating timescale, and the
major process of energy transport is the convection
(cf.~Ref.~\cite{LRP2005}). By the convective activity, the matter is
ejected to any direction except along the rotation axis (see
Sec.~\ref{sec3-4}).


We note that the viscous heating/angular momentum transport in the
high-density region of the disk also contribute to the expansion of
the entire region of the disk. Thus not only the convective motion but
also the continuous viscous effect and resulting disk expansion 
play a role for the mass ejection.  This is in particular the case for
the high viscous coefficients as indicated in Sec.~\ref{sec3-4}. In
addition, the thermal energy stored in the innermost region by the
viscous heating drives an intermittent wind. This contributes to
non-steady mass ejection in particular for the late time evolution of
the disk in which the neutrino cooling plays a negligible role and
viscous heating can be fully available for driving the wind.

To examine the significance of the neutrino heating (irradiation)
effect, we also performed a simulation without neutrino heating (model
K8n).  Our result shows that the neutrino heating does not play a
substantial role for the mass ejection; the mass ejection rate and
ejecta velocity are not substantially influenced by the neutrino
heating. The reason for this is that the neutrino luminosity is not
very high during the mass ejection stage in our model (see
Fig.~\ref{fig7}(d)): Only in the early stage of the disk evolution
with $t \alt 20$\,ms, the total luminosity exceeds $10^{53}\,{\rm
  erg/s}$, while for $t > 100$\,ms for which the mass ejection becomes
active, the luminosity drops to $< 10^{52}\,{\rm erg/s}$
exponentially, resulting in the small contribution to the mass
ejection by the neutrino heating.\footnote{We note that the maximum
  energy obtained from neutrinos per nucleon (neutron or proton) via
  the neutrino absorption is estimated approximately as $\Delta E_\nu
  \sigma_\nu/(4\pi r^2)$ where $\sigma_\nu$ is the cross section of
  nucleons with neutrinos and $\Delta E_\nu$ is the total energy
  emitted by each neutrino species (here electron neutrino or anti
  neutrino). In model K8, for each, $\Delta E_\nu \sim 5 \times
  10^{52}\,{\rm erg/s} \times 0.02\,{\rm s} \sim 10^{51}\,{\rm erg}$,
  $\sigma_\nu \sim 8\times 10^{-42}\,{\rm cm}^2$ for neutrinos of the
  energy of $\sim 10$\,MeV, and $4\pi r^2 \agt 10^{13}\,{\rm cm}^2$
  because $r \agt 2GM_{\rm BH}/c^2 \sim 9$\,km. If we do not consider
  the neutrino cooling, the maximum energy obtained from neutrinos
  would be $\sim 800(2GM_{\rm BH}/c^2r)^2$\,MeV per nucleon.  In
  reality the energy gain of each nucleon is much smaller than this
  because the cooling by the neutrino emission could be of the same
  order as the heating. However, even if we ignore the neutrino
  cooling, the obtained energy gain is at most as large as the
  gravitational potential energy of nucleons, $\sim GM_{\rm BH}m_n/r$,
  where $m_n c^2$ is the rest-mass energy of nucleons $\sim
  940$\,MeV. In particular, for $r \geq 10GM_{\rm BH}/c^2$, the
  estimated energy gain is by more than one order of magnitude smaller
  than the gravitational potential energy.}  This result is in broad
agreement with that of Ref.~\cite{Just2015}.

After the onset of the convective motion at $t \sim 0.5$\,s (for
$\alpha_\nu=0.05$ with/without neutrino irradiation), the viscous heating
and resulting convection continue to play a leading role for the mass
ejection, although the mass ejection rate gradually decreases with the
decrease of the disk density.  The total ejecta mass is 15--30\% of
the initial disk mass (see Fig.~\ref{fig7} and Table~\ref{table2}).
For a fixed viscous coefficient with $\alpha_\nu=0.05$, the ejecta
mass is $\approx 15$--$25$\% irrespective of the initial condition
(i.e., initial $Y_e$ distribution, density and velocity profiles,
compactness of the disk, and black-hole spin).  The fraction of the
ejecta mass agrees broadly with the results of the earlier viscous
hydrodynamics work~\cite{MF2013,MF2014,Just2015}.

The entropy and electron fraction in the disk increase with its
viscous and convective expansion, after the initial quick
matter accretion onto the black hole for $t \alt 100$\,ms ceases (see
Fig.~\ref{fig7}(b) and (c)). The typical average value of the specific
entropy is 10--$12k$ when the mass ejection is activated. After the
initial infall stage of the disk matter onto the black hole, the
average value of $Y_e$ monotonically increases with the decrease of
the disk density, and it is higher than $\sim 0.2$ at the onset of the
mass ejection irrespective of the models for $\alpha_\nu=0.05$. In the
late time, the average value of $Y_e$ settles to constants of $\sim
0.3$. The mechanism for this $Y_e$ evolution is summarized as follows:
During the disk expansion ongoing until the weak-interaction freezes
out (i.e., the temperature is $k T \agt 2$\,MeV and the
  electron degeneracy is not so strong that the weak interaction rates
  are determined predominantly by the temperature: see Appendix A),
the weak interaction processes determine the electron fraction in the
disk (the value of $Y_e$ is approximately determined by the
  equality of the rates of electron/positron capture on nucleons: see
  Appendix A for more details) However, for $k T \alt 2$\,MeV, the
weak interaction plays a negligible role because the timescale for the
weak interaction processes, $\tau_\beta$, becomes longer than the
viscous timescale, $\tau_{\rm vis}$~\cite{MF2013} (here $\tau_\beta$
is approximately equal to the neutrino emission timescale).  Thus the
settled value of $Y_e$ is determined approximately by the condition of
$\tau_\beta=\tau_{\rm vis}$.  The onset time of this freeze out of the
weak interaction agrees approximately with the time that the mass
ejection is activated.  In our numerical result, the relaxed average
values of $Y_e$ are not very small, i.e., $\agt 0.25$, irrespective of
the models for $\alpha_\nu=0.05$. This has an important implication
for the nucleosynthesis of lanthanide elements as we discuss in
Sec.~\ref{sec3-5}.


Figure~\ref{fig7}(e) shows that the total neutrino luminosity for
$\chi=0.8$ is $\sim 6$--7\% of the rest-mass energy accretion rate
onto the black hole at the maximum. This efficiency agrees with that in
Ref.~\cite{Just2015}.  The maximum efficiency is achieved in an early
stage of the disk evolution in which the neutrino luminosity is high
and weak interaction timescale is still shorter than the viscous
evolution timescale. This efficiency as well as the neutrino
luminosity is smaller for $\chi=0.6$ than for $\chi=0.8$. The reason
for this is that the radius of the innermost stable circular orbit is
larger for the smaller value of $\chi$, and thus, the depth of the
gravitational potential in the vicinity of the black hole is
shallower; i.e., the gravitational potential energy available for the
dissipation is smaller. For higher viscous coefficients, the
efficiency is higher, but the enhancement is not very
remarkable. Thus, the maximum efficiency is basically determined by
the spin of the black hole.  In the late stage of the disk evolution
in which the weak interaction freezes out, the efficiency approaches
zero because the emissivity of neutrinos exponentially drops. As
already mentioned, in this late stage, the mass ejection is enhanced.


Figure~\ref{fig5} displays the evolution of the profiles for model
K8h, i.e., for a larger viscosity model than model K8.  For
$\alpha_\nu=0.10$ and $0.15$ (K8h and K8s models), the mass ejection
sets in earlier than for model K8. This is clearly found by comparing
the profiles at $t=0.5$\,s in Figs.~\ref{fig4} and \ref{fig5}. The
reason for the earlier onset of the mass ejection is simply that the
viscous timescale becomes shorter, the density of the disk decreases
more quickly, and hence, the freeze out of the weak interaction occurs
earlier.  Another remarkable difference among three models K8, K8h,
and K8s is that the mass ejection from the inner edge of the disk as a
disk wind is more powerful for the higher viscosity models. This
occurs because the thermal pressure by the viscous heating and also
viscous angular momentum transport for $\alpha_\nu=0.10$ and 0.15 are
more enhanced than for $\alpha_\nu=0.05$, in particular in the
innermost region of the disk.

Associated with the stronger viscous heating, the specific entropy and
the average value of $Y_e$ in the disk increase more rapidly (see
Fig.~\ref{fig7}(b) and (c),early phases).  The reason for this is that the
temperature of the disk is higher and the electron degeneracy is
weakened at given time for the higher viscosity models. As we mentioned
above, however, the disk expansion occurs more quickly for the larger
viscous coefficients. As a result, the weak interaction freezes out
earlier, and also, the electron fraction settles to smaller values for
the larger viscous coefficients (compare the panels in
Figs.~\ref{fig4} and \ref{fig5} at $t=2$\,s and also see
Fig~\ref{fig7}(c)).  This modifies the $Y_e$ distribution of the
ejecta as discussed in Sec.~\ref{sec3-4}.

The larger viscous effect slightly suppresses the matter infall into
the black hole (see Fig.~\ref{fig6}(b)). Associated with this, the
total ejecta mass for $\alpha_\nu=0.10$ and 0.15 becomes larger than
that for $\alpha_\nu=0.05$ by $0.004M_\odot$ and $0.007M_\odot$,
respectively (see Fig.~\ref{fig7}(e) and (f) as well as
Table~\ref{table2}). All these results show that the magnitude of the
viscous coefficient significantly influences the disk evolution and
the quantitative properties of the ejecta.


Figure~\ref{fig7} clearly shows that the effect of the black-hole spin
is not substantial on the evolution of the system for the
astrophysically plausible values of $\chi$ as the merger remnants
(compare the results of models K8 and K6). We can find that the
average value of $Y_e$ and neutrino luminosity for model K6 are
slightly smaller than those for model K8, because the innermost edge
of the disk for model K6 is located slightly far outside the black
hole than for model K8, and as a result, the viscous heating
efficiency and resulting ejecta mass are slightly smaller. However,
this spin effect is minor for modifying the evolution of the system:
The effect in the change of the viscous coefficient by a factor of 2
has much stronger impact.

\begin{figure*}[t]
\includegraphics[width=85mm]{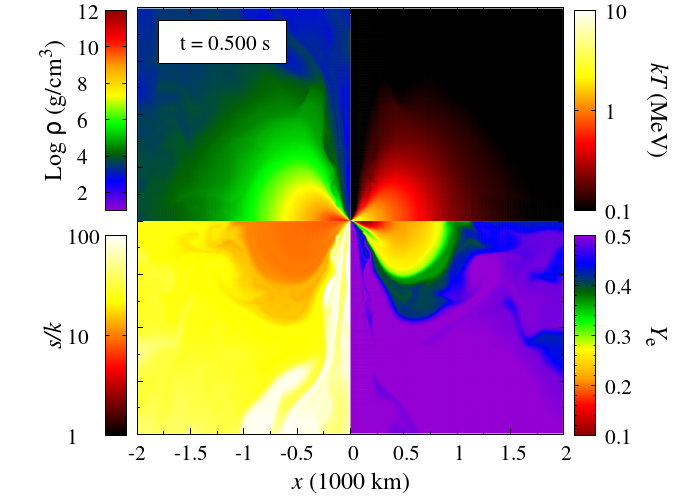} 
\includegraphics[width=85mm]{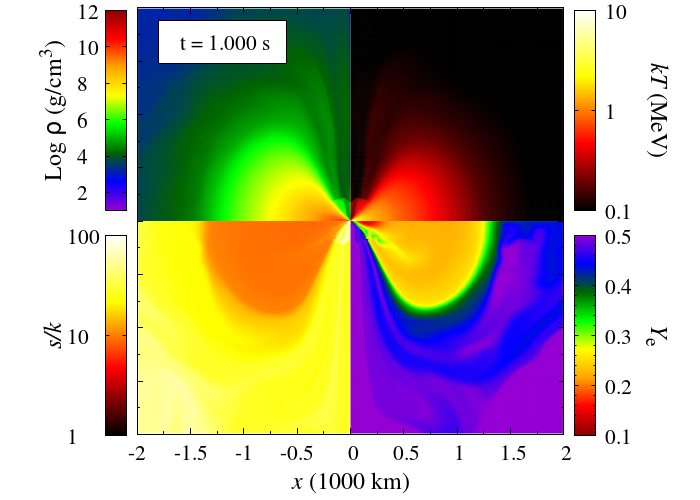} \\
\includegraphics[width=85mm]{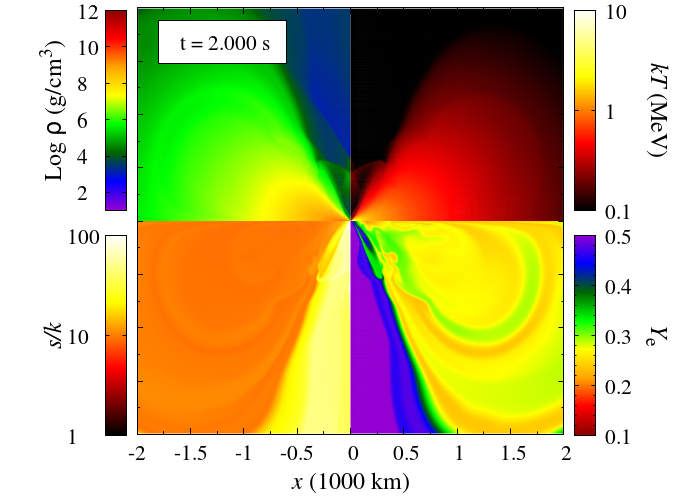} 
\includegraphics[width=85mm]{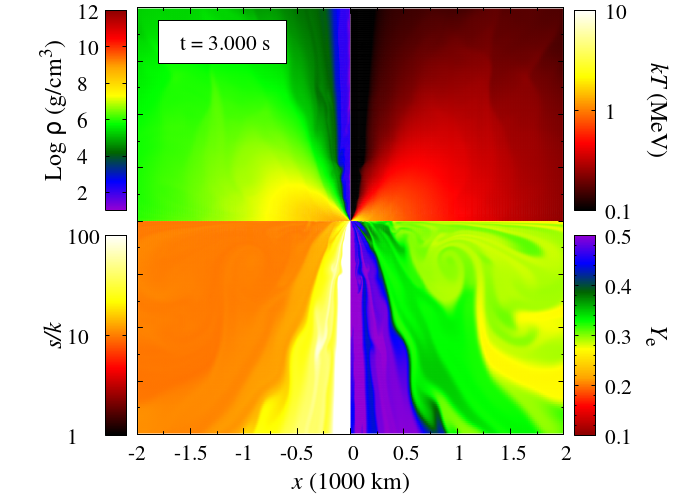} \\
\includegraphics[width=85mm]{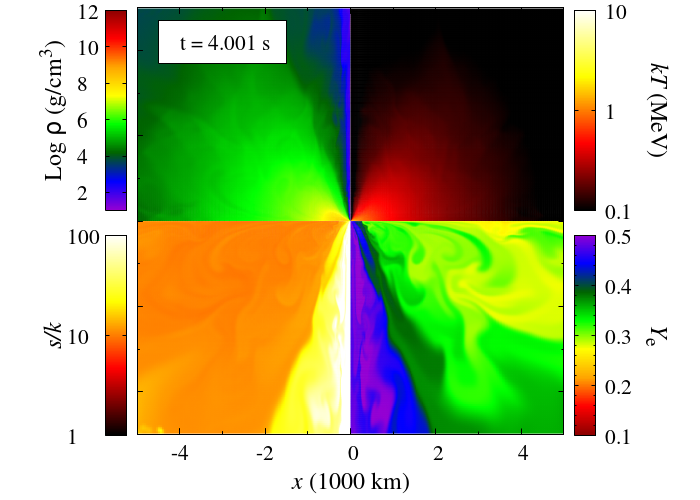}
\includegraphics[width=85mm]{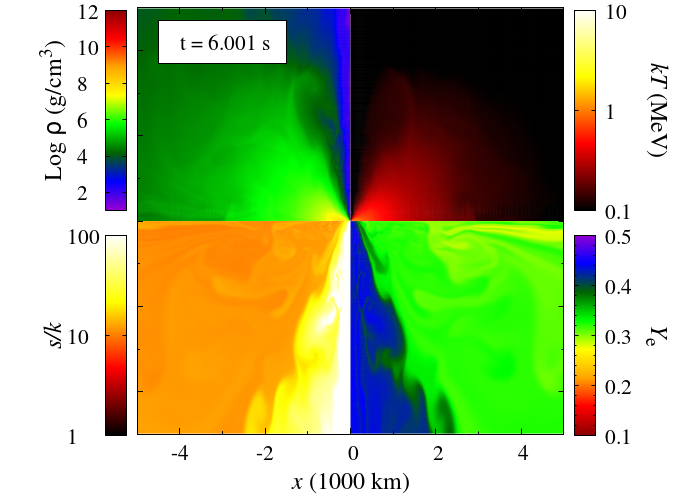}
\caption{The same as Fig~\ref{fig4} but for model C8.  For the last
  two panels at $t=4$ and 6\,s, the plots are made for an enlarged
  region of 5000\,km$\times 5000$\,km.
\label{fig8}}
\end{figure*}

\begin{figure*}[t]
(a)\includegraphics[width=84mm]{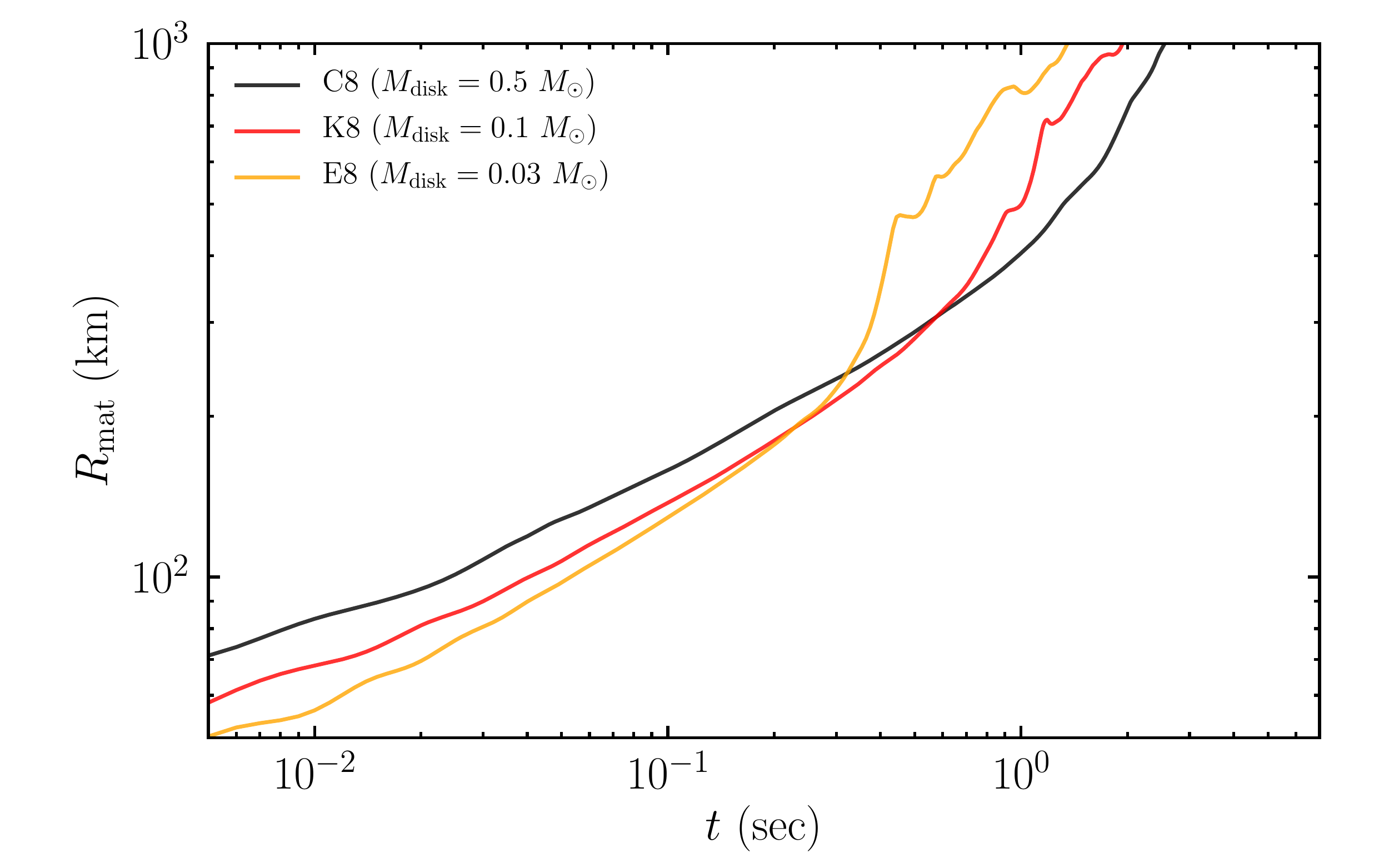}~~
(b)\includegraphics[width=84mm]{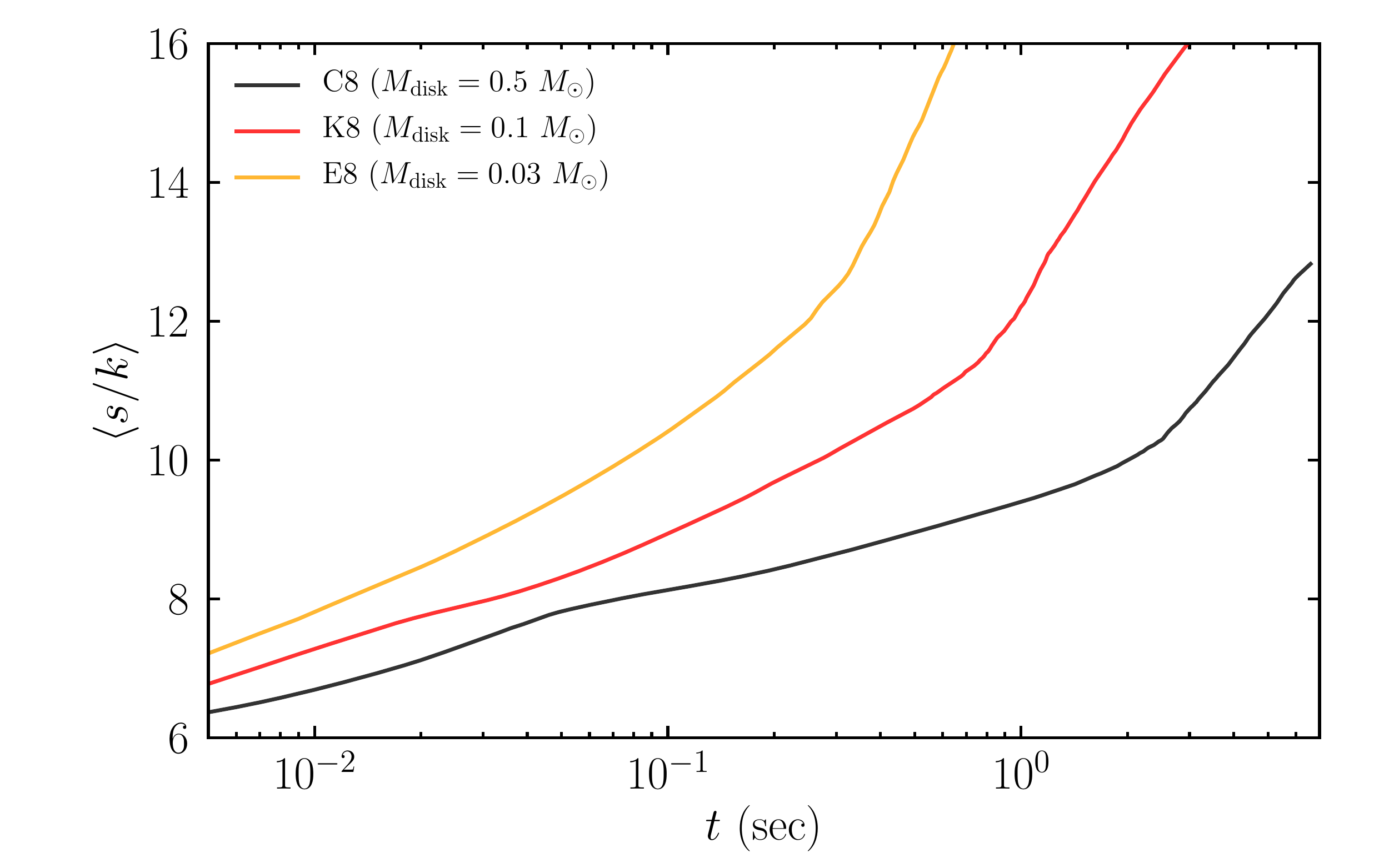} \\
(c)\includegraphics[width=84mm]{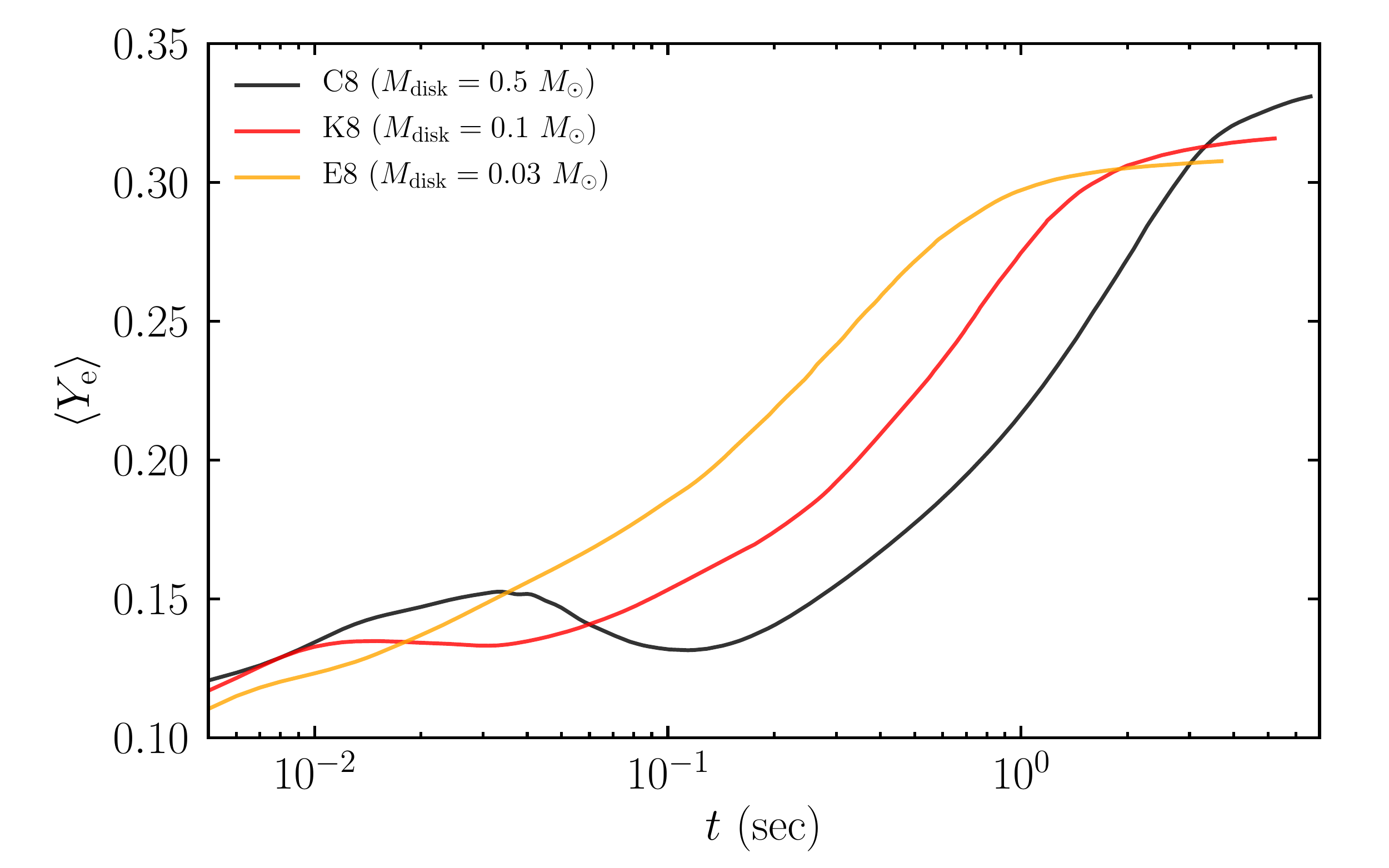}~~
(d)\includegraphics[width=84mm]{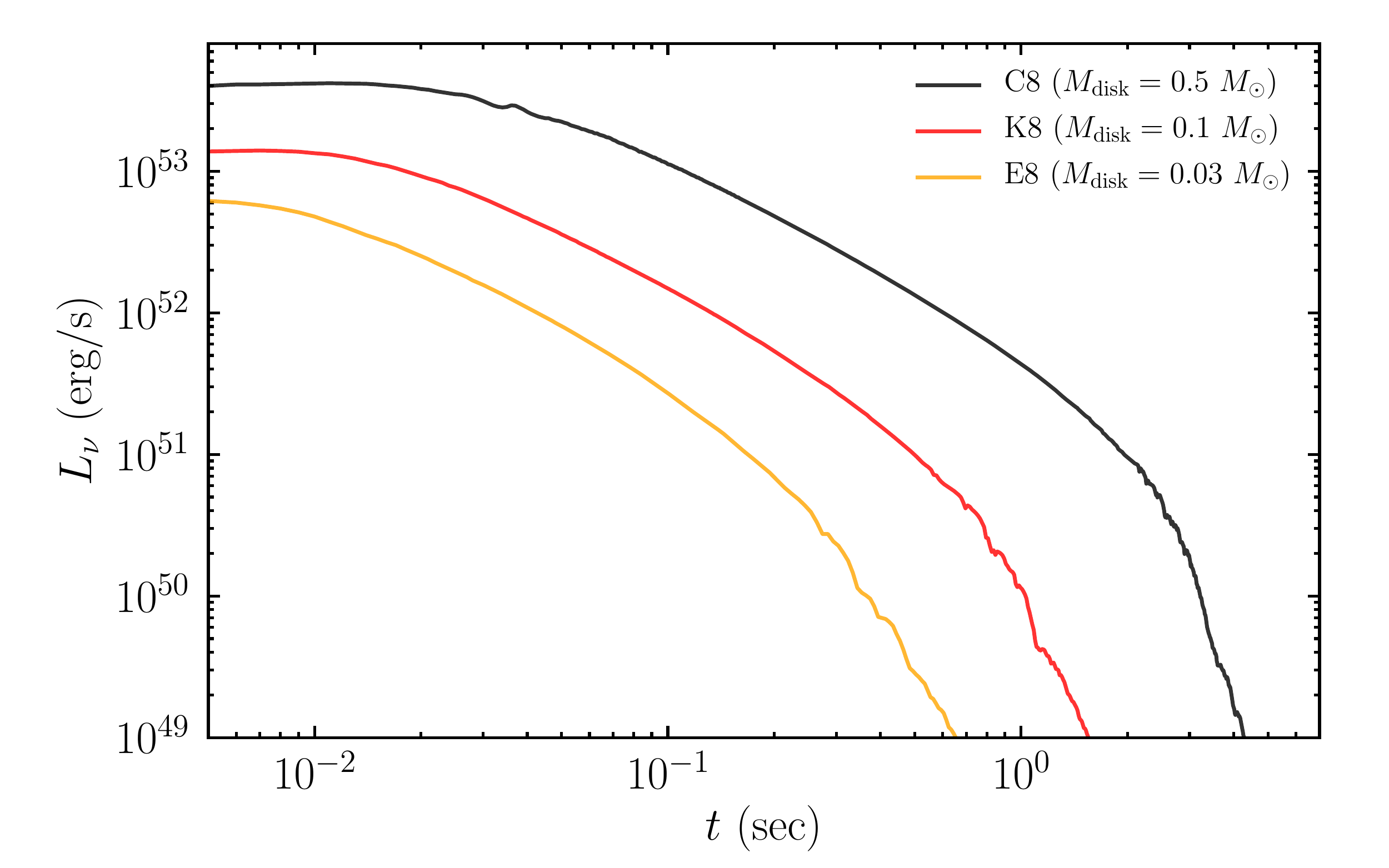} \\
(e)\includegraphics[width=84mm]{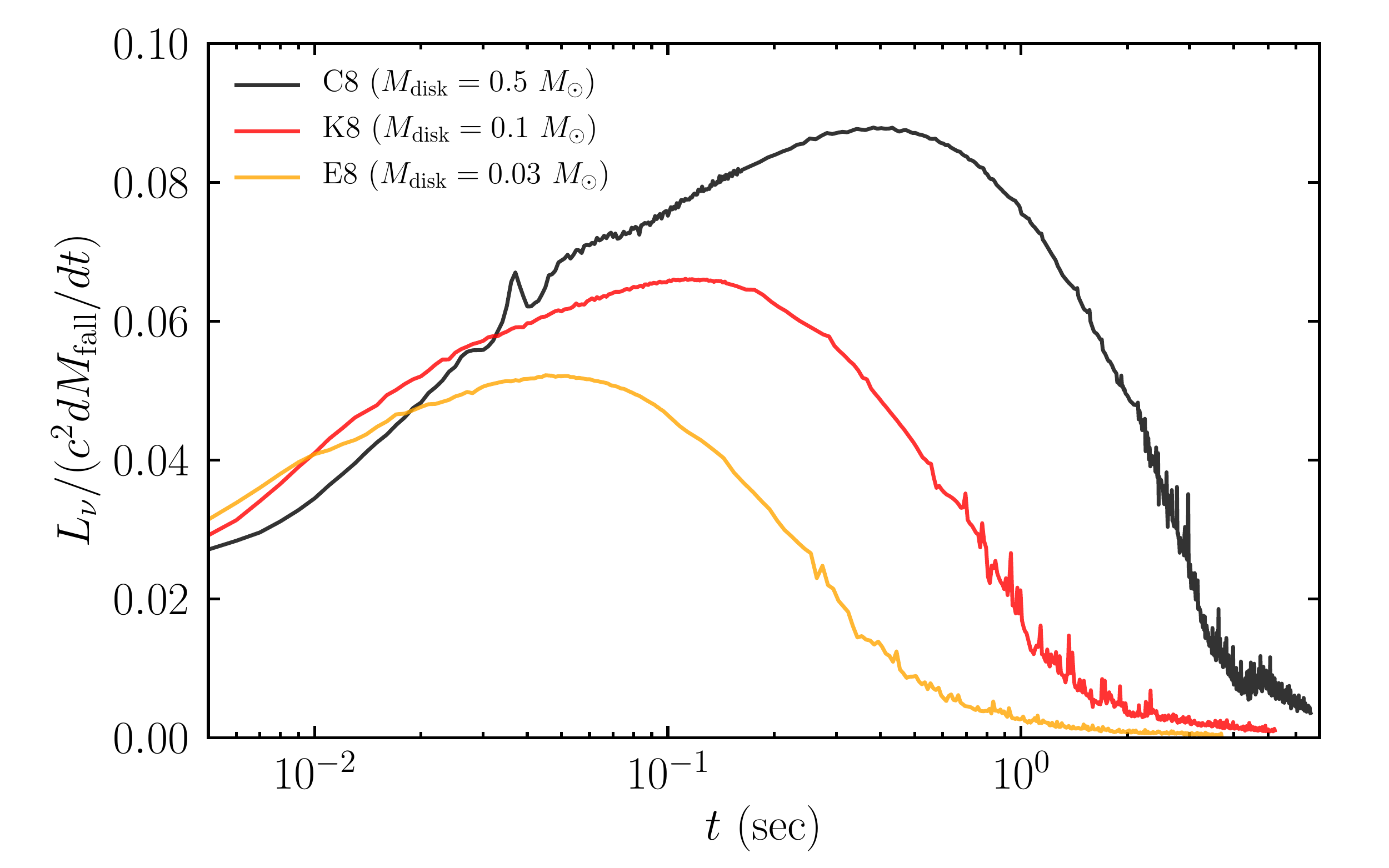}~~
(f)\includegraphics[width=84mm]{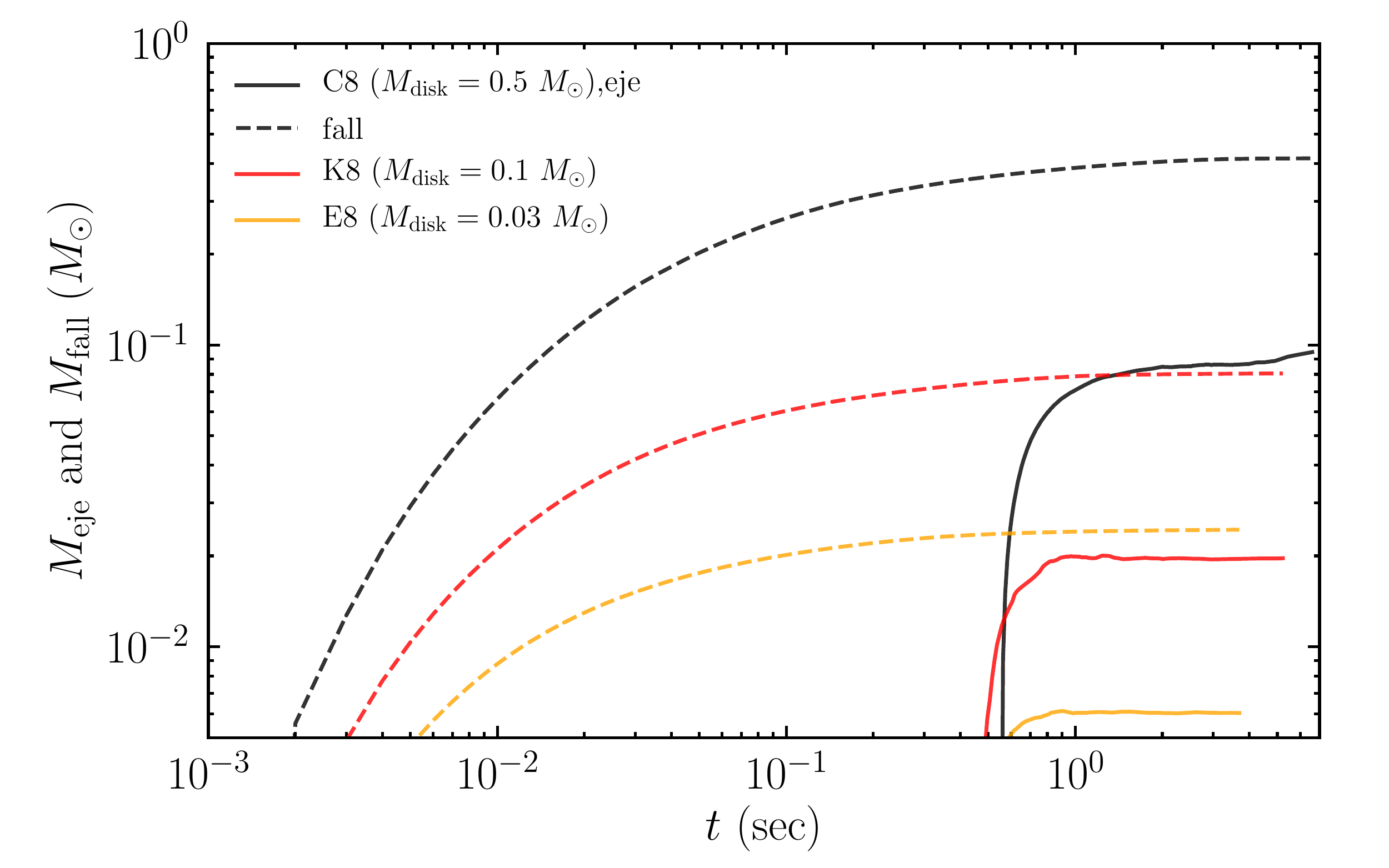} 
\caption{The same as Fig.~\ref{fig7} but for comparison among
  different disk mass models C8, K8, and E8.  In the bottom right
  panel, not only the total ejecta mass (solid curves) but also the
  total mass that falls into the black hole (dashed curves) are
  plotted.
\label{fig9}}
\end{figure*}

\subsection{Viscous hydrodynamics of disks: comparison among different disk mass models}\label{sec3-3}

The difference in the disk mass (compare models K8, C8, and E8)
results in the modification of the timescale to reach the freeze out
of the weak interaction and onset time of the mass ejection.
Figure~\ref{fig8} displays the snapshots of the profiles for the
rest-mass density, temperature, specific entropy per baryon, and
electron fraction for models C8 at selected time slices.
Figure~\ref{fig9} also compares the average cylindrical radius,
average specific entropy, average value of $Y_e$ for the matter
outside the black hole, total neutrino luminosity, an efficiency of
the neutrino emission, $L_\nu/(dM_{\rm fall}/dt)$, and ejecta mass as
well as the total mass swallowed by the black hole among models K8,
C8, and E8. These figures show that for the larger disk mass, the
density and temperature is always higher at given time and the freeze-out time of the
weak interaction comes later. As a result, for the more massive
models, the thermal energy generated by the viscous heating is released 
by the neutrino emission for a longer term, resulting in more luminous
and long-lasting neutrino emission. Then, the disk expansion timescale
becomes longer (see Fig.~\ref{fig9}(a)), the onset time of the
convection becomes later, and the mass ejection is
delayed~\cite{Just2015}. These facts are well reflected in the maximum
efficiency of the neutrino emission and the duration of the
high-luminosity neutrino emission (see Fig.~\ref{fig9}(c) and (e)).
Due to the delay of the freeze out of the weak interaction, the
average value of $Y_e$ in the disk at the freeze out of the weak
interaction becomes higher for more massive disk models (see
Fig.~\ref{fig9}(c)).  Figure~\ref{fig8} also illustrates that besides
the quantitative difference mentioned above, the evolution process of
the disk is qualitatively similar among different disk-mass models;
e.g., the mass ejection is primarily driven by the convective activity
caused by the viscous heating in the region of the disk close to the
black hole for all the models.

For the remnant of neutron-star mergers, disks with mass of $\agt
0.5M_\odot$ are not very likely. However, such a heavy disk
surrounding a spinning black hole could be formed as a result of
rotating massive stellar core collapses, in particular as a central
engine of long gamma-ray bursts~\cite{Woosley93}. Such heavy disks are
also likely to be evolved effectively by viscous-hydrodynamics
processes. Our present numerical results demonstrate that for large disk
mass, the electron fraction at which the weak interaction freezes out
could be large with $Y_e \agt 0.3$ (unless the viscous coefficient is
extremely large). Thus, if the viscous process is the dominant
mechanism of mass ejection, the matter ejected from the disk may not
be very neutron-rich for the remnant of stellar core collapses (see
also the discussion in the next sections).

\subsection{Properties of ejecta}\label{sec3-4}

\begin{figure*}[t]
(a)\includegraphics[width=84mm]{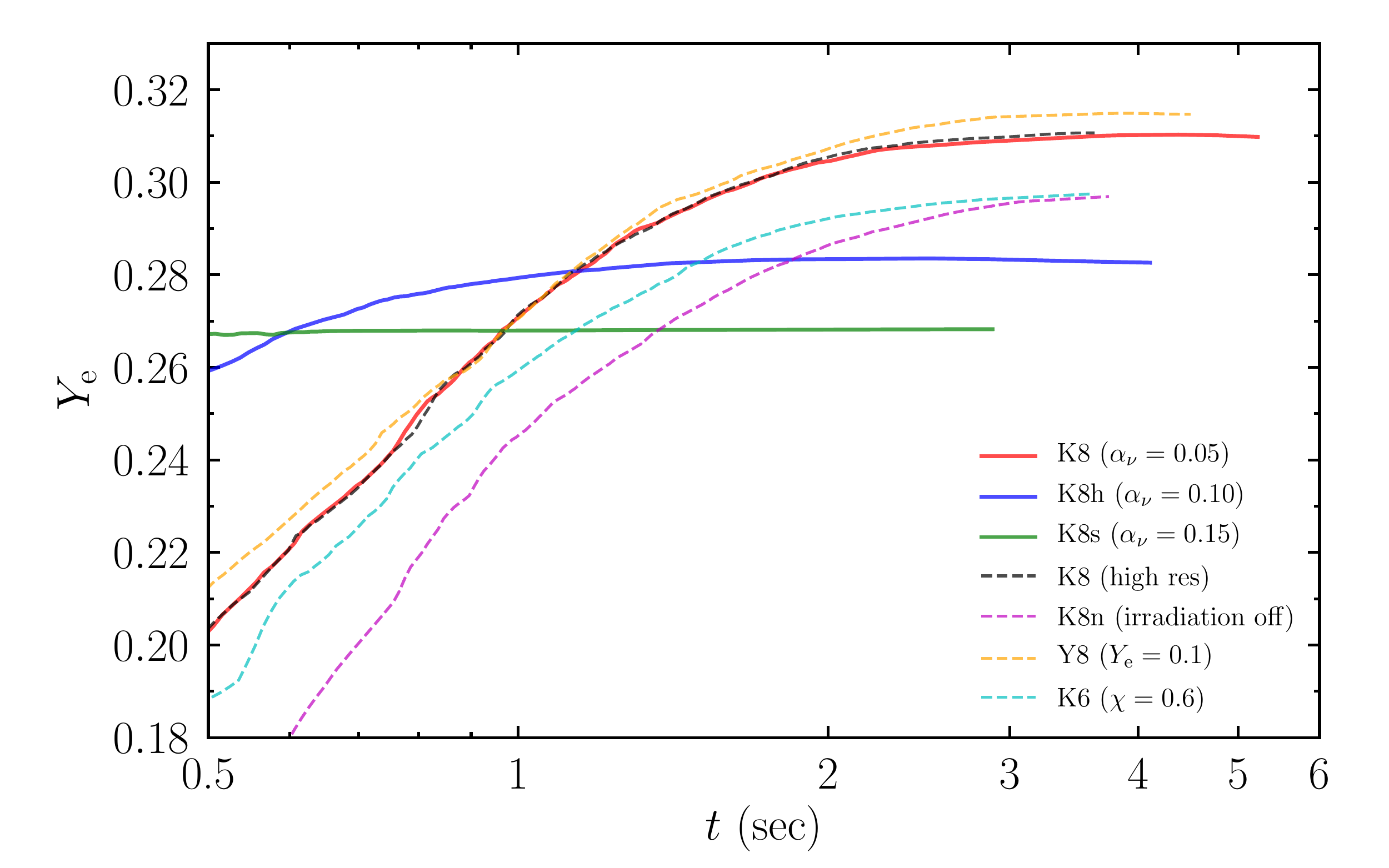}
(b)\includegraphics[width=84mm]{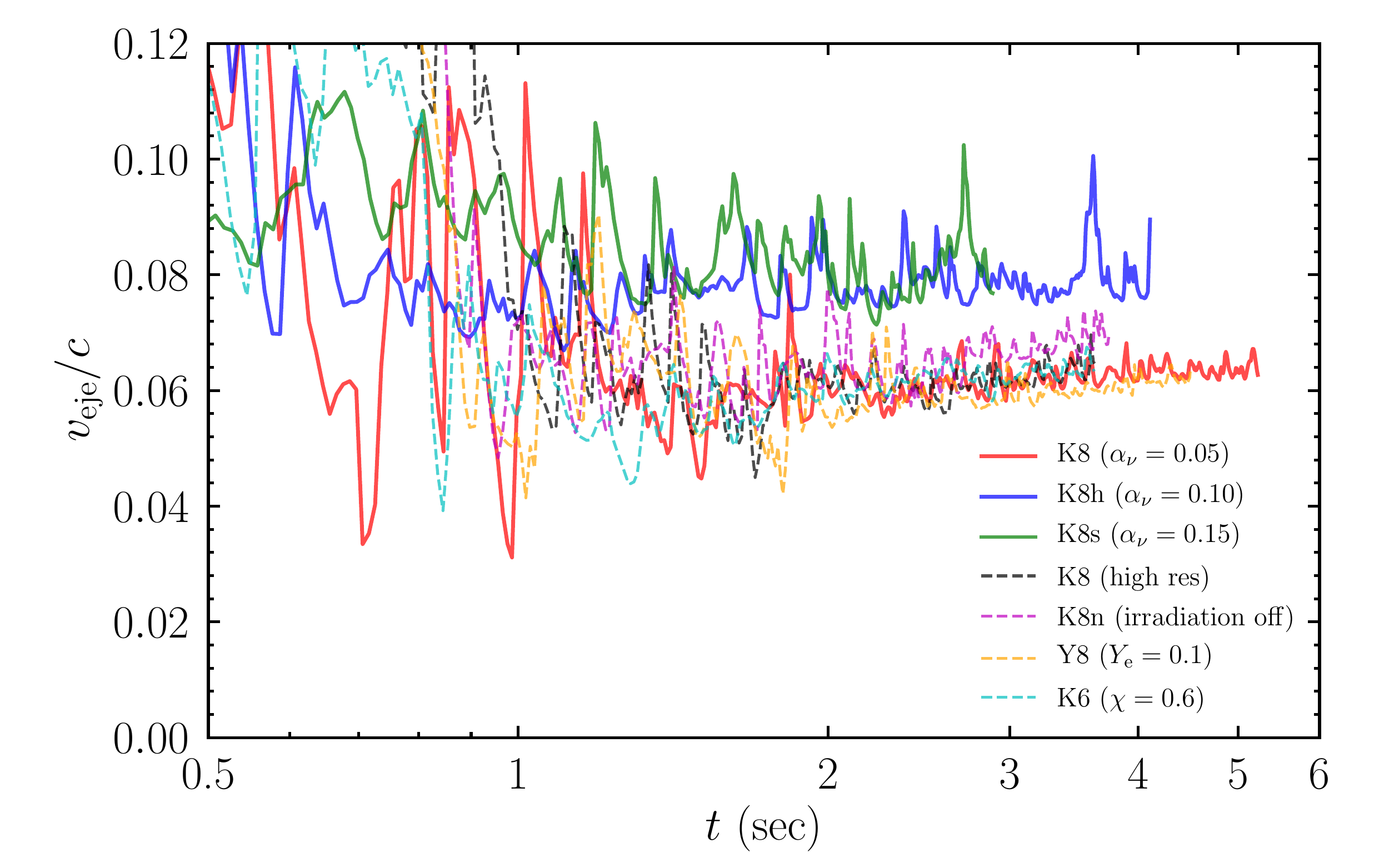}\\
(c)\includegraphics[width=84mm]{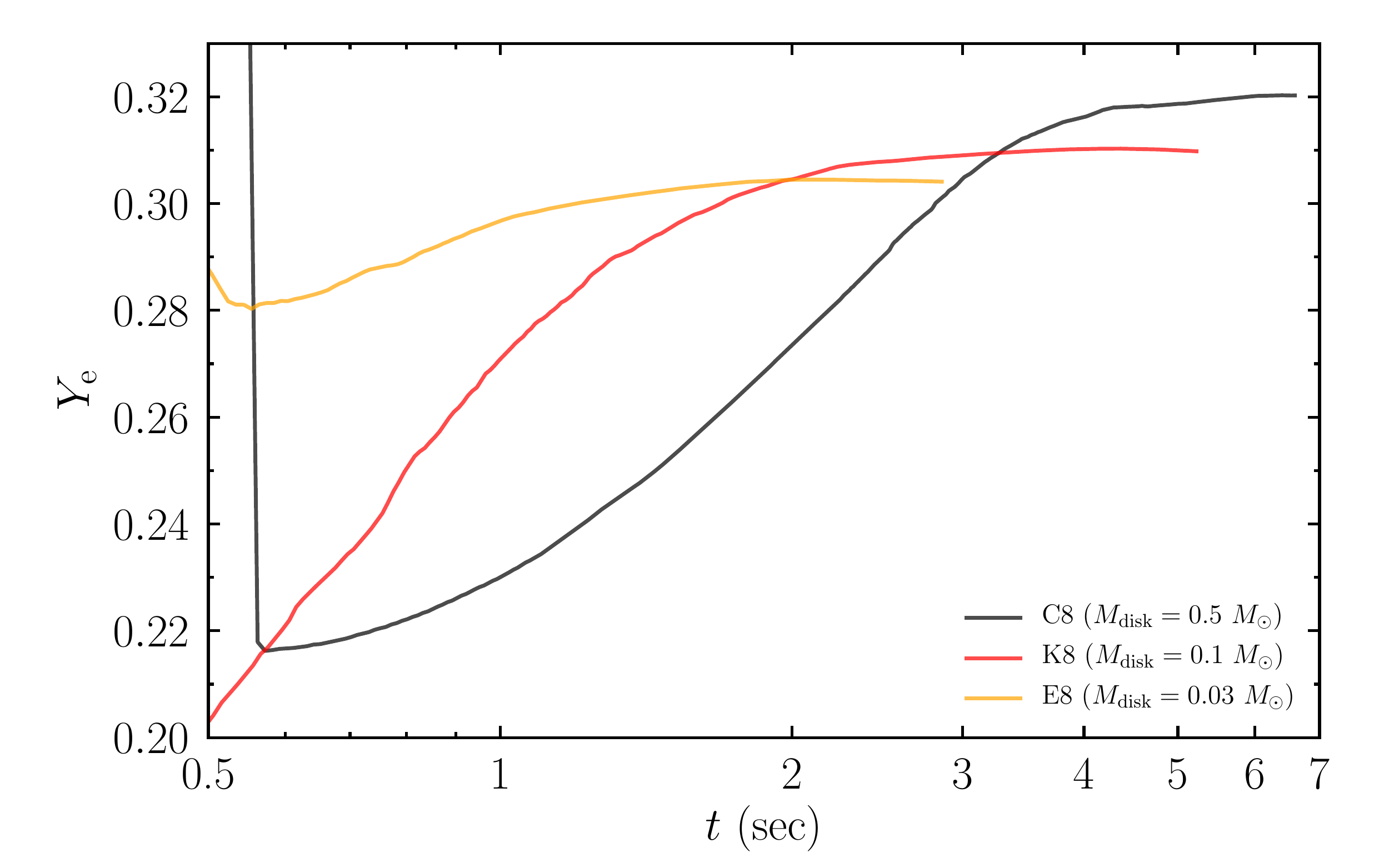}
(d)\includegraphics[width=84mm]{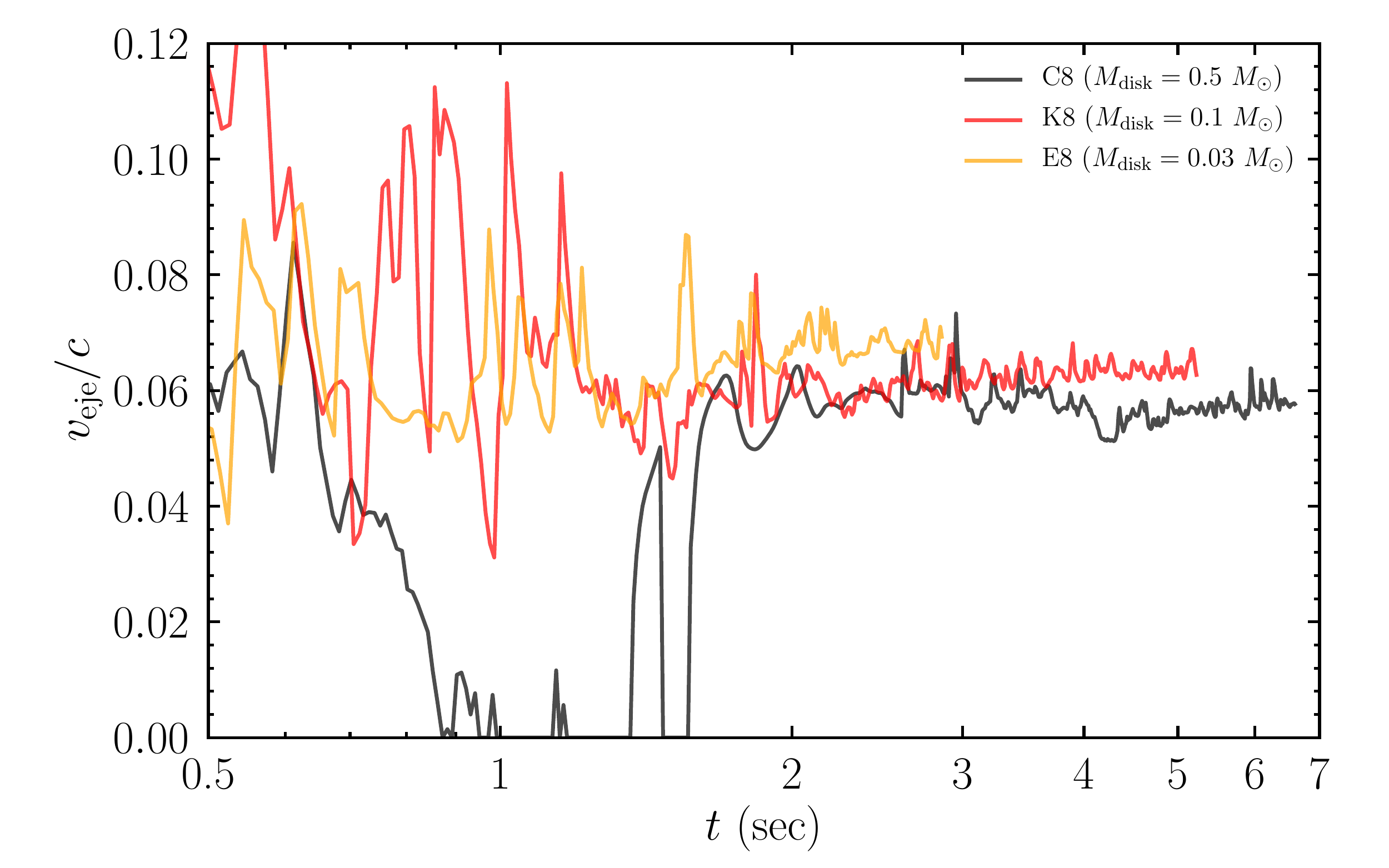}
\caption{Average value of $Y_e$ (left) and and average velocity
  (right) of the ejecta as functions of time for models with $M_{\rm
    disk}=0.1M_\odot$ (top) and for models C8, K8, and E8 (bottom).
  Here, the average velocity is determined only for the ejecta
  component that escapes from $r=2000$\,km.
\label{fig10}}
\end{figure*}

\begin{figure*}[t]
(a)\includegraphics[width=84mm]{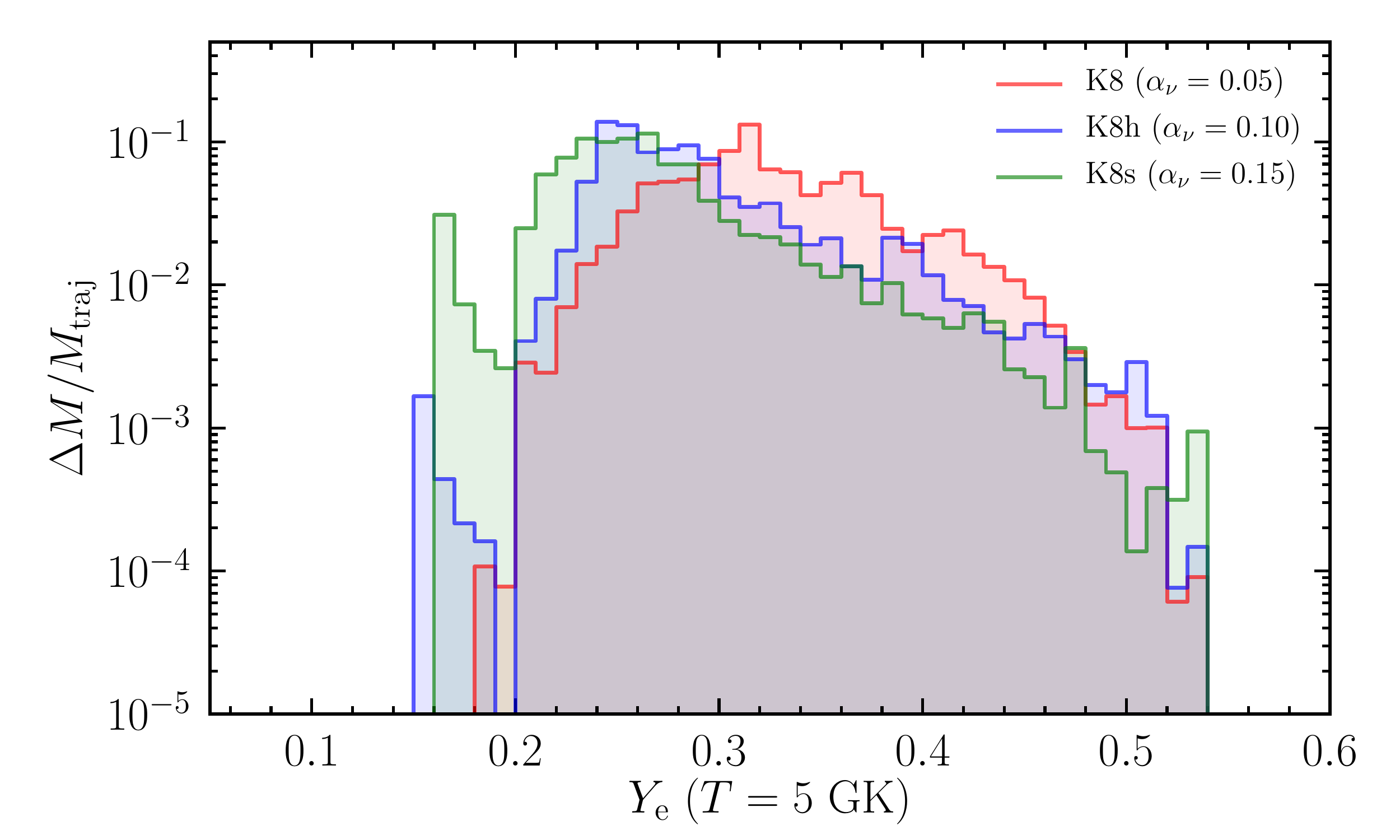}
(b)\includegraphics[width=84mm]{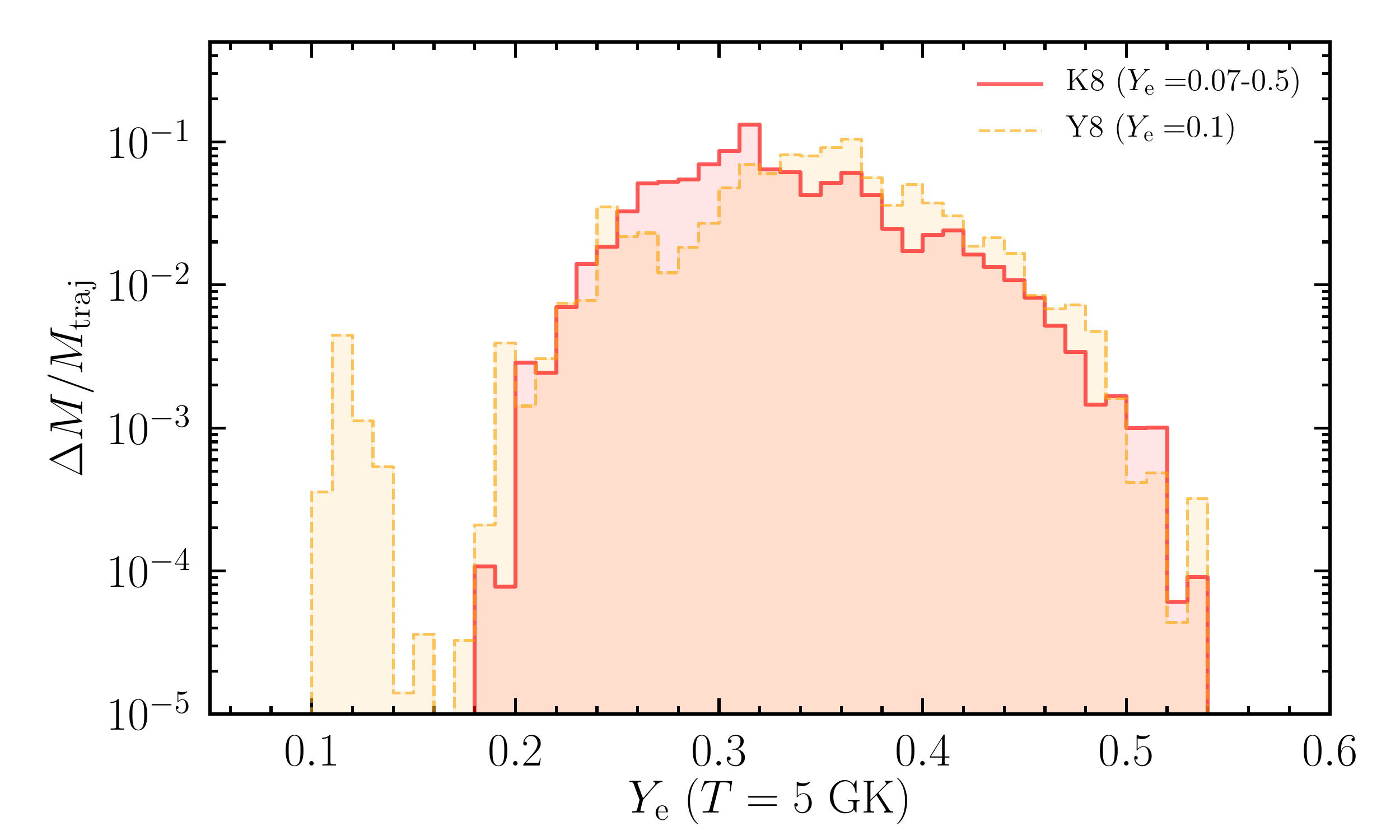}\\
(c)\includegraphics[width=84mm]{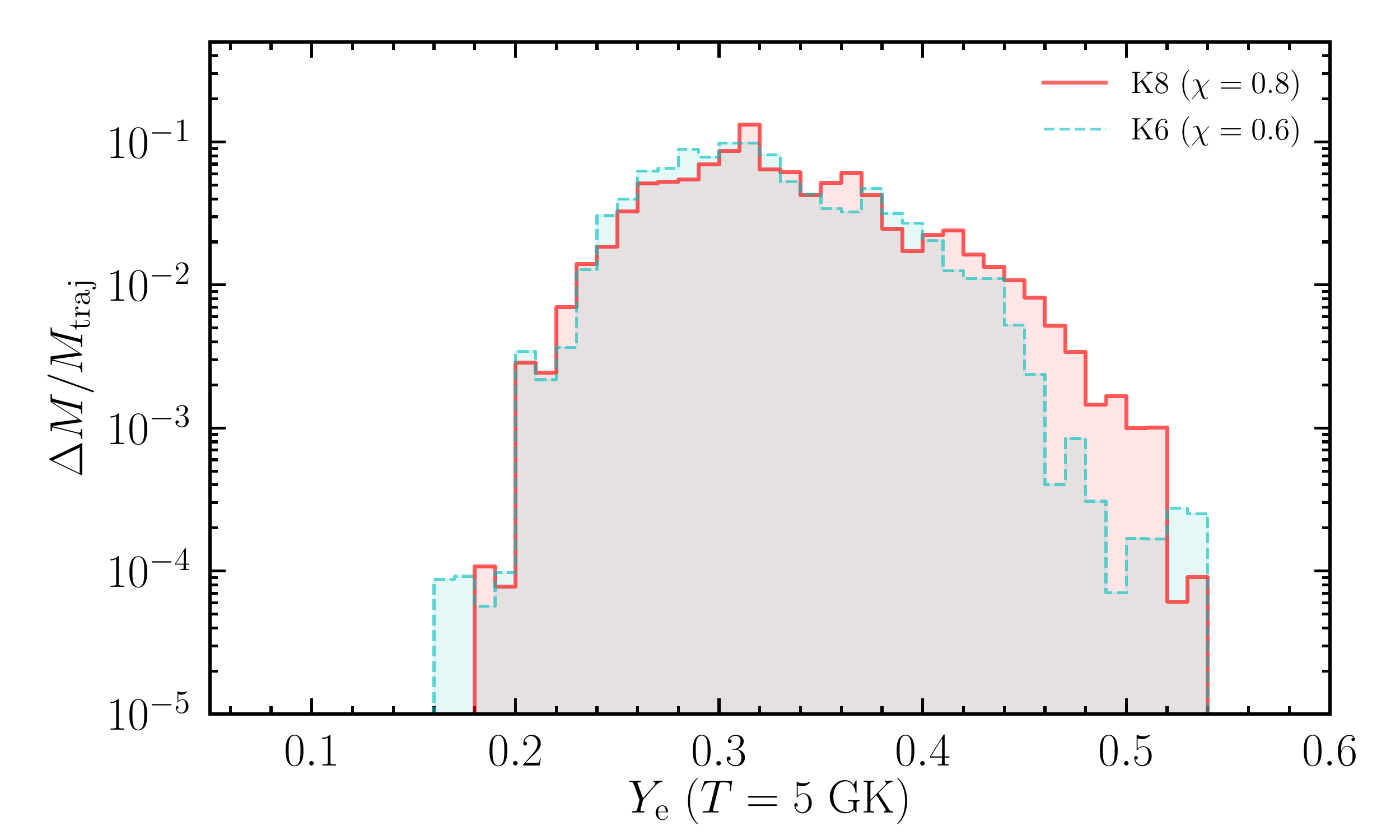}
(d)\includegraphics[width=84mm]{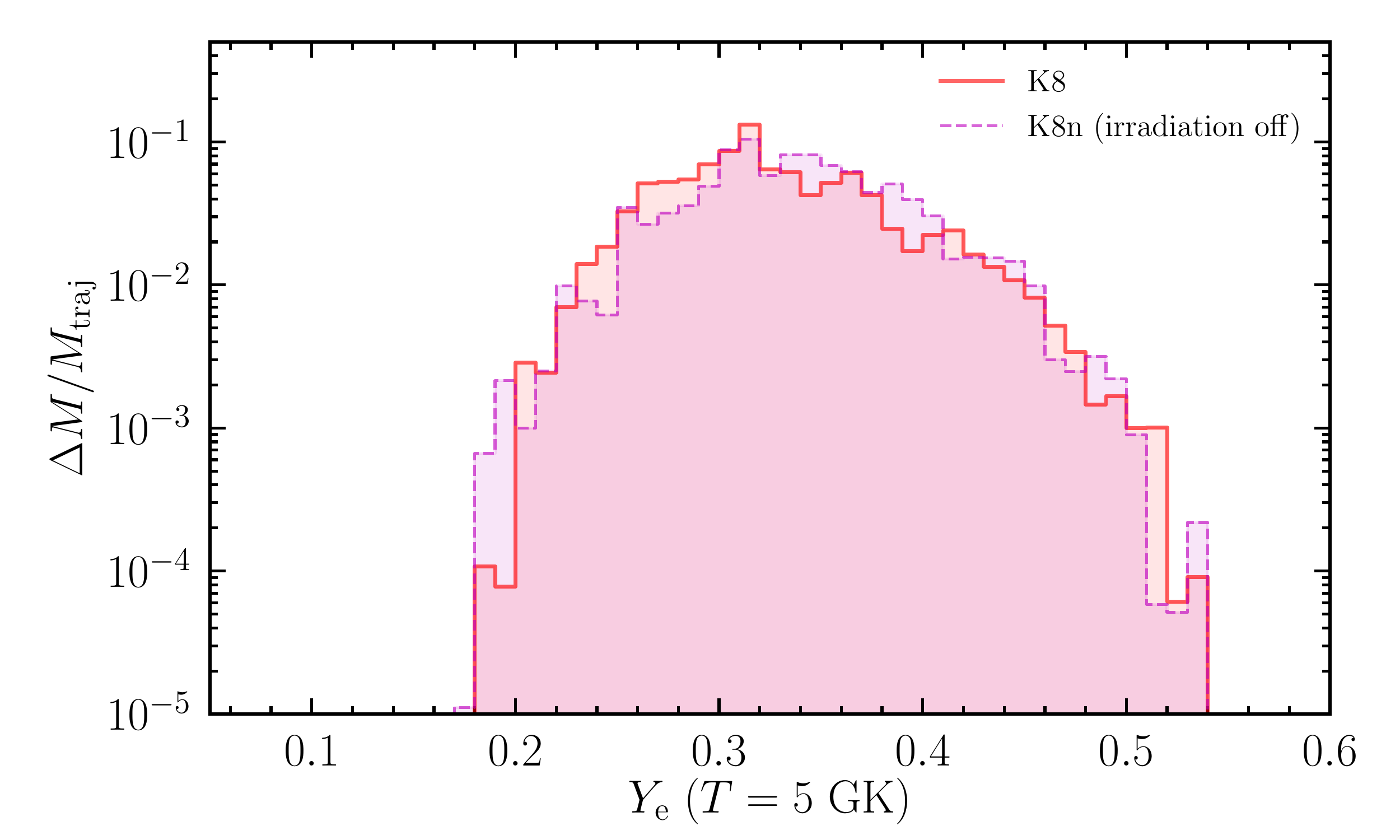}\\
(e)\includegraphics[width=84mm]{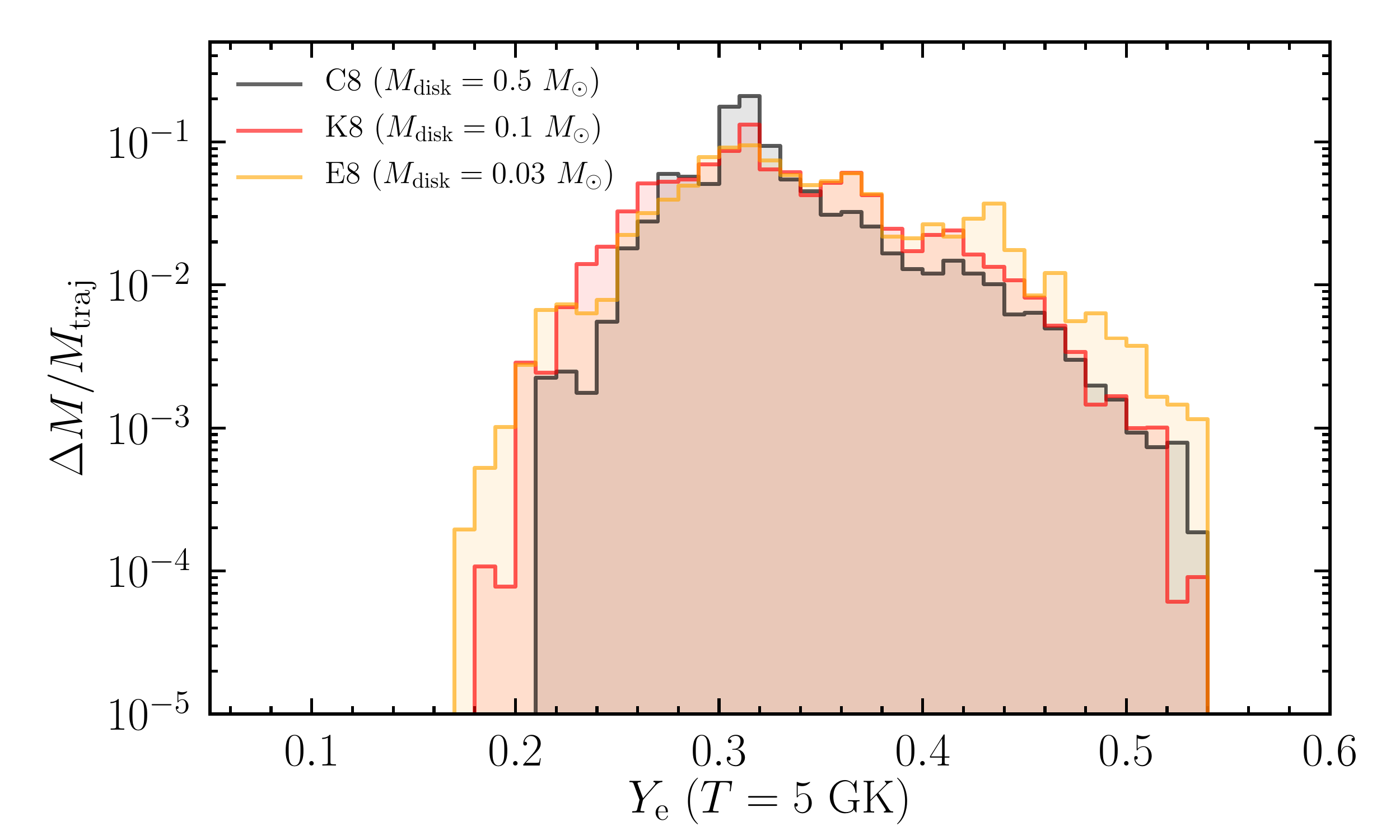}
(f)\includegraphics[width=84mm]{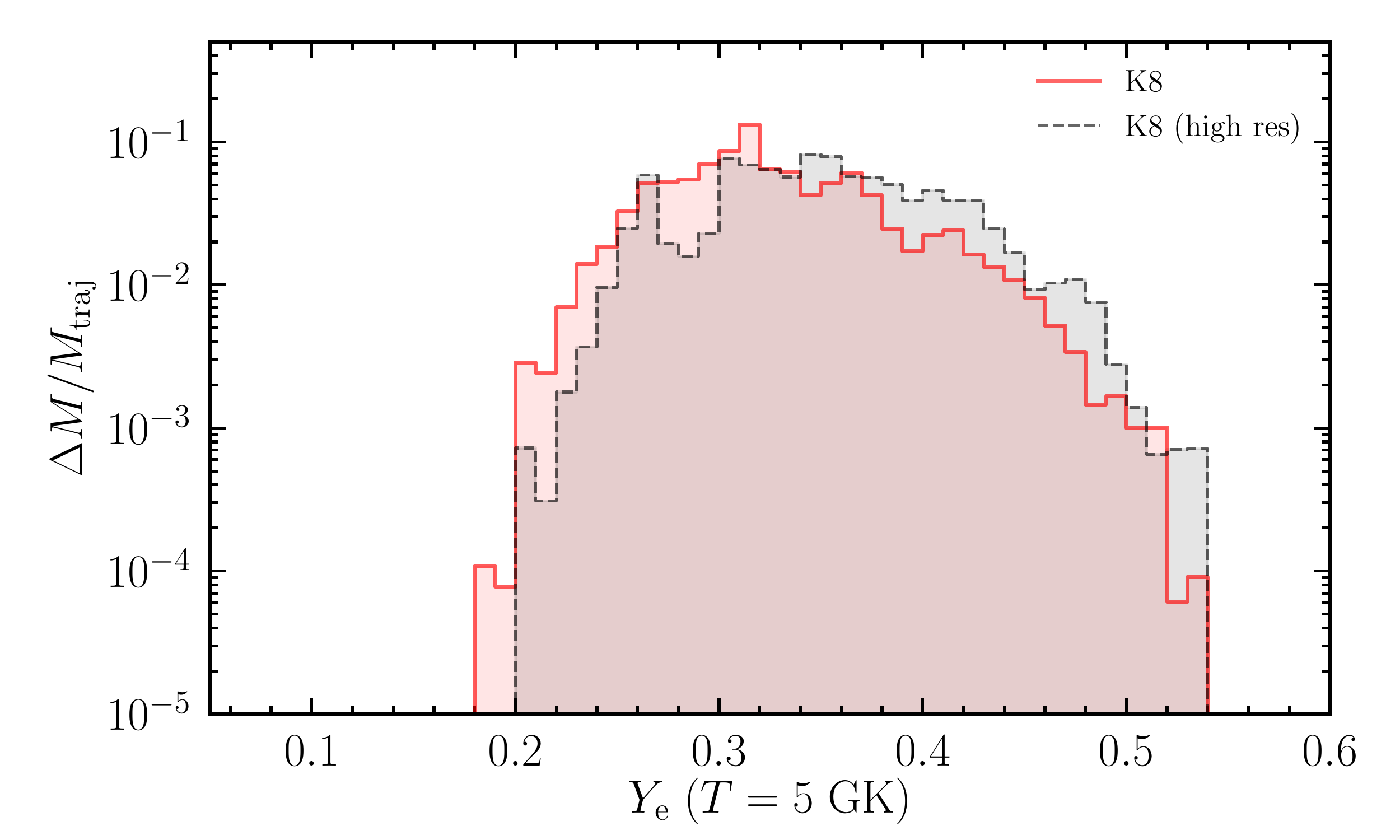}
\caption{Mass histogram (mass fraction) of the ejecta as a function of
  $Y_e$ for models (a) K8, K8h, and K8s, (b) K8 and Y8, (c) K8 and K6,
  (d) K8 and K8n, (e) C8, K8, and E8, and (f) K8 with two grid
  resolutions.  Here, the value of $Y_e$ is determined at the time
  when the temperature of each ejecta component decreases to $5 \times
  10^9$\,K (referred to as 5\,GK).
\label{fig11}}
\end{figure*}

\begin{table}[t]
\caption{Rest mass, average value of velocity, and average value of
  $Y_e$ for ejecta.  All the quantities are extracted at $t=3.5$\,s
  after the onset of the simulations except for models K8s and C8 for
  which the time for the extraction is $t=2.5$\,s and 6.0\,s,
  respectively.  The last column shows the lanthanide mass fraction determined 
  by the nucleosynthesis calculation for selected models. }
\begin{tabular}{ccccc} \hline
~Model~ & ~$M_{\rm ej}(M_\odot)$~ & ~$V_{\rm ave}/c$~ & ~$Y_{e,{\rm ave}}$~ 
& ~$X_{\rm lan}$~ \\
 \hline \hline
K8            & 0.020   & 0.06 & 0.31 & $2.8 \times 10^{-4}$\\
K8 (high res.)& 0.020   & 0.06 & 0.31 & \\
K8h           & 0.023   & 0.08 & 0.28 & $1.6 \times 10^{-3}$\\
K8s           & 0.027   & 0.08 & 0.26 & $8.4 \times 10^{-3}$\\
K8n           & 0.021   & 0.07 & 0.30 &\\
D8            & 0.015   & 0.06 & 0.31 &\\
Y8            & 0.023   & 0.06 & 0.32 &\\
K6            & 0.019   & 0.06 & 0.30 &\\
C8            & 0.092   & 0.06 & 0.32 & $1.2 \times 10^{-6}$\\
E8            & 0.006   & 0.07 & 0.31 & $1.5 \times 10^{-3}$\\
 \hline
 \end{tabular}
 \label{table2}
\end{table}

\begin{figure*}[t]
(a)\includegraphics[width=84mm]{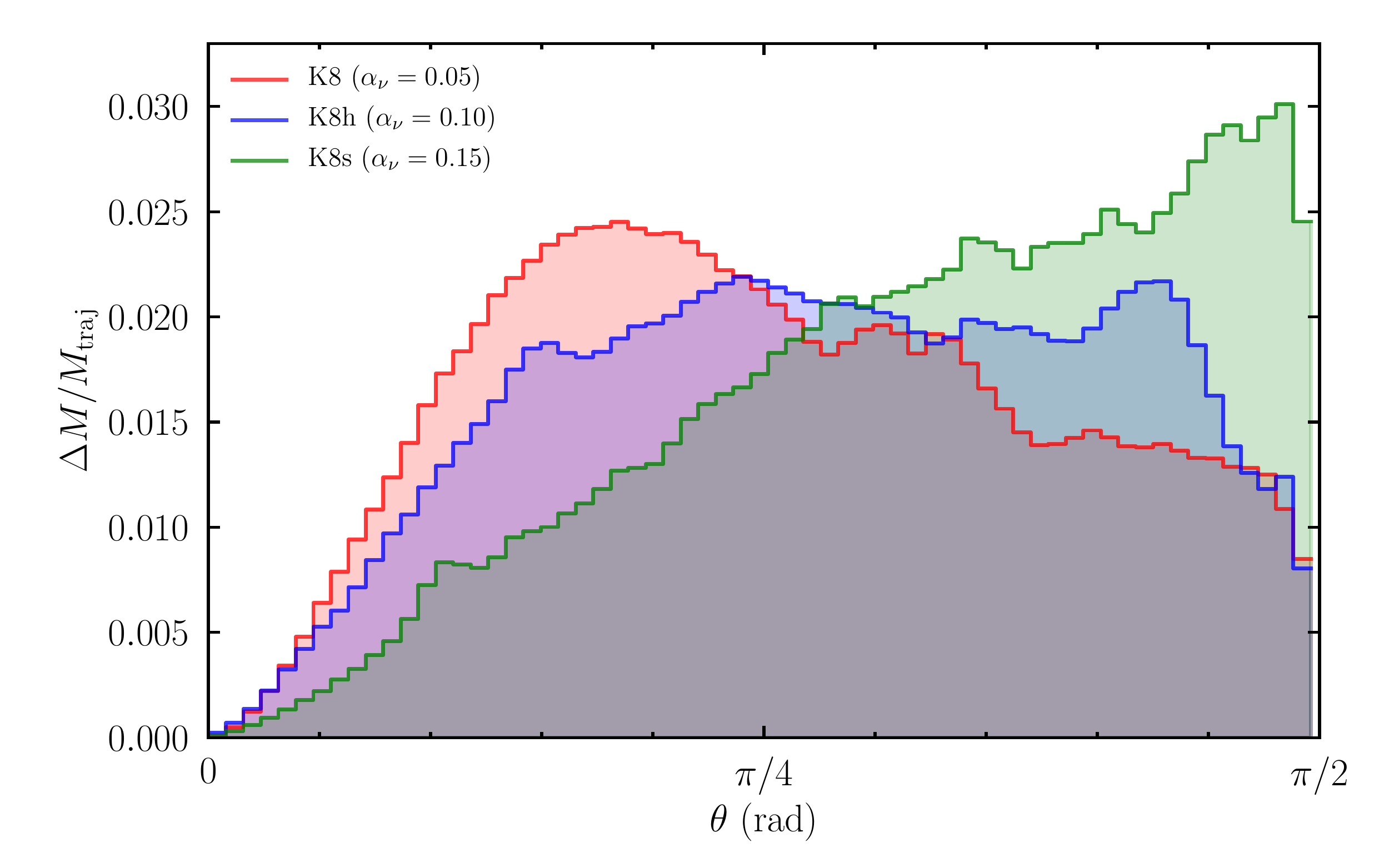}
(b)\includegraphics[width=84mm]{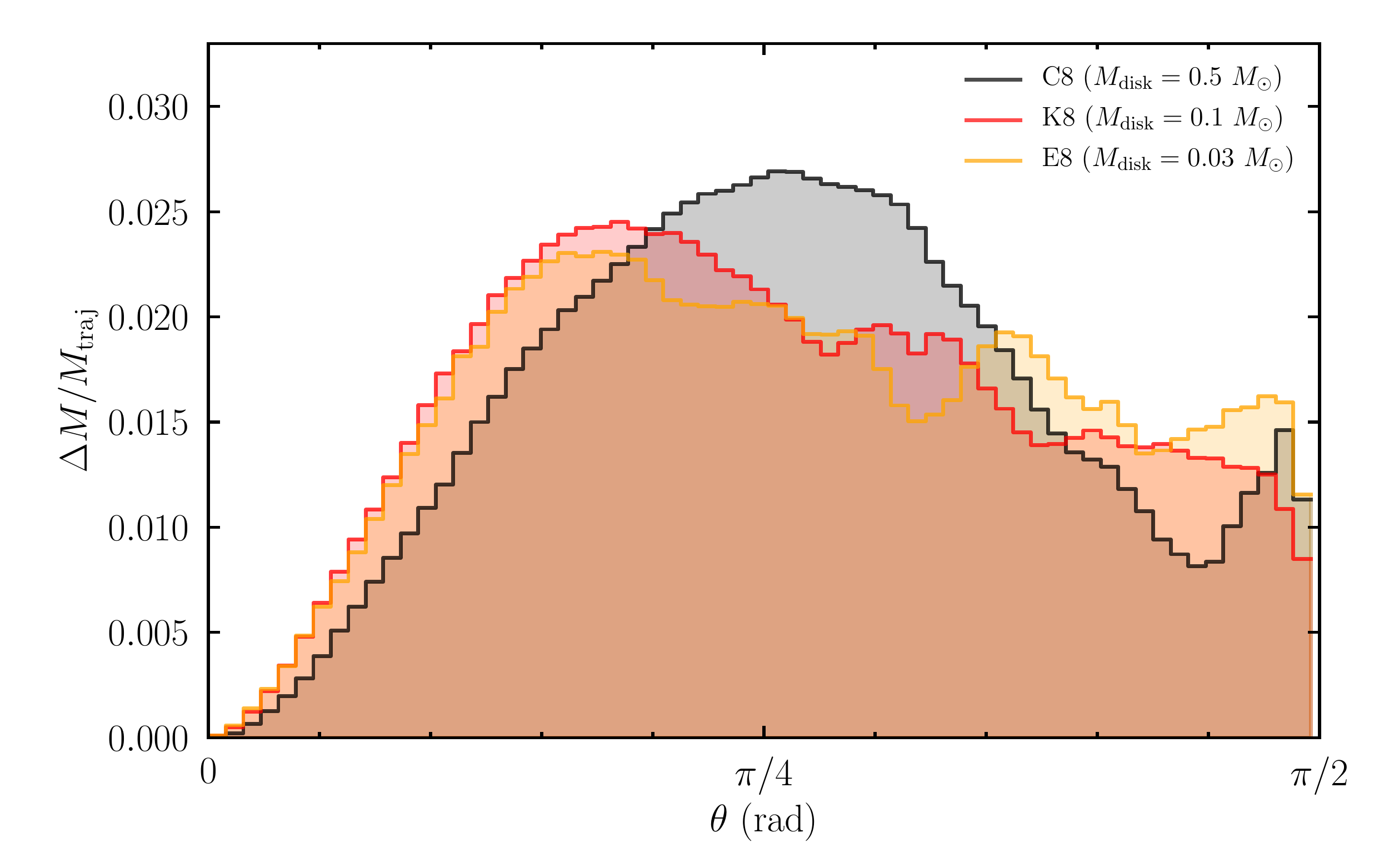}
\caption{ Mass fraction as a function of the polar direction of the
  mass ejection (a) for models K8, K8h, and K8s and (b) for models C8,
  K8, and E8.
\label{fig12}}
\end{figure*}

Figure~\ref{fig10} shows the average value of $Y_e$ and average
velocity of the ejecta as functions of time for all the models (except
for model D8) employed in this paper. Here, we plot the ejecta
velocity determined only for the ejecta component that escapes from
the sphere of $r=2000$\,km. We note that even if the extraction radius
is changed from $r=2000$\,km to 3000\,km and 4000\,km, the asymptotic
values of $Y_e$ and the velocity are varied only within 3\%.
Table~\ref{table2} lists the rest mass, average velocity, and average
value of $Y_e$ of the ejecta for all the models considered in this
paper. We also show the mass histogram (mass fraction) of the ejecta
as a function of $Y_e$ for models K8, K8h, K8s, K8n, Y8, K6, C8, and
E8 in Fig.~\ref{fig11}. Here, the value of $Y_e$ is determined at the
time when the temperature of the corresponding ejecta component
decreases to $5 \times 10^9$\,K (referred to as 5\,GK: note that the
temperature of the ejecta component in general decreases down to $\sim
3$ GK monotonically during the ejection process).

The average values of $Y_e$ for the ejecta with $\alpha_\nu=0.05$ are
$\sim 0.3$ and depend very weakly on the initial condition for the disk 
profiles of density, angular velocity, and $Y_e$.
On the other hand, it decreases significantly with the increase of the
viscous coefficient. As Fig.~\ref{fig11}(a) shows that the distribution
shifts to the lower side of $Y_e$ with the increase of $\alpha_\nu$.
This result can be expected from Fig.~\ref{fig7}(c), which shows that
for the matter outside the black hole, the asymptotic average values of
$Y_e$ is $\sim 0.32$, 0.29, and 0.27 for $\alpha_{\nu}=0.05$, 0.10,
and 0.15, respectively.
These values agree approximately with the average
value of $Y_e$ for the ejecta. This agreement indicates that the
matter in the outer part of the disk that expands by the viscous
effect and the resulting convective heating eventually becomes the
ejecta component.  The average value of $Y_e$ for the ejecta increases
with the increase of the disk mass (compare the results for models C8,
K8, and E8). This correlation also agrees with that found in
Fig.~\ref{fig9}(c).

The asymptotic average velocity of the ejecta, $v_{\rm eje}$, is $\sim
0.06c$ for $\alpha_\nu=0.05$ and again depends very weakly on the
initial condition for the disk profiles of density, angular velocity, and
$Y_e$, and the black-hole spin.  For higher viscous coefficients,
$v_{\rm eje}$ is increased, and for $\alpha_\nu=0.10$ and 0.15, it is
$\approx 0.08c$. This reflects an efficient acceleration of the matter
in the outer part of the disk for the higher viscous coefficients.

Although the results for the mass and velocity of the ejecta agree
broadly with those of the earlier numerical studies by other
groups~\cite{MF2013,MF2014,Just2015}, our result for the mass
distribution of $Y_e$ shows a noticeable difference from the previous
results. In the previous results, a substantial fraction of the ejecta
has rather small values of $Y_e$ between $0.1$ and 0.2. By contrast,
our results show that such low $Y_e$ components are rare, and the peak
is located approximately between $Y_e=0.25$ and 0.30. We note that the
low $Y_e$ component is present in the disk for the early stage in
which the disk density is high, $\agt 10^{9}\,{\rm g/cm^3}$ (see,
e.g., Fig.~\ref{fig4} at $t=0.5$\,s), but with the decrease of the
disk density the value of $Y_e$ is increased and the ejecta does not
have the component with $Y_e \geq 0.2$. One possible reason for the
difference of our results from the previous ones is that the equation
of state and/or treatment of the weak interaction (e.g., electron and
positron capture) that we employ may be different from those in the
previous studies. Another reason is that in our simulation we do not
have any mechanism for the mass ejection which is more efficient than
the viscosity driven ejection.  In the following, we describe these
points in more detail.

In our simulation, the mass ejection occurs only after the viscous
heating and angular momentum transport proceed at least for several
hundreds milliseconds, which causes the matter in the disk to expand
to $r \agt 10^3$\,km.  We do not find any other major components for
mass ejection like the neutrino-wind component, which is found in
Ref.~\cite{Just2015}. As we already mentioned in Sec.~\ref{sec3-2}, in
this viscous evolution process, associated with the decrease of the
density and with the decrease of the degree of electron degeneracy,
the value of $Y_e$ in the disk gradually increases from a low value of
$Y_e \sim 0.1$ to higher values $Y_e \agt 0.2$ until the weak
interaction freezes out (i.e., the temperature of the disk decreases
below $k T \sim 2$\,MeV; e.g., see Fig.~\ref{fig7} and Appendix
A). The value of $Y_e$ is determined approximately by the condition
that electron and positron capture rates are identical.  After the
freeze out of the weak interaction, the average of $Y_e$ approaches
asymptotically $\sim 0.3$ (for $\alpha_\nu=0.05$).  This asymptotic
value is slightly larger than that in a previous
study~\cite{Just2015}, and this suggests that the difference in the
equation of state, treatment of the weak interaction, and initial
thermodynamical condition (e.g., temperature) might result in the
difference in the values of $Y_e$ among different groups.

For the late stage with lower temperature of $k T \alt 2$\,MeV, the
weak interaction does not play a role any longer because the timescale
for the weak interaction process becomes longer than the viscous
timescale (see Appendix A). This results in the freeze out of the
value of $Y_e$ in the disk.  Since the $Y_e$ distribution of the
ejecta is just the reflection of that in the disk, the value of $Y_e$
for the ejecta thus determined is not very small but rather large as
$\agt 0.25$ (for $\alpha_\nu=0.05$). We note that in our simulation,
most of the ejecta component experiences the weak interaction process
during the viscous evolution of the disk, resulting in the relatively
high value of $Y_e$. This indicates that the low $Y_e$ ejecta with
$Y_e \sim 0.1$ found in the previous studies would be driven before
experiencing the weak interaction processes sufficiently.

For higher values of $\alpha_\nu$, the weak interaction freezes out
earlier (much earlier than 1\,s) because faster expansion occurs for
the disk. Since the duration to increase the value of $Y_e$ is
shorter, $Y_e$ becomes smaller for larger values of $\alpha_\nu$. This
tendency is clearly found by comparing the mass histogram as a
function of $Y_e$ for models K8, K8h, and K8s (see
Fig.~\ref{fig11}(a)). This suggests that in the presence of a mass
ejection process more efficient than the viscosity-driven mechanism,
like a magnetohydrodynamics mechanism, which would be primarily not
the MRI but the magnetic winding and resulting strong Lorentz force
with a hypothetically very large poloidal magnetic field aligned with
the black-hole spin direction~\cite{FTQFK19}, the matter with even
lower $Y_e$ components may be ejected. However, our present
simulations show that in the absence of such an efficient mass
ejection process with the ejection timescale much shorter than 1\,s
(i.e., only with the viscous process), the value of $Y_e$ for the
ejecta can be quite large as $Y_e \agt 0.2$ even in the absence of
strong neutrino irradiation sources, like a remnant massive neutron
star~\cite{Fujiba2018}.

Figure~\ref{fig11}(b) compares the mass histogram for models K8 and
Y8.  This shows that for model Y8, a small fraction of low-$Y_e$
ejecta components with $Y_e \leq 0.2$ is present, reflecting its
initial condition and the fact that a small fraction of matter is
spuriously ejected in the initial transient phase until the disk
relaxes to a quasi-stationary state.  However, for the major part of
the ejecta, the distribution for $Y_e$ is quite similar between these
two models. This indicates that the $Y_e$ distribution of the ejecta
depends only weakly on its initial condition, if we focus only on the
viscosity-driven ejecta. We note, however, that if a very efficient
mechanism of mass ejection is present in the very early stage of the
disk evolution for $t \ll 100$\,ms, the resulting mass histogram as a
function of $Y_e$ may reflect the difference of the initial condition. 
This point should be explored in the simulation community. 

Figure~\ref{fig11}(c) compares the mass histogram for models K8 and
K6, i.e., for different black-hole spin models.  This figure shows
that for the smaller black-hole spin (K6), the $Y_e$ distribution
shifts slightly to a lower $Y_e$ side. Our naive interpretation for
this is that for the smaller spin, the disk mass slightly becomes
smaller because a larger fraction of the disk matter falls into the
black hole, resulting in slight shortening until the freeze out of the
weak interaction is reached.  Alternative possibility is that a
difference in the initial profile of the disk is reflected.  However,
the difference in the $Y_e$ distribution is quite small, and hence, we
may conclude that the difference of the spin is not very important for
the properties of the viscosity-driven ejecta, as far as we focus on
the astrophysically plausible values of $\chi$.

Figure~\ref{fig11}(d) compares the mass histogram for models K8 and
K8n to understand the unimportance of the neutrino irradiation for the
ejecta in the black hole-disk system. This figure shows that two
results are quite similar each other, and hence, the neutrino
irradiation indeed does not play an important role. This is quite
reasonable in our viscous evolution models because the mass ejection
is activated only after the neutrino luminosity drops (i.e., after the
weak interaction freezes out), whereas the neutrino irradiation could
be important only for the case that the neutrino luminosity is still
high. Thus, the neutrino irradiation would be important only for the
case that mass ejection occurs in an early stage in which the neutrino
luminosity is still high (e.g., Ref.~\cite{Miller19}). 

Figure~\ref{fig11}(e) compares the mass histogram for models C8, K8,
and E8; different initial disk mass models.  It is found that for
the higher disk mass, the lowest end of $Y_e$ is larger.  As already
mentioned in the previous section, for the higher disk mass, the time at
which the weak interaction freezes out comes later, and the value of
$Y_e$ for the disk component becomes higher.  Associated with this
effect, the lowest value of $Y_e$ in the ejecta component also
increases for the larger disk mass model. This result suggests that
for the case that the remnant disk mass is large, e.g., in the remnant
of stellar core collapse to a spinning black hole~\cite{Woosley93},
the ejecta from the disk may not be very neutron rich and cannot
synthesize a substantial fraction of heavy elements like lanthanide
(see Sec.~\ref{sec3-5}).

Figure~\ref{fig12} displays the mass fraction as a function of the
polar angle of the mass ejection (a) for models K8, K8h, and K8s and
(b) for models C8, K8, and E8.  This shows that the mass ejection
occurs to a wide range of angles except for the direction of the rotation
axis irrespective of the viscous coefficient and disk mass.  The
absence of the mass ejection toward the rotational axis is natural
because the ejecta, which are driven in a region far from the black
hole, have a substantial angular momentum.  An interesting point is
that the primary direction of the mass ejection depends on the
magnitude of the viscous coefficient. For $\alpha_\nu=0.05$, the mass
ejection occurs most strongly toward the direction of $\theta \sim
\pi/6$ where $\theta$ denotes the polar angle. However, with the
increase of $\alpha_\nu$, the primary angle of the mass ejection
increases, and for model K8s, the mass ejection occurs primarily to
the direction of the equatorial plane, $\theta \sim \pi/2$. This
indicates that not only the convective activity but also the outward
angular momentum transport plays an important role for the mass
ejection with larger values of $\alpha_\nu$.


\subsection{Nucleosynthesis in the ejecta}\label{sec3-5}

\begin{figure*}[t]
(a)\includegraphics[width=82mm]{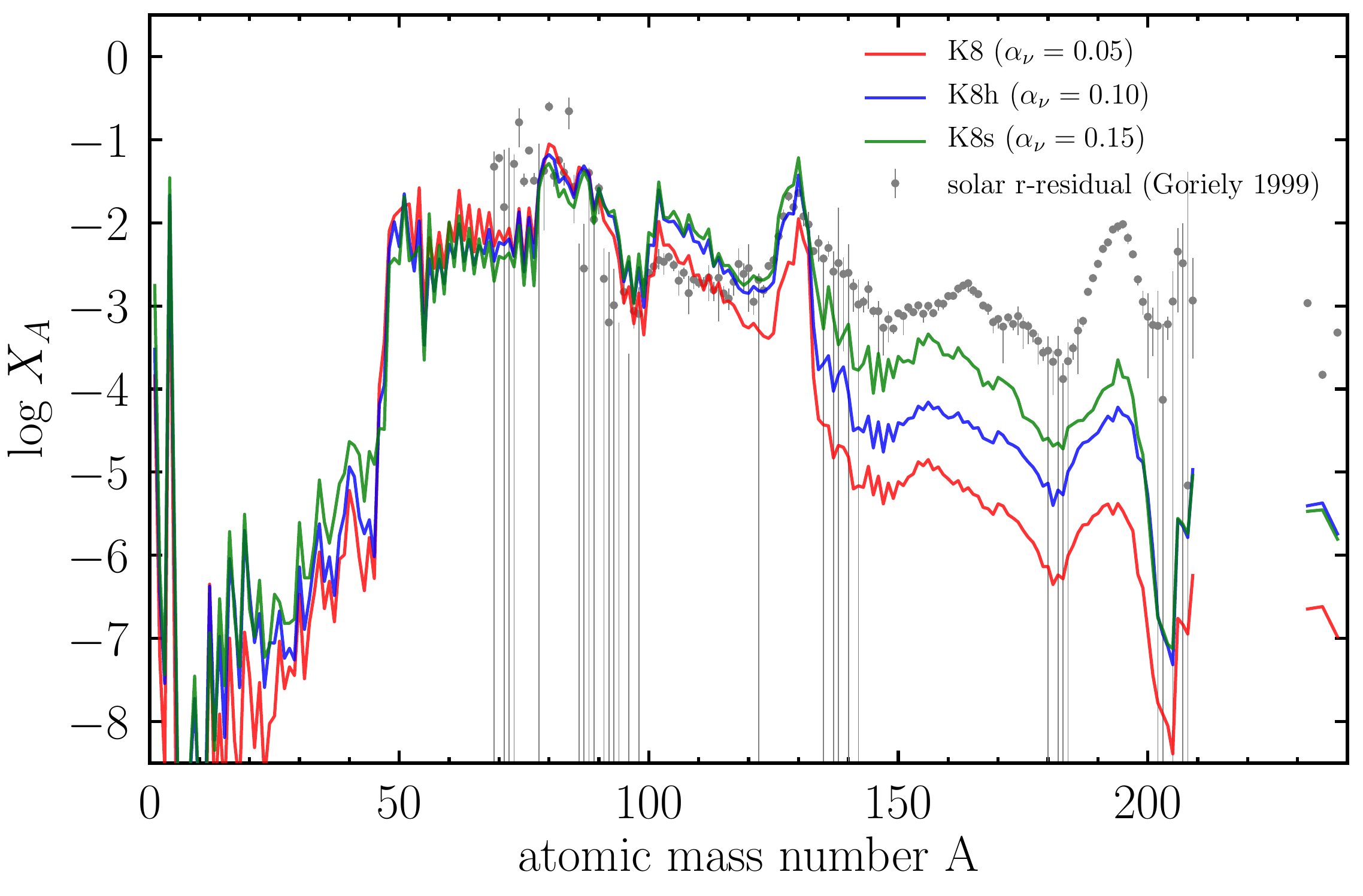}~~
(b)\includegraphics[width=82mm]{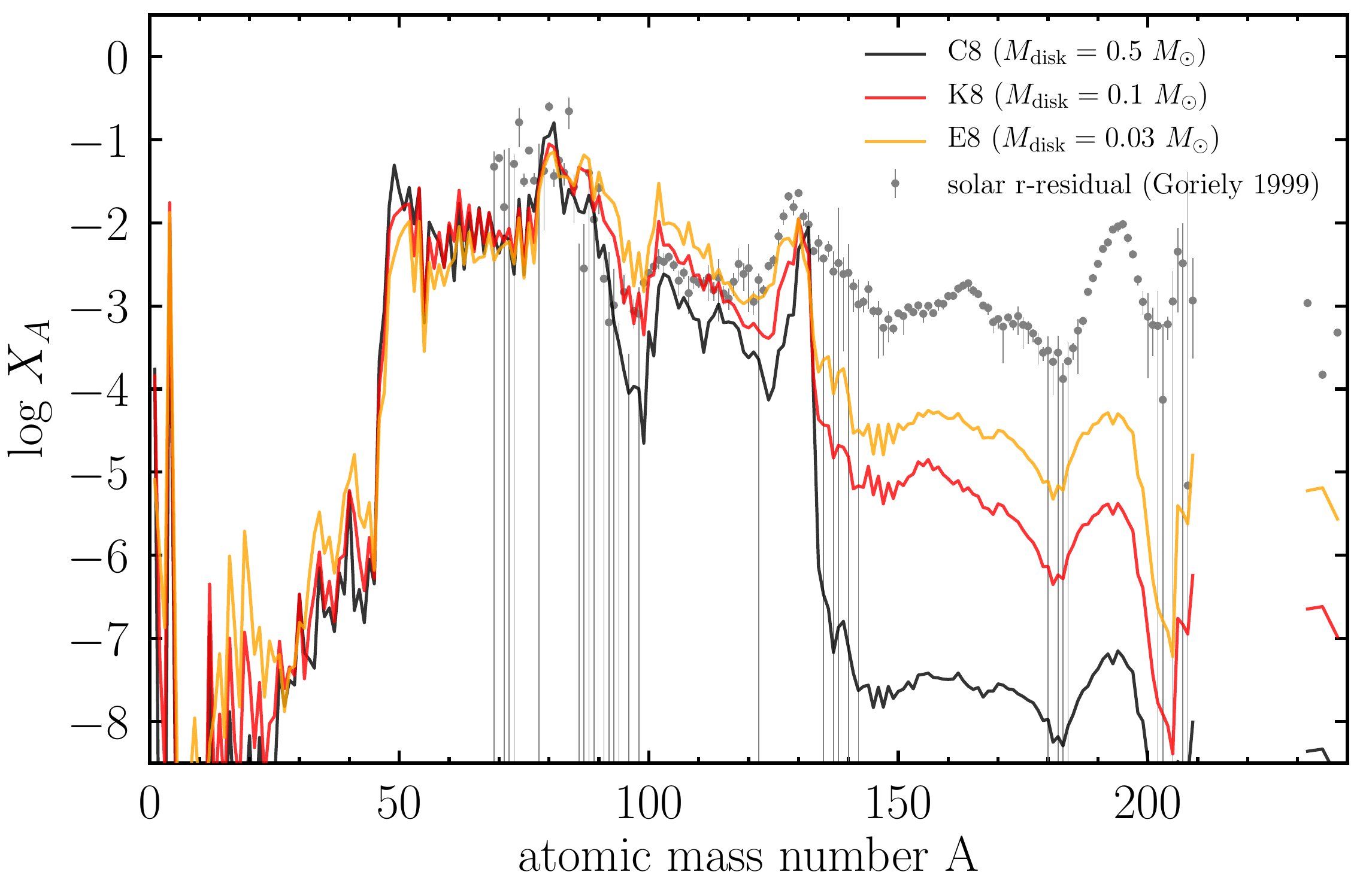}
\caption{Patterns of mass fraction obtained in the nucleosynthesis
  calculation for the ejecta (a) for models K8, K8h, and K8s and (b)
  for models C8, K8, and E8.  The filled circles with the error bar
  denote the $r$-process residual to the solar system abundance for
  $A\geq 69$ \cite{Gorie99}. The data are shifted to match the
  calculated mass fraction for model K8 at $A=83$. 
\label{fig13}}
\end{figure*}

A nucleosynthesis calculation is performed for models K8, K8h, K8s,
E8, and C8. For each model, a few ten thousands of tracer particles are
arranged by the method described in Ref.~\cite{Fujiba2019}, and in
each tracer particle, the nucleosynthesis is explored as a
post-processing step by using the reaction network
code~\texttt{rNET}~\cite{Wanajo2018a}. The reaction network consists
of 6300 species from single neutron and proton up to the isotopes with
$Z = 110$ (see Ref.~\cite{Wanajo2018a} for more details). For those
relevant to the $r$-process, the rates of both neutron capture
(TALYS~\cite{Goriely2008a}) and $\beta$-decay
(GT2~\cite{Tachibana1990a}) are based on the microscopic prediction of
nuclear masses, HFB-21~\cite{Goriely2010a}. In this work, we exclude
fission process from the network, which is relevant only for $Y_e <
0.15$. Neutrino-induced reactions are not included, either, which are
unimportant as described in previous sections. Each nucleosynthesis
calculation is started when the temperature decreases to $10^{10}$\,K
with the initial compositions of free neutron and proton to be $1 -
Y_e$ and $Y_e$, respectively. Note that nuclear statistical
equilibrium (NSE) immediately establishes at such high temperature.

Figure~\ref{fig13} displays the mass fraction of each element as a
function of the atomic mass number, $A$.  In our ejecta data, the peak
in $Y_e$ comes around $Y_e \sim 0.25$--$0.3$ and mass fraction with
$Y_e \alt 0.2$ is tiny except for model K8s (see
Fig.~\ref{fig11}). For such mass distribution with respect to $Y_e$,
the heavy elements with $A \agt 130$ are not synthesized
substantially.  By contrast, elements with $A \sim 80$--90 and $A \sim
100$--110 are synthesized significantly through the nucleosynthesis
process. The resulting abundance pattern is different from the solar
abundance pattern of $r$-process elements~\cite{Gorie99}.  This
suggests that low-mass black hole-disk systems might be a
subdominant site for the $r$-process nucleosynthesis, assuming
that the solar abundance pattern is universal~\cite{Cowan19} in the
universe and that the viscosity-driven mechanism is the main channel
of the mass ejection.


For model K8 (as well as model C8), the mass fraction of lanthanide
elements is quite small (see the last column of Table~\ref{table2}) as
expected from the mass histogram shown in Fig.~\ref{fig11}.  By
contrast, for the higher viscous coefficient cases (models K8h and
K8s), a fraction of the lanthanide elements is synthesized, although
the amount of the lanthanide and heavier elements are still smaller
than that in the solar abundance. The reason for this dependence is
that for the higher viscous coefficients, the ejecta contains a
fraction of neutron-rich components with $Y_e \alt 0.2$ due to the
earlier mass ejection (see Fig.~\ref{fig11}(a)). Thus, the mass fraction of
the lanthanide depends on the magnitude of the viscous coefficient,
or in other words, on the onset time of the mass ejection.

For model K8s, the mass fraction of the lanthanide synthesized is
about 0.84\%, while for models K8 and K8h, it is only 0.03\% and
0.16\%, respectively. For the small lanthanide fraction, the opacity
of the ejecta is not extremely enhanced, whereas for the lanthanide
fraction of $\agt 10$\%, the opacity is $\sim 10^2$ times higher than
that for lighter elements like irons~\cite{opacity,TH2013,T2018}.  Our
present results suggest that even for the ejecta from black hole-disk
system, the lanthanide fraction could be minor, in contrast to the
previous understanding (e.g.,~Refs.~\cite{MF2013,MF2014,Just2015}), if
the mass ejection sets in for $\agt 0.3$\,ms after the merger: A
kilonova associated with the ejecta from black hole-disk systems may
shine in an optical band for the early time after the merger (at $\sim
1$\,d) in contrast to the previous belief~\cite{Kasen2015}. On the
other hand, for the high viscosity model K8s, the mass ejection occurs
earlier than for lower viscosity models so that the mass fraction of
the lanthanide can be $\sim 1$\%. For this case, the enhancement of
the opacity would not be negligible~\cite{Kasen2015}. This obviously
shows it very important to quantitatively understand the typical onset
time of the mass ejection.

In the astrophysical context, the actual viscous effects should result 
effectively from turbulence induced by magnetohydrodynamics
processes. This implies that for more physical modeling of the mass
ejection, we need a magnetohydrodynamics simulation that can
accurately capture the nature of the turbulence.  For preserving the
turbulence in the disk orbiting a stellar-mass black hole in
magnetohydrodynamics simulations for seconds, we have to perform a
high-resolution non-axisymmetric simulation. Here, the high resolution
implies that both the black hole spacetime and inner region of the
disk are well resolved; in particular, the fastest growing mode of the
MRI has to be resolved, and in addition, the resulting turbulence has
to be maintained with a sufficient grid resolution for the entire disk
evolution~\cite{BH98,Hawley11,Kiuchi18}.  Although such a simulation
is much more expensive than that in the axisymmetric viscous
hydrodynamics simulation that we performed in this work, it is not
totally impossible in the current computational resources to perform a
small number of the simulations (e.g.,
Refs.~\cite{FTQFK18,FTQFK19}). We plan to perform a general
relativistic neutrino-radiation magnetohydrodynamics simulation to
understand the magnitude of the effective viscous coefficient in
future work.  

Figure~\ref{fig13}(b) displays the nucleosynthesis results for models 
C8, K8, and E8. This shows that for more massive disk models, the
lanthanide fraction is smaller as expected from Fig.~\ref{fig11}(e).
A remarkable point is that for model C8, only a tiny fraction of the
heavy elements with $A > 132$ like lanthanide is synthesized (see also
Table~\ref{table2}). This result suggests that if the viscous process
is the dominant mechanism of mass ejection and if the viscous
coefficient is not extremely large, the ejecta from the massive disk
around black holes may not be the source for the $r$-process
nucleosynthesis of the heavy elements like lanthanide and third peak
elements (e.g., gold). Reference~\cite{Miller19a} has recently
illustrated that if the mass ejection sets in within $\sim 100$\,ms
after the merger, in which the neutrino luminosity is still high, the
neutron richness of the ejecta is significantly decreased by the
neutrino irradiation. For the massive disks
(Reference~\cite{Miller19a} considered the disk of mass $\ll
0.1M_\odot$), the neutrino luminosity should be higher as our present
work shows, and hence, the neutrino irradiation would be even more
enhanced.  Thus, if the mass ejection occurs in late time (only in the
presence of viscous mass ejection), the weak interaction process in
the disk enhances the value of $Y_e$ and if the mass ejection occurs in
early time (by some powerful process; e.g., by a poloidal magnetic
field aligned with the black-hole spin direction), the neutrino
irradiation process enhances the value of $Y_e$.  This suggests that
for the massive disk around a black hole to be a site for the
nucleosynthesis of the heavy $r$-process elements, a fine tuning for the
timing of the mass ejection would be necessary.

\subsection{Convergence on the grid resolution}\label{sec3-6}

Before closing Sec.~\ref{sec3}, we comment on the convergence of our
numerical results with respect to the grid resolution. As found from
Fig.~\ref{fig6}(a), the accuracy for following the evolution of
rapidly spinning black holes depends strongly on the grid
resolution. Due to the truncation error, the black-hole spin
spuriously decreases and the decrease rate can be too high for
insufficient grid resolutions to be acceptable.  For $\Delta
x=0.020M_{\rm BH}$ of model K8, the dimensionless spin decreases by
$\approx 0.06$ in 1\,s. Such a large error is not acceptable.
However, the decrease rate is suppressed significantly with improving
the grid resolution. For $\Delta x=0.016M_{\rm BH}$ and $0.0133M_{\rm
  BH}$, the decrease rates are $\sim 0.02/$s and 0.01/s, respectively,
so that the effect of the rapid spin for these simulations can be
taken into account for $t \alt 2$\,s during which the mass ejection is
driven. 


In Figs.~\ref{fig7} and \ref{fig10}, the time evolution of various
quantities is also compared for model K8 with two different grid
resolutions. These figures show that the quantities for the matter
outside the black hole and ejecta well achieve convergence with
respect to the grid resolution. All these results confirm that 
our present choice of the grid resolution with $\Delta x=0.016M_{\rm BH}$ 
is acceptable for a reliable numerical simulation. 

Figure~\ref{fig11}(f) compares the mass histogram as a function of
$Y_e$ for model K8 with two different grid resolutions. We find that
two histograms agree broadly with each other, although the mass
fraction at each value of $Y_e$ does not exactly agree. 


\section{Summary}\label{sec4}

This paper presents our first numerical results for a viscous
neutrino-radiation hydrodynamics simulation of accretion disks
surrounding a spinning black hole in full general relativity as models
for the evolution of a merger remnant of massive binary neutron stars
or low-mass black hole-neutron star binaries.  We reconfirm the
following results found by previous viscous hydrodynamics studies by
other groups~\cite{MF2013,Just2015}: About 15--30\% of the disk mass
is ejected from the system with the average velocity of $\sim 5$--10\%
of the speed of light for the plausible profile of the disks as merger
remnants. In our simulation, the main driving force of the mass
ejection is the viscous heating in the innermost region of the disks
and the resulting convection that starts relatively in a late stage
of the disk evolution, as well as the viscous angular momentum
transport.  Our new finding is that for the not extremely high viscous
coefficient case, the neutron richness of the ejecta does not become
very high, because the weak interaction in the disk enhances the
electron fraction during its viscous expansion until the weak
interaction freezes out and mass ejection sets in. This results in the
suppression of the lanthanide synthesis in the ejecta, and as a
result, the opacity of the ejecta may not be very high even for the
ejecta from black hole-disk system~\cite{opacity,TH2013,T2018}: A
kilonova associated with the ejecta from black hole-disk systems may
shine in an optical band for an early stage (e.g., at $\sim 1$\,d
after the merger)~\cite{Kyutoku20}.

As we described in Sec.~I, in the popular interpretation, we believe
that a massive neutron star is temporarily formed after the merger of
a binary neutron star in the GW170817
event~\cite{EM2017,MM2017,shibata17,Perego17}. The reason for this
interpretation is that in the presence of the massive neutron star,
the system can have a strong and long-lasting neutrino irradiation
source, by which the electron fraction of the ejecta can avoid being
neutron-rich, its opacity can be decreased significantly, and the
ejecta can shine for an early stage in the optical band: The last
point agrees with the observational results.  However, the results of
our present simulation suggest that if the mass ejection from the disk
surrounding a spinning black hole occurs after a long-term viscous
expansion of the disk (for the duration longer than several hundreds
ms) in which the weak interaction plays an important role, the ejecta
can be weakly neutron rich and avoid synthesizing a lot of lanthanide
elements. For producing such weakly neutron-rich ejecta, the mass
ejection process must not be efficient in $\alt 0.1$\,s after the
formation of accretion disks.

We also find that the total mass, the average velocity, and the
electron fraction of the ejecta depend on the magnitude of the viscous
coefficient.  For the higher viscous coefficients, the mass ejection
sets in earlier, its mass and velocity are larger, and the electron
fraction is smaller for the ejecta.  By contrast, we find that the
quantitative properties of the ejecta depend only weakly on the
initial profile of the density, angular velocity, and electron
fraction for the disk as well as the black-hole spin (of
astrophysically plausible values).

For high-mass disks, the viscous expansion timescale is increased due
to an enhanced dissipation by the neutrino emission (i.e., the
timescale until the freeze out of the weak interaction becomes
longer), and hence, the electron fraction of the ejecta becomes
larger. Thus, for synthesizing heavy elements, lower-mass disks (as
well as high viscous coefficient) are favorable.  Our result suggests
that for the case that the remnant disk mass is large, e.g., in the
remnant of stellar core collapse to a spinning black
hole~\cite{Woosley93}, the ejecta from the disk may not be so neutron
rich that the matter in the ejecta cannot synthesize a substantial
fraction of heavy elements like lanthanide. If a mechanism, which
ejects matter in a short timescale, is present, a substantial fraction
of neutron-rich matter could be ejected~\cite{SBM19}.
Magnetohydrodynamics effects may be the mechanism for the efficient
mass ejection, but at present, it is not very clear whether this is
indeed the case because the results of the magnetohydrodynamics
simulation depends very strongly on the initial condition of the
magnetic field strength and profile~\cite{FTQFK19} and long-term
magnetohydrodynamics simulations with a variety of the initial
magnetic-field profiles have not yet been performed taking into
account detailed microphysics (equation of state and neutrino
processes).

In the present work, the viscous effect is the driving force for the
mass ejection. In reality, the viscous effects should result from
turbulence induced by magnetohydrodynamics processes in the
astrophysical context. This implies that for more physical modeling of
the mass ejection, we obviously need a magnetohydrodynamics
simulation.  It should be particularly emphasized that the onset time
of the mass ejection essentially determines the neutron richness and
resulting lanthanide fraction of the ejecta. We find that if the onset
time is longer than $\sim 0.3$\,s, the lanthanide synthesis is
significantly suppressed.  Magnetohydrodynamics simulations show that
the early mass ejection is possible if a strong poloidal magnetic
field is present at the formation of the
disk~\cite{SM17,FTQFK18,FTQFK19}. However, it is not clear at all
whether such a magnetic field favorable for the early mass ejection is
present for the remnant disk of neutron-star merger. A 
magnetohydrodynamics simulation from the merger stage throughout the
post merger stage is required.

For enhancing and preserving the turbulence in the disk orbiting a
stellar-mass black hole in magnetohydrodynamics simulations for
seconds, we have to perform a high-resolution non-axisymmetric
simulation.  Such simulations have not been done yet (but see
Refs.~\cite{FTQFK18,FTQFK19}) since the high-resolution radiation
magnetohydrodynamics simulation in general relativity is much more
expensive than that in the axisymmetric viscous hydrodynamics
simulation that we performed in this work.  However, it is not
impossible in the current computational resources to perform a small
number of the simulations. We plan to perform a general relativistic
neutrino-radiation magnetohydrodynamics simulation to examine the
difference between the viscous effect and magnetohydrodynamics effect
and to understand the magnitude of the effective viscous coefficient
in future work.  If we can determine the magnitude of the effective
viscous coefficient in the magnetohydrodynamics simulation, the
results of the viscous hydrodynamics simulations can be used more
robustly for predicting the nucleosynthesis in the ejecta from the
black hole-disk system.

\acknowledgments

We thank Kyohei Kawaguchi and Masaomi Tanaka for helpful discussions.
This work was in part supported by Grant-in-Aid for Scientific
Research (Grant Nos.~JP16H02183, JP16H06342, JP17H01131, JP18H01213,
JP18H04595, JP18H05236, JP19K14720, JP16K17706, JP16H06341, JP17H06363, and JP18K03642) of Japanese MEXT/JSPS.  We
also thank the participants of a workshop entitled ``Nucleosynthesis
and electromagnetic counterparts of neutron-star mergers'' at Yukawa
Institute for Theoretical Physics, Kyoto University (No.~YITP-T-18-06)
for many useful discussions. Numerical computations were performed on
Sakura and Cobra clusters at Max Planck Computing and Data Facility,
on Oakforest-PACS at Information Technology Center of the University
of Tokyo, on XC50 at National Astronomical Observatory of Japan, and
on XC40 at Yukawa Institute for Theoretical Physics.

\appendix

\section{Weak interaction in the disk}

\begin{figure*}
\includegraphics[width=85mm]{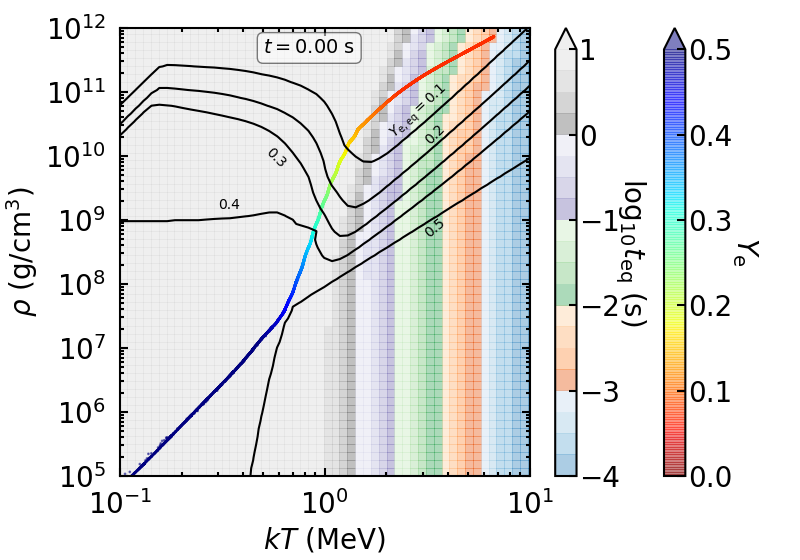}~~~
\includegraphics[width=85mm]{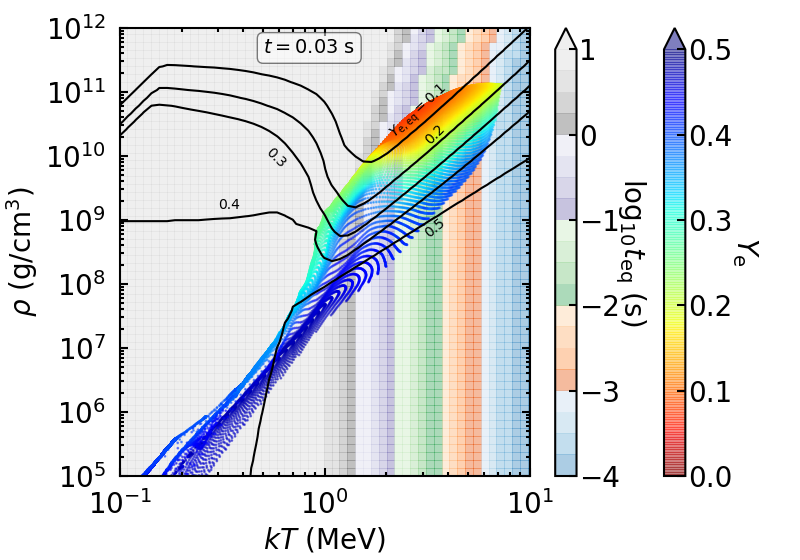} \\
\vspace{3mm}
\includegraphics[width=85mm]{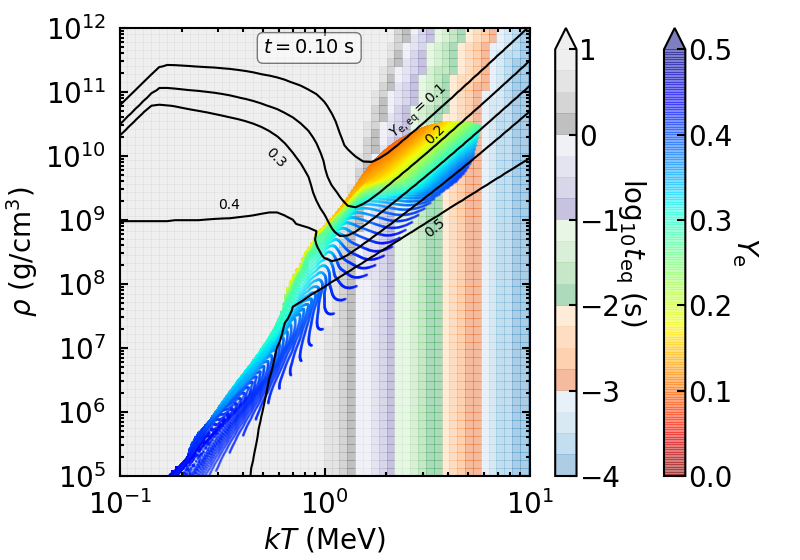}~~~
\includegraphics[width=85mm]{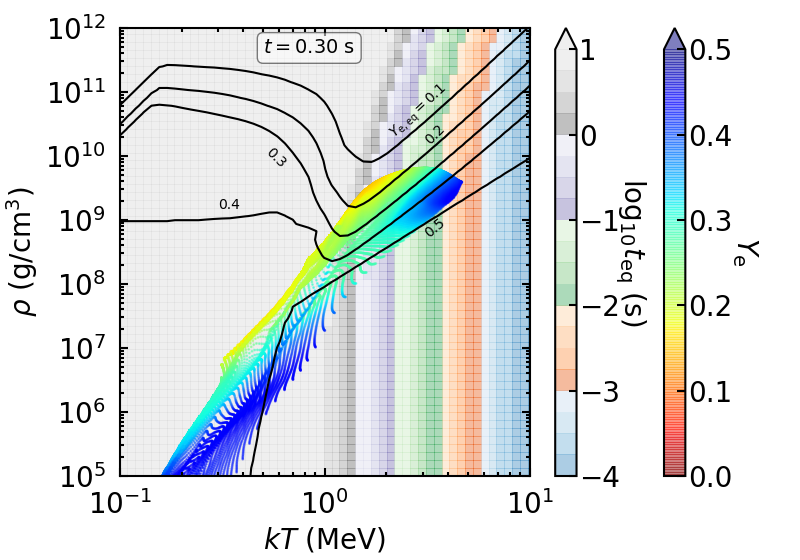}
\caption{Distribution of disk matter in the $\rho$-$T$ plane for
  model K8 at $t=0$ (upper left), 0.03\,s (upper right), 0.1\,s (lower
  left), and 0.3\,s (lower right).  Each point shows the density and
  temperature at a certain grid point in the simulation result.
  $Y_{e}$ at each grid point is displayed in color points.  The black
  curves show the contours of $Y_{e,\mathrm{eq}}$.  Behind the
  $\rho$-$T$ distribution of the simulation, we also plot (by shaded
  color) the weak interaction timescale at each bin, $t_{\rm eq}$,
  which is defined as the inverse of Eq.~\eqref{eq:ec_rate} with
  $Y_{e} = Y_{e,\mathrm{eq}}$.  In the plane, there is a region for
  which $Y_{e,\mathrm{eq}}$ should be higher than
  0.6~\cite{Beloborodov2003a}.  We cannot calculate the
  electron/positron capture rates in such a region because $Y_{e} >
  0.6$ is out of range of the tabulated equation of state, and thus,
  the timescale there is defined as the geometric mean of
  Eqs.~\eqref{eq:ec_rate} and \eqref{eq:pc_rate} with $Y_{e} = 0.6$ .
}
\label{fig:ye_eq}
\end{figure*}

In this Appendix, we describe how the electron fraction in the disk is
determined by the weak interaction processes. Here, we consider the
electron/positron capture on nucleons and nuclei as the major reaction
that determines the electron fraction. We suppose that the neutrino
absorption process plays only a minor role for the evolution of black
hole-disk systems, in particular for the late stage of the disk
evolution, because its luminosity is too low to significantly change
the electron fraction in the disk.

The reaction rates of the electron and positron capture on free
nucleons and heavy nuclei are written as (see, e.g.,~Ref.~\cite{Fuller1985a})
\begin{eqnarray}
\mathcal{R}_\mathrm{ec} &=& X_\mathrm{p}
\lambda_\mathrm{ec}^\mathrm{f} + \frac{X_\mathrm{h}}{\langle A
  \rangle_\mathrm{h}}
\lambda_\mathrm{ec}^\mathrm{h},\label{eq:ec_rate}\\ \mathcal{R}_\mathrm{pc}
&=& X_\mathrm{n} \lambda_\mathrm{pc}^\mathrm{f} +
\frac{X_\mathrm{h}}{\langle A \rangle_\mathrm{h}}
\lambda_\mathrm{pc}^\mathrm{h},\label{eq:pc_rate}
\end{eqnarray}
where $X_\mathrm{n}$, $X_\mathrm{p}$, and $X_\mathrm{h}$ are the mass
fractions of neutron, proton, and heavy nuclei, $\langle
A\rangle_\mathrm{h}$ is the average mass number of heavy nuclei, and
$\lambda_\mathrm{ec/pc}^\mathrm{f/h}$ is the rate of each reaction
which has the unit of (time)$^{-1}$.  Here, the superscripts, ``f" and
``h", indicate the capture on free nucleons and heavy nuclei,
respectively.  The reaction rates are written by the integration with
respect to the energy as~\cite{Beloborodov2003a}
\begin{eqnarray}
\lambda_\mathrm{ec/pc}^\mathrm{f/h} &=& \frac{\ln 2}{\langle ft
  \rangle^\mathrm{f/h} m_{e}^5c^{10}} \int_0^\infty d\omega\,
\omega^2 (\omega-Q_\mathrm{ec/pc}^\mathrm{f/h})^2 \nonumber \\
&&~~~ \times
\sqrt{1-\frac{m_{e}^2c^4}{(\omega-Q_\mathrm{ec/pc}^\mathrm{f/h})^2}}
F_{e}(\omega-Q_\mathrm{ec/pc}^\mathrm{f/h}) \nonumber \\
&&~~~ \times
\Theta(\omega-Q_\mathrm{ec/pc}^\mathrm{f}-m_{e}c^2),\label{eq:rate}
\end{eqnarray}
where $\langle ft \rangle$ is the so-called $ft$-value of these
reactions, $F_{e}$ is the distribution function of electrons,
which is assumed to be the Fermi-Dirac form, and $Q$ is the Q-value of
these reactions. For the capture processes by free nucleons, $\langle
ft \rangle_\mathrm{ec/pc}^\mathrm{f}\approx 1035$ s,
$Q_\mathrm{ec}^\mathrm{f}=(m_\mathrm{p}-m_\mathrm{n})c^2$, and
$Q_\mathrm{pc}^\mathrm{f}=(m_\mathrm{n}-m_\mathrm{p})c^2$.  For the
capture processes by heavy nuclei, we follow the approximations in Ref.~\cite{Fuller1985a}
for $ft$-values and Q-values with
\begin{eqnarray}
Q_\mathrm{ec}^\mathrm{h} &=& \mu_\mathrm{p} - \mu_\mathrm{n},\nonumber\\
Q_\mathrm{pc}^\mathrm{h} &=& \mu_\mathrm{n} - \mu_\mathrm{p},
\end{eqnarray}
and
\begin{eqnarray}
&&\log_{10}\left(\langle ft \rangle_\mathrm{ec/pc}^\mathrm{h}/{\rm s}\right) \nonumber\\
&=&
\begin{cases}
3.2 & (\mbox{unblocked and } \mu_e < Q_\mathrm{ec/pc})~~\\
2.6 & (\mbox{unblocked and } \mu_e > Q_\mathrm{ec/pc})~~\\
2.6 + 25.9/T_9 & (\mbox{blocked}),
\end{cases}
\end{eqnarray}
where $T_9=T/10^9\,{\rm K}$. $\mu_n$ and $\mu_p$ denote the relativistic (including the mass) chemical
potential of neutrons and protons, respectively. Here, ``blocked''
and ``unblocked'' cases imply that $\langle N\rangle \geq 40 $ or
$\langle Z\rangle \leq 20 $ and that $\langle N\rangle < 40 $ and
$\langle Z\rangle > 20 $, respectively. $\langle N \rangle$ and
$\langle Z \rangle$ denote the average neutron and proton numbers 
of the heavy nuclei, respectively. 
Note that we do not consider the updates for the electron/positron capture on heavy nuclei in \cite{Langanke2000a}, which play a minor role in the present case owing to the freeze out of weak interaction at sufficiently high temperature (see below).

The reaction rates are functions of $\rho$, $T$, and $Y_{e}$.  We can
derive the electron fraction in the equilibrium, $Y_{e,\mathrm{eq}}$,
at each density and temperature by equating the electron and positron
capture rates as
\begin{eqnarray}
&&\mathcal{R}_\mathrm{ec}^\mathrm{f}(\rho,T,Y_{e,\mathrm{eq}})
+\mathcal{R}_\mathrm{ec}^\mathrm{h}(\rho,T,Y_{e,\mathrm{eq}}) \nonumber\\
&=&
\mathcal{R}_\mathrm{pc}^\mathrm{f}(\rho,T,Y_{e,\mathrm{eq}})
+\mathcal{R}_\mathrm{pc}^\mathrm{h}(\rho,T,Y_{e,\mathrm{eq}}).
\end{eqnarray}
Here we do not consider the blocking by neutrinos, because it is
important only for the case that the optical depth to neutrinos is
large. Note that the disk material is supposed to be optically thin to
neutrinos except for the very early epoch of the disk evolution (i.e., 
$t \alt 0.1$\,s).

Figure~\ref{fig:ye_eq} plots the contour of $Y_{e,\mathrm{eq}}$ (black
curves) together with the distribution of $Y_{e}$ for the disk
material (color plots) in the $\rho$--$T$ plane at selected time
slices, $t=0$, 0.03, 0.1, and 0.3\,s of the simulation for model K8.
Here, we employ the DD2 equation of state to determine the mass
fractions, average mass number, and chemical potentials in
Eqs.~\eqref{eq:ec_rate}--\eqref{eq:rate} as functions of $\rho$, $T$,
and $Y_e$.  Due to the viscous heating/angular momentum transport and
resulting expansion, the density and temperature in the disk decrease.
On the other hand, the specific entropy increases. Thus, the
distribution of the disk matter moves basically to the bottom left
region gradually.
Figure~\ref{fig:ye_eq} also displays the weak interaction timescale at
each point, $t_{\rm eq}$, which is defined as the inverse of
Eq.~\eqref{eq:ec_rate} with $Y_{e} = Y_{e,\mathrm{eq}}$.

In an early stage of the disk evolution, the
condition of $t > t_{\rm eq}$ is satisfied for the majority of the
disk matter so that the values of $Y_e$ approach $Y_{e,\mathrm{eq}}$.
For $t \alt 0.1$\,s, a part of the disk matter has high density 
with $\agt 10^{10}\,{\rm g/cm^3}$, and hence, 
until $t=0.1$\,s, we still have the material with $Y_e<0.1$.  However,
because of the expansion of the disk, the density goes below $\sim
10^{10}\,{\rm g/cm^3}$ and then the values of $Y_e$ have to be larger 
than $\sim 0.2$ at $t=0.3$\,s.  
Also the increase of the entropy by the viscous heating helps
leptonization of the disk.

It is still possible to eject the low $Y_{e}$ material from the disk
in the presence of the mass ejection mechanism with a short timescale
of $\alt 0.1$\,s.  As discussed in Sec.~\ref{sec3-4}, one possibility
is the Lorentz force by the aligned magnetic field, which is not taken
into account in this work.  If such a mechanism works for $t\lesssim
0.1$\,s, a fraction of the low $Y_{e}$ material may be
ejected. However, for the purely viscous mass ejection with the
timescale $\agt 0.3$\,s, the ejecta cannot have the small values of
$Y_e \alt 0.2$. This agrees totally with the results of our numerical
simulations.

The dependence of $t_{\rm eq}$ on the temperature shows
that for $k T \agt 2$\,MeV, $t_{\rm eq} \alt 0.2$\,s and for $k T
\agt 1.4$\,MeV, $t_{\rm eq} \alt 1$\,s. Thus, when the temperature
decreases below $\sim 2$\,MeV/$k$, the neutrino cooling timescale, which has the same order of magnitude as $t_{\rm eq}$ in the region where the baryon dominates the internal energy,
becomes longer than the viscous timescale, and as a result, the freeze
out of the weak interaction occurs. This effect is also observed in
our numerical results.


\section{Dependence on the disk compactness}

\begin{figure*}[t]
(a)\includegraphics[width=84mm]{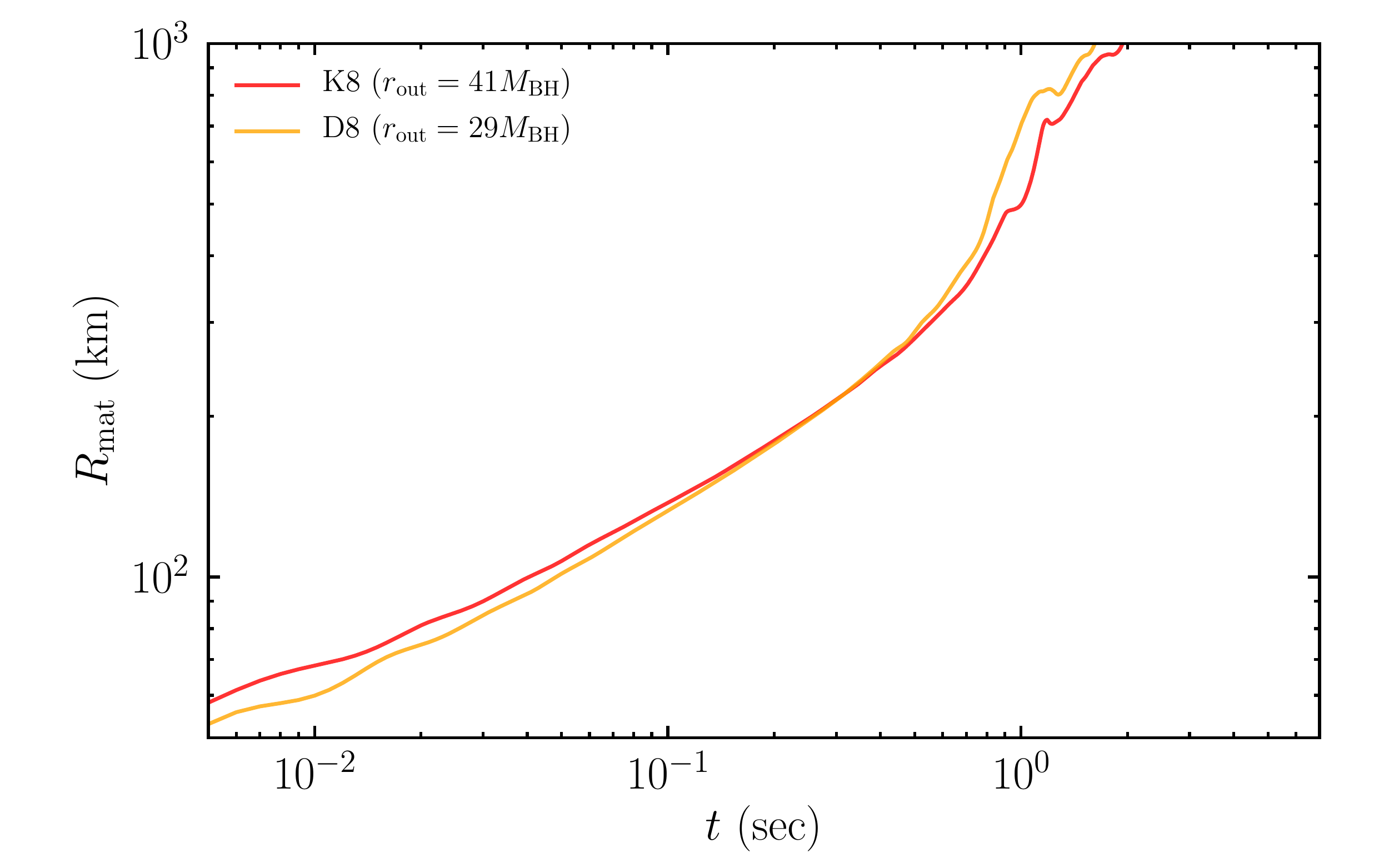}~~
(b)\includegraphics[width=84mm]{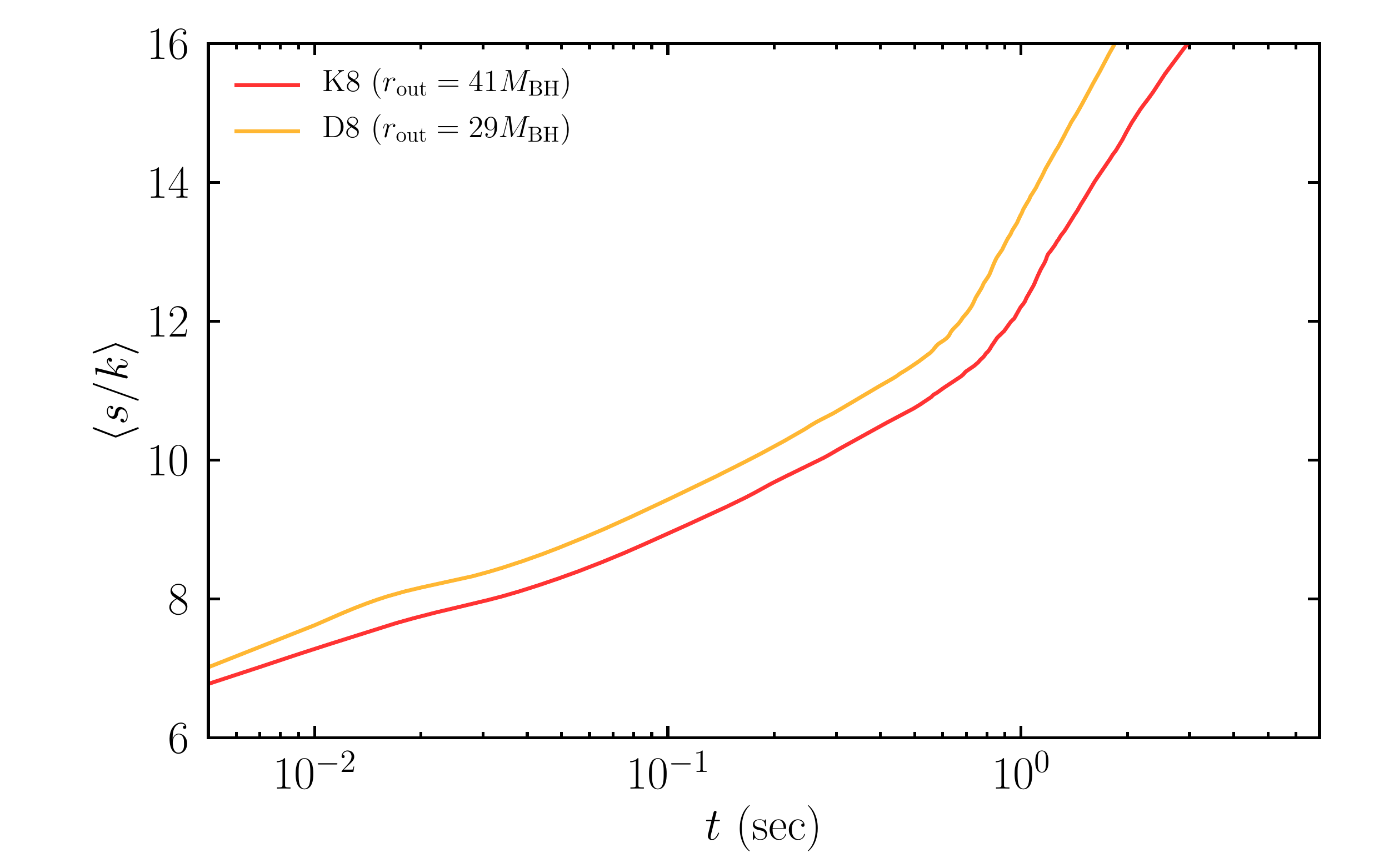} \\
(c)\includegraphics[width=84mm]{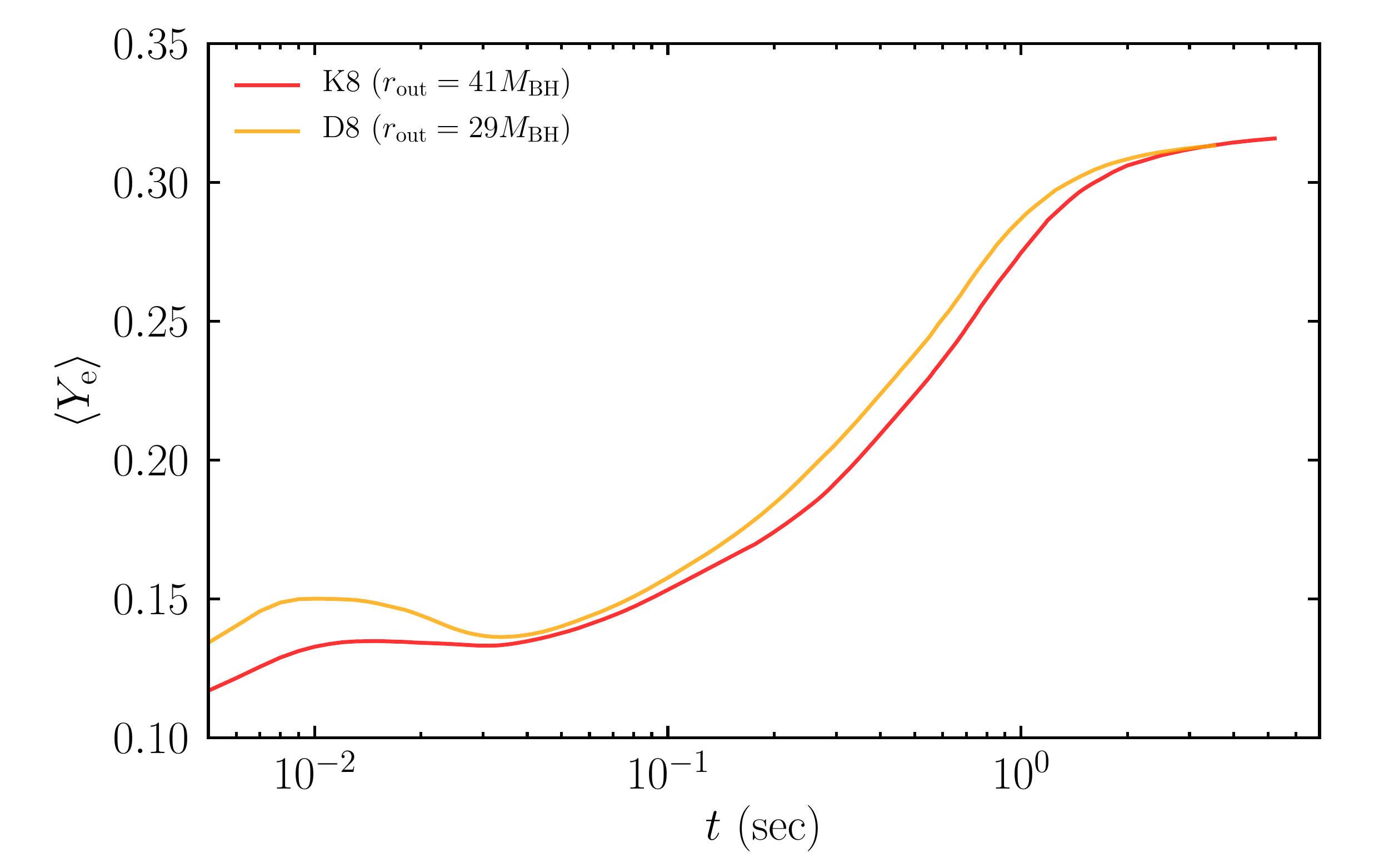}~~
(d)\includegraphics[width=84mm]{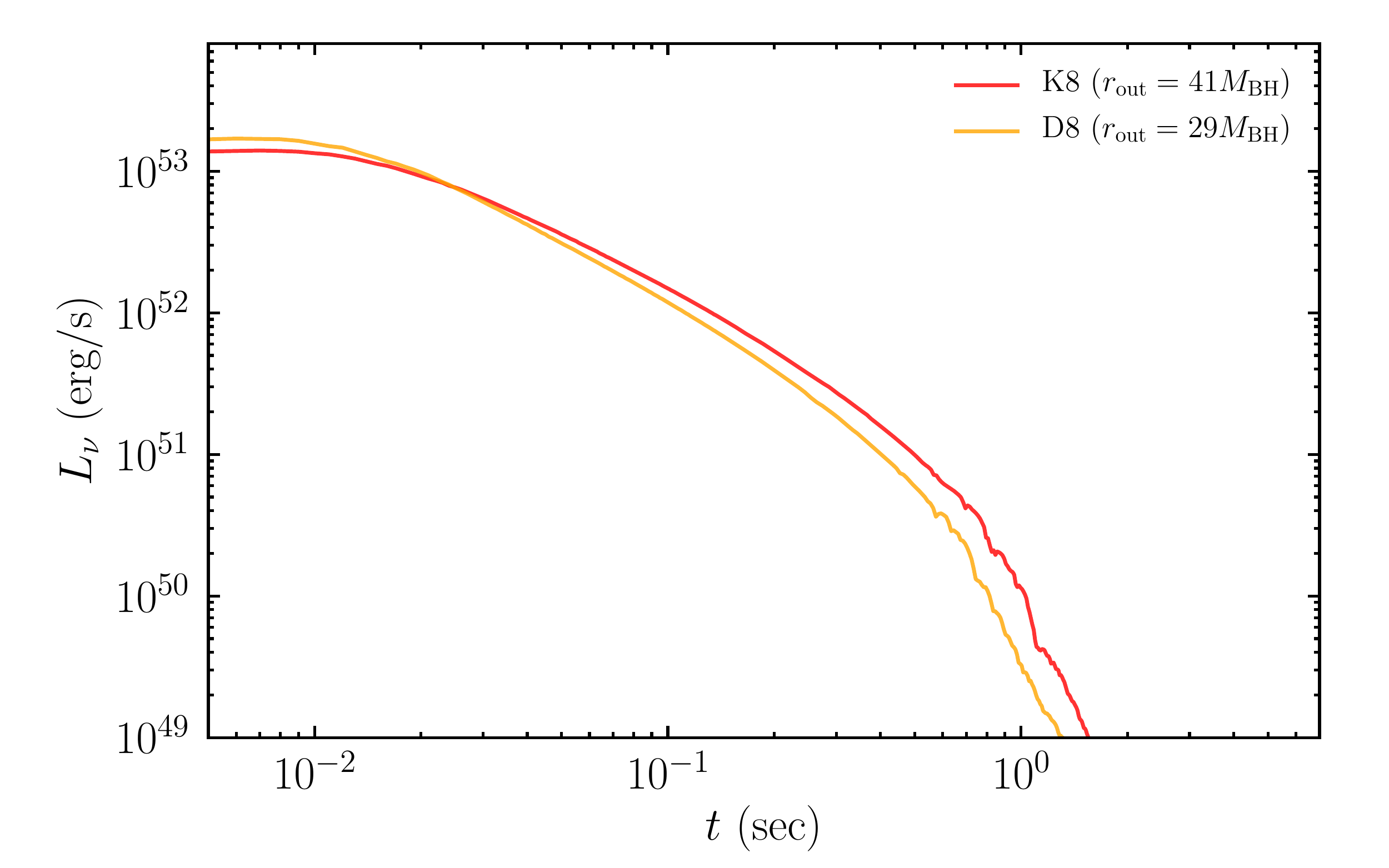} \\
(e)\includegraphics[width=84mm]{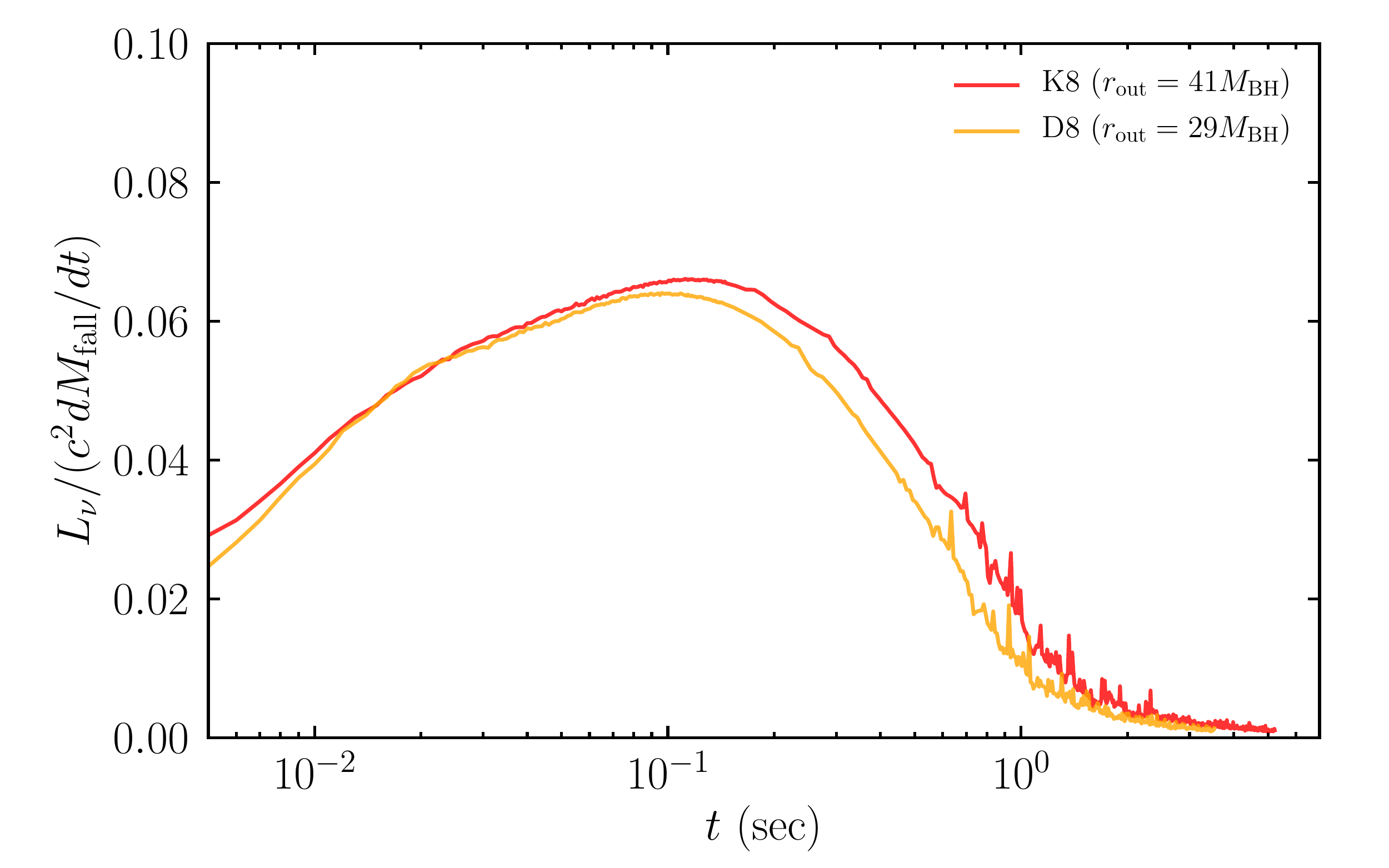}~~
(f)\includegraphics[width=84mm]{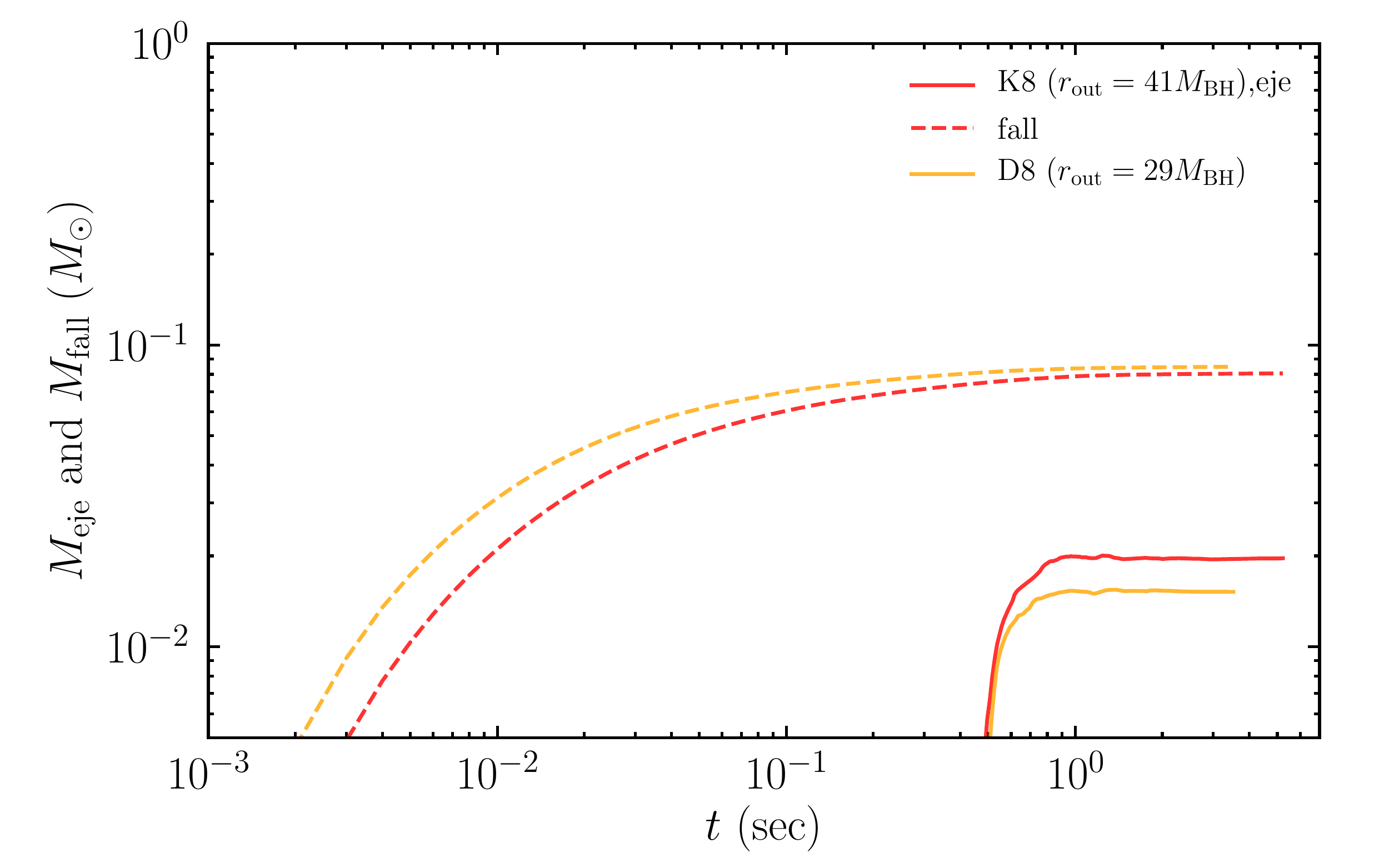} 
\caption{The same as Fig.~\ref{fig9} but for comparison between models K8 and D8. 
\label{figB}}
\end{figure*}

To show the dependence of the viscous evolution on the initial
compactness of the disk, we here compare the results for models K8 and
D8. Figure~\ref{figB} displays the evolution of the same quantities
for the matter located outside the black hole as in Fig.~\ref{fig9}
bur for models K8 and D8.  It is found that the values of $M_{\rm
  fall}$ and $M_{\rm eje}$ for model D8 are larger and smaller than
those for model K8, respectively. For model D8, the ejecta mass is
$\approx 15$\% of the initial disk mass, which is appreciably smaller
than that for models with less compact disk like K8 and Y8 for which 
the ejecta mass is $\agt 20$\% of the initial disk mass (see Table~\ref{table2}).

Because the larger fraction of the disk mass falls into the black hole
in the early stage of the viscous evolution, the disk mass for model
D8 becomes smaller than for model K8 during the long-term viscous
evolution. Due to this reason, the timescale to reach the freeze out
of the weak interaction for model D8 is slightly shorter than for
model K8 (see Sec.~\ref{sec3-3}). However, besides this small
difference, the evolution process for two models is quite similar 
each other. The final average value of $Y_e$ for model D8 is only
slightly smaller than for model K8. As a result, the average value of
$Y_e$ and the mass histogram for the ejecta is not very different
between two models.  The average velocity of the ejecta is also
approximately the same for two models (see Table~\ref{table2}).


\end{document}